\documentclass[journal]{IEEEtran}
\IEEEoverridecommandlockouts
\usepackage{cite}
\usepackage{amsmath,amssymb,amsfonts}
\usepackage{algorithmic}
\usepackage{lipsum,adjustbox}
\usepackage{verbatim}
\usepackage{graphics}
\usepackage{nicefrac}
\usepackage[colorlinks]{hyperref}
\usepackage{stackengine}
\usepackage{pgfplots}
\pgfplotsset{width=7cm,compat=1.8}
\usepackage{multirow}
\usepackage{soul}
\usepackage{amssymb}
\usepackage{amsmath}
\usepackage{algorithm}
\usepackage{algorithmic}
\usepackage{graphicx}
\usepackage[acronym]{glossaries}
\usepackage{subcaption}
\usepackage{textcomp}
\usepackage{booktabs}
\usepackage{stfloats}
\usepackage{scalefnt}
\usepackage{mathrsfs}          % provides math script font \mathttps://www.overleaf.com/project/5f5232de1e41b0000189f930hscr{}
\def\BibTeX{{\rm B\kern-.05em{\sc i\kern-.025em b}\kern-.08em
    T\kern-.1667em\lower.7ex\hbox{E}\kern-.125emX}}
\usepackage[nolist]{acronym}
    
 \usepackage{bm}
\newcommand{\F}{\mathbf{F}}
\newcommand{\Ht}{^{\rm H}}

\newcommand{\figref}[1]{Figure~\ref{#1}}

\DeclareMathOperator{\Tr}{Tr}
\DeclareMathOperator{\SINR}{SINR}

\newcommand{\compl}{\mathbb{C}}         % complex number field, e.g., x \in \compl
\newcommand{\ma}  [1]{ \bm{#1} } % matrix (upper case) and vector (lower case)
\newcommand{\Norm}[1]  { \| #1 \|  }
\newcommand{\kronProd} [2] {#1\otimes #2 }% kronecker product
\newcommand{\set} [1]{{\mathcal {#1}}} % define set
\newenvironment{Delay Minimization}[2][Delay Minimization]{\begin{trivlist}
\item[\hskip \labelsep {\bfseries #1}\hskip \labelsep {\bfseries #2.}]}{\end{trivlist}}

\begin{acronym}
	\acro{OFDM}{orthogonal frequency division multiplexing}
    \acro{UAV}{unmanned aerial vehicle}
    \acro{CSI}{channel state information}
    \acro{SC}{single carrier}
    \acro{OTFS}{orthogonal time frequency space}
    \acro{OCDM}{orthogonal chirp division multiplexing}
    \acro{FER}{frame error rate}
    \acro{CP}{cyclic prefix}
    \acro{ISI}{intersymbol interference}
    \acro{LMMSE-PIC}{linear minimum mean square error parallel interference cancellation}
    \acro{FDD}{frequency division duplex}
    \acro{FD}{full duplex}
    \acro{LTV}{linear time-variant}
    \acro{DOA}{direction of arrival}
    \acro{AGC}{automatic gain control}
    \acro{RF}{radio frequency}
    \acro{LMMSE}{linear minimum mean square error}
    \acro{LNA}{low noise amplifier}
    \acro{PA}{power amplifier}
    \acro{E2E}{end-to-end}
    \acro{ADC}{analog-to-digital}
    \acro{DAC}{digital-to-analog}
    \acro{AI}{artificial intelligence}
    \acro{MIMO}{multiple-input, multiple-output}
    \acro{PAPR}{peak-to-average power ratio}
    \acro{AWGN}{additive white Gaussian noise}
    \acro{MLD}{maximum likelihood detection}
    \acro{NLSE}{stochastic nonlinear Schr\"odinger equation}
    \acro{NFT}{nonlinear Fourier transform}
    \acro{FT}{Fourier transform}
    \acro{MMSE}{minimum mean square error}
    \acro{eMBB}{enhanced mobile broadband}
    \acro{URLLC}{ultra-reliable low latency communications}
    \acro{mMTC}{massive machine type communications}
    \acro{MIMO}{multiple-input multiple-output}
    \acro{KPI}{key performance indicator}
    \acro{D2D}{device-to-device}
    \acro{NFV}{network functions virtualization}
    \acro{MEC}{mobile edge computing}
    \acro{SDN}{software-defined networking}
    \acro{UCBC}{uplink centric broadband communication}
    \acro{RTBC}{real-time broadband communication}
    \acro{HCS}{harmonized communication and sensing}
     \acro{OLED}{organic light-emitting diode}
    \acro{AVC}{advanced video coding}
    \acro{DSP}{digital signal processing}
    \acro{CPU}{central processing unit}
    \acro{GPU}{graphics processing unit}
    \acro{TPU}{tensor processing unit}
    \acro{FPGA}{field programmable gate array}
    \acro{ASIC}{application-specific integrated circuit}
    \acro{PU}{processing unit}
    \acro{NU}{network unit}
    \acro{IoT}{Internet of Things}
    \acro{I/O}{input/output}
    \acro{DL}{deep learning}
    \acro{BM}{block-multiplexed}
    \acro{WHT}{Walsh-Hadamard transform}
    \acro{SWH}{sparse Walsh-Hadamard}
    \acro{FFT}{fast Fourier transform}
    \acro{ISAC}{integrated sensing and communications}
    \acro{DFRC}{dual-functional radar and communication}
    \acro{DoA}{direction of arrival}
    \acro{LDPC}{low-density parity-check code}
    \acro{BCC}{binary convolutional code}
    \acro{ToA}{time of arrival}
    \acro{RIS}{reconfigurable intelligent surface}
    \acro{SINR}{signal-to-interference-noise ratio}
    \acro{ML}{machine learning}
    \acro{XR}{extended reality}
    \acro{VR}{virtual reality}
    \acro{AR}{augmented reality}
    \acro{OAM}{orbital angular momentum}
    \acro{SAM}{spin angular momentum}
    \acro{EIT}{electromagnetic information theory}
    \acro{IFFT}{inverse fast Fourier transform}
    \acro{QAM}{quadrature amplitude modulation}
    \acro{ANN}{artificial neural network}
    \acro{RMT}{random matrix theory}
    \acro{URLLC}{ultra-reliable low-latency communication}
    \acro{DML}{distributed ML}
    \acro{SNR}{signal-to-noise ratio}
    \acro{AE}{autoencoder}
    \acro{TBM}{Tensor-based modulation}
    \acro{BCH}{Bose–Chaudhuri–Hocquenghem}
    \acro{QoS}{quality-of-service}
    \acro{QoE}{quality-of-experience}
    \acro{LDPC}{low-density parity-check}
    \acro{TDD}{time division duplexing}
    \acro{TR}{time reversal}
\end{acronym}

\usepackage[utf8]{inputenc}
\usepackage[english]{babel}
\usepackage{amsthm}
\usepackage{commath}
\usepackage{subcaption}

\begin{document}

\title{
Twelve Scientific Challenges for 6G: \\ Rethinking the Foundations of Communications Theory   
\thanks{
%Authors acknowledge the \dots for the support of the project.
M. Chafii is with the Engineering Division, New York University (NYU) Abu Dhabi, 129188, UAE and NYU WIRELESS, NYU Tandon School of Engineering, Brooklyn, 11201, NY (e-mail: marwa.chafii@nyu.edu).} 
\thanks{L. Bariah is with the Technology Innovation Institute, 9639 Masdar City, Abu Dhabi, UAE, and with the Electrical and Computer Engineering Department, University at Albany, Albany, NY 12222 USA (e-mail:lina.bariah@ieee.org).}
\thanks{S. Muhaidat is with the KU Center for Cyber-Physical Systems, Department of Electrical Engineering and Computer Science, Khalifa University, Abu Dhabi 127788, UAE, and also with the Department of Systems and Computer Engineering, Carleton University, Ottawa, ON K1S 5B6, Canada (e-mail: muhaidat@ieee.org ).}
\thanks{M. Debbah is with the Technology Innovation Institute, 9639 Masdar City, Abu Dhabi, UAE and also with Centrale Supelec, University Paris-Saclay, 91192 Gif-sur-Yvette, France (e-mail:merouane.debbah@tii.ae).
}
}

 \author{Marwa Chafii,~\IEEEmembership{Member,~IEEE,}
         Lina Bariah,~\IEEEmembership{Senior Member,~IEEE,}
         Sami Muhaidat,~\IEEEmembership{Senior Member,~IEEE,}
         and~Merouane Debbah,~\IEEEmembership{Fellow,~IEEE}}

\maketitle

\begin{abstract}
The research in the sixth generation of wireless networks needs to tackle new challenges in order to meet the requirements of emerging applications in terms of high data rate, low latency, high reliability, and massive connectivity. To this end, the entire communication chain needs to be optimized,  including the channel and the surrounding environment, as it is no longer sufficient  to control  the transmitter and/or the receiver only. Investigating large intelligent surfaces, ultra-massive multiple-input-multiple-output, and smart constructive environments will ultimately contribute to this direction. In addition, to allow the exchange of high dimensional sensing data between connected intelligent devices, semantic and goal-oriented communications need to be considered for a more efficient and context-aware information encoding. In particular, for multi-agent systems, where agents are collaborating together to achieve a complex task, emergent communication, instead of hard-coded communication, can be learned for more efficient task execution and communication resources use. 
Moreover, new physical phenomena,  such as the thermodynamics of communication and the interaction between information theory and electromagnetism should be exploited to better understand the physical limitations of different technologies, e.g., holographic communications.
Another new communication paradigm is to consider the end-to-end communication system optimization instead of block-by-block optimization, which requires exploiting machine learning theory, non-linear signal processing theory, and non-coherent communications theory. Within this context, we identify and investigate twelve scientific challenges for rebuilding the theoretical foundation of communications. Furthermore, we present an overview of each of the challenges, along with their respective research opportunities and associated challenges.
\end{abstract}

\begin{table*}
\caption{List of Acronyms} 
\begin{tabular}{llll} 
 \hline
 
 ADC & Analog-to-Digital &  LTV & Linear Time-Variant\\ 
 AE & Auto-Encoder & MEC & Mobile Edge Computing\\ 
 AGC & Automatic Gain Control & MIMO & Multiple-Input Multiple-Output \\ 
 AI & Artificial Intelligence &  ML & Machine Learning\\ 
 ANN & Artificial Neural Network &  MLD & Maximum Likelihood Detection\\ 
 AR & Augmented Reality &  MMSE & Minimum Mean Squared Error\\
 ASIC & Application-Specific Integrated Circuit &  mMTC & massive Machine Type Communications\\
 AVC & Advanced Video Coding & NFT & Nonlinear Fourier Transform \\
 AWGN & Additive White Gaussian Noise &  NFV & Network Functions Virtualization \\
 BCC & Binary Convolutional Code & NLSE & Stochastic Nonlinear Schr\"odinger Equation \\
BCH & Bose–Chaudhuri–Hocquenghem & NU & Network Unit \\
 BM & Block-Multiplexed & OAM & Orbital Angular Momentum \\
 CP & Cyclic Prefix & OCDM & Orthogonal Chirp Division Multiplexing\\
 CPU & Central Processing Unit & OFDM & Orthogonal Frequency Division Multiplexing \\
 CSI & Channel State Information & OLED & Organic Light-Emitting Diode \\
 D2D & Device-to-Device & FD & Full Duplex \\
 DAC & Digital-to-Analog & OTFS & Orthogonal Time Frequency Space \\
 DFRC & Dual-Functional Radar and Communication & PA & Power Amplifier \\
 DL & Deep Learning & PAPR & Peak-to-Average Power Ratio \\
 DML & Distributed ML & PU & Processing Unit \\
 DOA & Direction of Arrival & QAM & Quadrature Amplitude Modulation \\
 DSP & Digital Signal Processing & QoE & Quality-of-Experience \\
 E2E & End-to-End & QoS & Quality-of-Service \\
 EIT & Electromagnetic Information Theory & RF & Radio Frequency \\
 eMBB & enhanced Mobile Broadband & RIS & Reconfigurable Intelligent Surface\\
 FDD & Frequency Division Duplex &  RMT & Random Matrix Theory \\ 
 FER & Frame Error Rate & RTBC & Real-Time Broadband Communication \\
 FFT & Fast Fourier Transform & SAM & Spin Angular Momentum \\ 
 FPGA & Field Programmable Gate Array & SC & Single Carrier \\
 FT & Fourier Transform & SDN & Software-Defined Networking \\
 GPU & Graphics Processing Unit & SNR & Signal-to-Noise Ratio \\
% HCS & Harmonized Communication and Sensing & SINR & Signal-to-Interference-Noise Ratio \\
 IFFT & Inverse Fast Fourier Transform & SWH & Sparse Walsh-Hadamard \\
 I/O & Input/Output & TBM & Tensor-Based Modulation \\
 IoT & Internet-of-Things & ToA & Time of Arrival \\
 ISAC & Integrated Sensing and Communications & TPU & Tensor Processing Unit \\
 ISI & Intersymbol Interference & TR & Time Reversal  \\
 KPI & Key Performance Indicator & UAV & Unmanned Aerial Vehicle \\
 LDPC & Low-Density Parity-Check Code & UCBC & Uplink Centric Broadband Communication \\
 LMMSE & Linear Minimum Mean Squared Error & URLLC & Ultra-Reliable Low-Latency Communications \\
 LMMSE-PIC & Linear Minimum Mean Squared Error Parallel Interference Cancellation & VR & Virtual Reality \\
 LNA & Low Noise Amplifier & WHT & Walsh-Hadamard Transform \\
 & & XR & Extended Reality  \\
\hline
\end{tabular}
\end{table*}

\section{Introduction} \label{introduction}

%\hl{Marwa}

\IEEEPARstart{W}{hile} the deployment of 5G networks is still ongoing globally, the research has started to shape the speculated vision of 6G, by exploring  new emerging applications and services, defining new requirements, and identifying disruptive  enabling technologies.  According to the conducted research and the current deployment \cite{tong20226g}, three main paradigms have been identified as the pillars of 5G networks, namely, (i) \ac{eMBB}, which includes services that require fast connections and high data rates, such as video streaming applications and mobile augmented reality. In this regard, 5G networks are anticipated to guarantee a peak data rate of 20 Gbps in the downlink and 10 Gbps in the uplink. Moreover, in mobility scenarios, users experience an average data rate of 100 Mbps in the downlink and 50 Mbps in the uplink; (ii) \ac{URLLC}, which targets 1 ms latency and supports  mission-critical applications, such as autonomous driving and remote robotic surgery; (iii) \ac{mMTC}, which envisions the deployment of 1 Million devices per $\text{km}^2$, and requires low-cost, low-power and long-range devices. 

In order to meet the 5G \acp{KPI} for the different cited use cases, several technologies have been developed, which include  massive \ac{MIMO}, mmWave frequencies, \ac{NFV} which relies on replacing network hardware with virtual machines, \ac{MEC} to process in real-time large amounts of data produced by edge devices, \ac{D2D} communications, and \ac{SDN}. Today, 5.5G is already on the road, and new use cases have been introduced that combine features of the original 5G use cases; (i) \ac{UCBC} which refers to massive things with broadband abilities such as HD video uploading and machine vision; (ii) \ac{RTBC} which combines broadband features with high reliability such as extended reality applications and holograms; (iii) \ac{ISAC} which integrates both communications and sensing capabilities in applications like positioning, spectroscopy, and imaging.

Owing to this, it is difficult to predict exactly what 6G will be, but the research community seems to agree that 6G will be the seed for enabling extremely immersive experiences, haptics, industry 4.0 with connected intelligence, 3D full coverage of the earth, and native \ac{AI}-empowered wireless communication. Furthermore, fog computing enjoys several advantages that can be leveraged to complement edge and cloud computing in 6G. First, for scenarios where edge devices lack the needed storage and computing capabilities, fog devices can play an important role in bringing additional storage in close proximity to edge devices, compared to cloud storage. By doing so, communication overhead, security, and privacy can be maintained at a desirable level. Furthermore, fog computing paves the way for enabling distributed systems, by overcoming the limitations experienced at centralized, cloud-based systems. This includes increased latency, resource drainage, and compromised privacy \cite{ji2021survey}. As analyzed in \cite{tong20226g}, such applications necessitate the development of novel schemes to achieve peak data rates of 1 Tbps, reliability of $99.99999\%$, the position accuracy of $50$cm outdoor and $1$ cm indoor, device density of $10$ million/$\text{km}^2$, 20 years of sensing battery life, air-interface latency of $0.1$ ms and network coverage of 167 dB. Such stringent \acp{KPI} impose further challenges on future wireless networks, where traditional schemes will fail to meet the needed performance. This requires a radical departure from conventional communication paradigms to more innovative methods and motivates the research community to approach the enhancement of wireless networks from different perspectives by revisiting the fundamentals of communication theory.

%The new communications paradigms
%\begin{itemize}
%	\item Optimize the environment and not only the Transmitter and the Receiver
%	\begin{itemize}
%		\item Smart Environments, Ultra Massive MIMO, Large Intelligent Surfaces, etc
%	\end{itemize}
%\item Memory is key in the model and exploit it to build better communication systems
%\begin{itemize}
%	\item Semantic Communication, Distributed AI, Multi agent Wireless Systems
%\end{itemize}
%\item Exploit new physics phenomenon's
%\begin{itemize}
%	\item OAM, Bessel beams, holographic MIMO, thermodynamics of communication
%\end{itemize}
%\item Do not compensate wireless discrepancies but exploit them!
%\begin{itemize}
%	\item Sparse communications, non coherent communication, non linear signal processing theory
%\end{itemize}
%\end{itemize}

\subsection{Surveys on 6G in the Literature}
There are several papers available in the literature on visions, requirements, and use cases of 6G. In \cite{jiang2021road}, the authors present a comprehensive discussion on the state-of-the-art of 6G networks. Furthermore, they investigated the key drivers that motivated the need for beyond 5G by identifying the emerging use cases and scenarios, including ubiquitous mobile broadband, ultra-reliable low-latency broadband communication, and massive ultra-reliable low-latency communication. In addition to the key requirements, they discussed in detail enabling technologies, focusing on new air interfaces, new spectrum, and new network architectures. On the other hand, the survey in \cite{alsabah20216g} studies the future 6G networks from enabling technology perspective, including massive \ac{MIMO}, holographic radio communication, intelligent surfaces, multi-access and modulation schemes, and backscatter communication, to name a few. The paper \cite{uusitalo20216g} presents the 6G vision of the European 6G Flagship project Hexa-X, from a joint industry and academic perspective. The authors in \cite{uusitalo20216g} further discussed the standardization activities taking place over physical, network, and link layers. The work in \cite{tomkos2020toward} focuses on the \ac{IoT} and \ac{AI} technologies for 6G networks. In specific, the authors discussed the evolution of mobile edge computing into \ac{AI} at the edge. They also shed light on distributed and federated intelligence between 5G and 6G. The survey in \cite{tataria20216g} presents a top-down approach to 6G starting from the motivations and societal changes that drive the need for 6G, to the 6G applications and their technical requirements. In specific, they thoroughly studied various use cases for different paradigms, including, network and computing convergence, holographic communication, tactile internet and haptics, and space-terrestrial integrated communication. Six challenges that include using subterahertz spectrum, semiconductor technologies, and transceiver designs for high data rate and low latency have been studied in \cite{tataria2022six}, and some practical solutions to realize these challenges have been presented. The paper in \cite{bhat20216g} introduces use cases and perspectives that drive the vision of 6G. Also, in \cite{ji2021several}, the authors articulate the key characteristics of 6G networks, and provide a thorough discussion on the key differences between 6G and 5G in terms of reliability, throughput, energy, latency, etc. The authors further discuss the main 6G technologies that are identified as enablers for 6G, including terahertz communication, reconfigurable intelligent surfaces, and blockchain. For each technology, the authors elaborated on its key characteristics, applications, and limitations. Table~\ref{tab:soa} provides an overview of the most important surveys on 6G in the literature and their approach and focus on dealing with 6G vision.

\begin{table}
\caption{State-of-the-art Surveys} \label{tab:soa}
\begin{tabular}{ | c | p{25em} |  } 
  \hline
  Reference & Topics covered\\
  \hline 
  \cite{jiang2021road} & \begin{itemize}
      \item Key driving factors of 6G.
      \item Technical requirements of 6G and envisioned KPIs.
      \item Summary of specification and standardization activities.
  \end{itemize}  \\ 
  \hline
  \cite{alsabah20216g} & \begin{itemize}
      \item Deep discussion on the key enabling technologies of 6G: massive MIMO, RIS, Holographic-type communications, etc. 
  \end{itemize}  \\ 
  \hline
  \cite{uusitalo20216g} & \begin{itemize}
      \item 6G vision from an industrial point of view.
      \item Standardization and regulation initiatives on 6G.
      \item Detailed specification of identified 6G use-cases by academia and industry.
      \item Key 6G research challenges: network protocols, intelligence, trustworthiness, and sustainability.
  \end{itemize}  \\ 
  \hline
  \cite{tomkos2020toward} & \begin{itemize}
      \item ML for communication and communication for 6G.
      \item Mobile-edge computing to edge intelligence.
      \item Distributed and federated learning in 6G.
  \end{itemize} \\ 
  \hline
  \cite{tataria20216g} & \begin{itemize}
      \item Societal and environmental impact on future wireless generations: Holographic Society, ubiquitous intelligence, time-sensitive applications.
      \item 6G use cases with the corresponding technical requirements.
  \end{itemize}  \\ 
  \hline
  \cite{tataria2022six} & \begin{itemize}
      \item Critical challenges of 6G networks with emphasis on the terahertz communication: 1) Subterahertz frequencies, 2) advancements in semiconductor technologies, 3) integrated transceiver design at high frequencies, 4) achieving Tbps rates, 5) submillisecond latency, and 6) backward compatibility.
Rates
  \end{itemize}  \\ 
  \hline
  \cite{bhat20216g} & \begin{itemize}
      \item 6G trends.
      \item 6G use-cases and KPIs.
      \item 6G business model.
      \item Sustainability in 6G networks.
      \item Edge intelligence.
  \end{itemize}  \\ 
  \hline
  \cite{ji2021several} & \begin{itemize}
      \item Key characteristics of 6G, and the main differences between 5G and 6G.
      \item Enabling technologies for 6G: Terahertz communication, reconfigurable intelligent surface, and blockchain.
      \item Applications, characteristics, and limitations of considered technologies.
  \end{itemize}  \\ 
  \hline
  \hline
\end{tabular}
\end{table}

Different than existing articles that tackle the 6G vision from applications, use cases and technologies perspectives, in this survey, we lay down the foundation for rethinking the communication theories, which have underpinned earlier wireless generations from 1G to 5G. In particular, we dig deeply into the critical scientific challenges that need to be addressed in order to characterize the fundamental limits of what 6G can achieve, and provides the roadmap to answering the essential question of what will 6G be?
\subsection{Organization}
In this article, we aim to dive deep into the fundamentals of wireless communications systems and identify the roots of the highlighted challenges anticipated to be experienced in future wireless generations. In particular, we cover the following scientific challenges needed to revisit the foundations of communication theory:
\begin{enumerate}
    \item Electromagnetic information theory: In Section~\ref{sec:electromagnetic}, we discuss the limitations imposed by the laws of electromagnetism on the Shannon information capacity and the number of degrees of freedom.  We further discuss several cutting-edge technologies that are envisioned to realize the full potential of 6G. 
    \item Non-linear signal processing: in Section~\ref{sec:nonlinear}, we explain the source of nonlinearities in communications, the limitations of the linear model, and the need of designing signal processing algorithms beyond the linear assumption. We also discuss possible approaches in modeling the non-linearities and designing non-linear transceivers, either based on non-linear signal processing tools such as non-linear Fourier transform or an end-to-end approach using machine learning algorithms. We discuss challenges and open questions for each approach.
    \item Multi-agent learning systems: in Section~\ref{sec:multiagent}, we overview the drawbacks of conventional centralized learning systems, and we point out the motivation behind the need for distributed/federated learning architecture. We explore the distributed AI theory, with emphasis on multi-agent systems, its advantages, and shortcomings of such a learning scheme on the individual node as well as the system-wide levels. We also overview the concept of emergent communications in multi-agent systems and discuss the main challenges of this emerging topic.
    \item Super-resolution theory: in Section~\ref{sec:resolution}, we well-define the task of super-resolution in reconstructing the fine details of an observation from coarse-scale information. We review the classical methods of addressing this problem and show their limitations. We then focus on the frequency response of the multipath channel and provide the limitations of state-of-the-art solutions in estimating the different channel parameters, especially when ignoring the noise and non-linearity effects. We then explain some challenges and research opportunities such as considering realistic assumptions and fusing different measurements at different bands and from different sensors.
    \item Thermodynamics of computation and communication: the thermodynamics of communication and computation theory may help the development of novel algorithms for energy-efficient 6G networks. In Section~\ref{sec:thermo},  besides investigating the energy required for information processing, we also describe how it is essential to understand the fundamental energy limits of information processing and how they are connected to the fundamental limits of communication performance. For example, what is the optimal trade-off between communication performance and network energy consumption?
    \item Signals for time-varying systems theory: in Section~\ref{sec:varying}, we explain how signals should be designed in time-varying systems, in particular doubly selective wireless channels. We review the proposed waveforms in the literature within the general framework of spreading waveforms that spread the symbol energy in the domain of selectivity of the channel. We then explain the limitations and challenges of spreading waveforms in time-varying systems and provide future research directions.
    \item Semantic communication theory: Section~\ref{sec:semantic} digs deeply into Shannon's information theory, and rethinks the classical communication theory, in which the meaning and the semantic aspect of transmitted messages are ignored. We further identify the limitation of such classical definitions in the era of native-AI, where machines enjoy a considerable level of intelligence. Also, we thoroughly discuss the fundamentals of semantic communications and its role in future 6G networks, redefining the concept of information structure. 
    \item Signals for integrated sensing and communication: in Section~\ref{sec:sensing}, we first explain the benefits of the integration of sensing and communications, and then we focus on waveform design challenges for different joint communications and radar scenarios e.g. monostatic active radar, bistatic active and passive radar. Then we describe other challenges beyond the waveform design, such as performance indicators, channel modeling, and representation and sharing of sensing data.
    \item Large-scale communication theory: in Section~\ref{sec:large}, we shed light on the challenges anticipated to be experienced in current and future wireless networks, which are attributed to the modeling and optimization of large-scale networks. We further explore the limitations of current computer simulation and artificial neural network-based approaches. Subsequently, we discuss three major tools that have manifested themselves as promising, generalized yet resilient methods for modeling, analyzing, and optimizing large-scale dynamic wireless networks. This includes random matrix theory, decentralized stochastic optimization, tensor algebra, and low-rank tensor decomposition.
   \item Non-equilibrium information theory: information theory deals with fundamental limits of communication (or information processing in general), and it allows to development of strategies to achieve these limits. These fundamental limits may then be used to assess the performance of actual communication systems, and the strategies to construct frameworks for the system design. Hidden in the mathematical proofs is the assumption of infinite-length sequences, such that asymptotic techniques can be applied. Real-world applications, however, operate on finite lengths. In this finite-length regime, classic information-theoretic results still apply as ultimate limits but they are usually not achievable, and thus bounds may have little value. In recent years, finite-length information has been developed, aiming at fundamental information-theoretic limits (and strategies) subject to finite-length constraints. In Section~\ref{sec:nonequilibirum}, we review the main concepts related to Non-equilibrium Information Theory.
    \item Combining queuing and information theory: in Section~\ref{sec:queuing}, we shed light on the impact of information theory on networking.  In particular, we  review several topics related to communication networks with queuing and information-theoretic aspects, including multi-access protocols and effective bandwidth of bursty traffic. We also discuss how layering can be formulated based on different optimization techniques including the optimization decomposition for systematic network design. 
    \item Non-coherent communication theory: in Section~\ref{sec:noncoherent}, We revisit the challenge of channel state information acquisition within the context of large-scale networks, and emphasize the shortcomings of coherent and blind-based approaches in achieving the needed reliability, latency, and spectral and energy efficiency requirements of such networks. Also, we will reveal the role of Grassmannian modulation and non-coherent tensor modulation as potential candidates for fulfilling the massive-scale network vision of 6G.
\end{enumerate}

%\textcolor{blue}{maybe we should add something like: unlike other surveys (cite them), we don't focus on the applications/use case but on the scientific challenges...?}

%\textbf{Notations}: Throughout the paper, vectors are defined with lowercase bold symbols $\ma{x}$ whose $k$-th element is $\ma{x}[k]$. Time and frequency domain vectors are represented by $\ma{x}$ and $\tilde{\ma{x}}$ respectively. Matrices are written as uppercase bold symbols $\ma{X}$. $\Ex{.}$ is the expectation operator. The trace of a square matrix $\ma{X}$ is $\trace{\ma{X}}$. The notation $\odot$ and $\oslash$ denote the  element-wise multiplication and division operations, respectively. Finally. the pseudo inverse and conjugate matrices of $\ma{X}$ are denoted by $\ma{X}^{\dagger}$ and $\ma{X}^{\text{H}}$ respectively. 

\section{Synergy between 6G Scientific Challenges}
\begin{figure*}[ht]
\centering
  \includegraphics[width=0.9\textwidth,height=18cm]{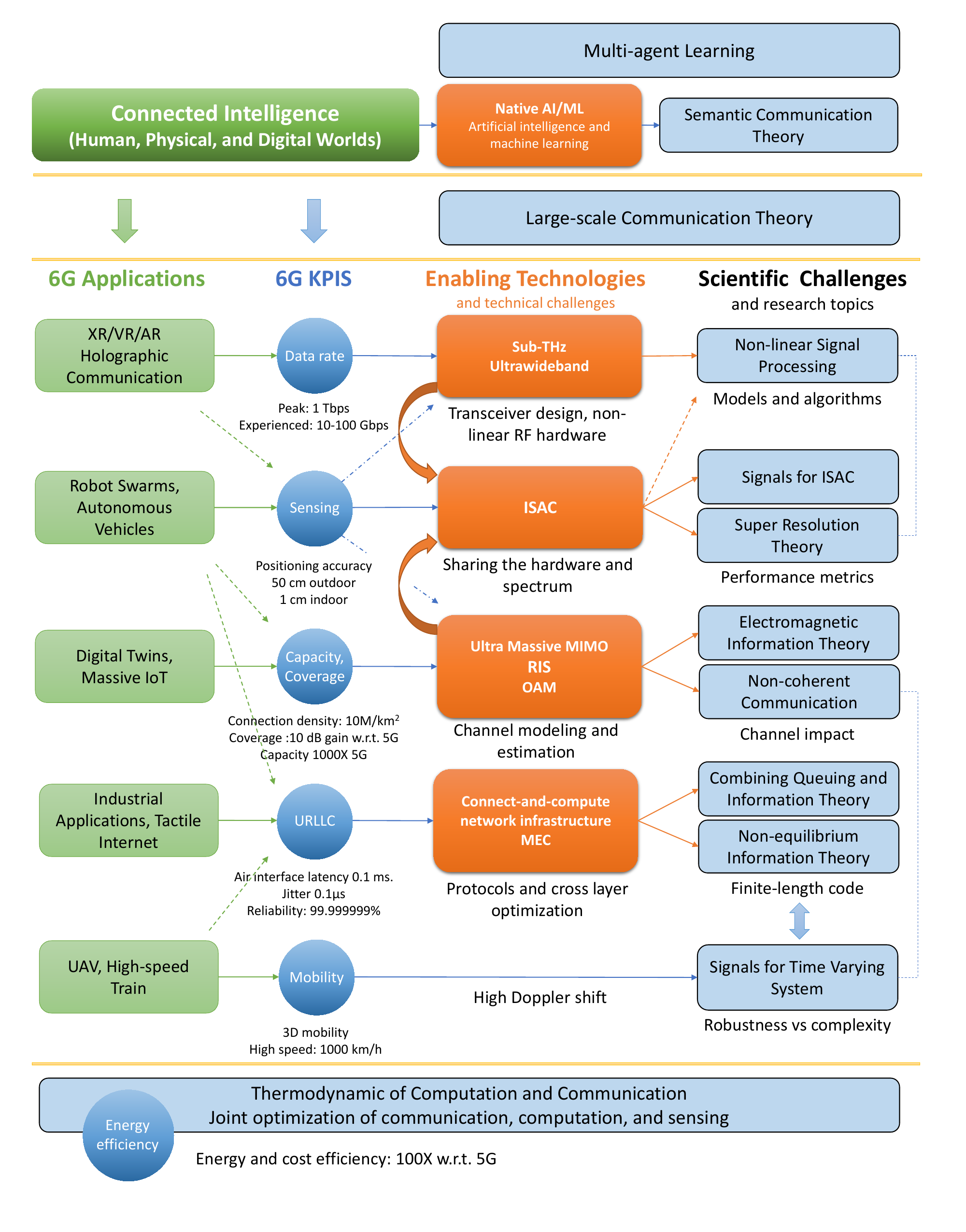}
  \caption{Synergy between 6G Scientific Challenges.}
  \label{fig:synergy}
\end{figure*}
Figure~\ref{fig:synergy} provides an overview of the synergy between the scientific challenges discussed in this paper, and 6G KPIs, applications, and enabling technologies.

It is envisioned that 6G will provide connectivity and computation to interconnect the three worlds, digital, physical, and human, under the term connected intelligence, where different types of sensors, actuators, and computation platforms are interconnected through a high-performance network, that can deliver ultra-high data rate, low-latency, high reliability, and connect all things. Moreover, the 6G system will natively integrate \ac{AI} and \ac{ML} techniques, that are exploited at different levels, for application processing, network optimization, and transmission techniques. In addition to the communication and computation services, 6G will integrate sensing as a service to exploit the interaction between radio waves and the environment. Accordingly, the 6G system vision allows the realization of a wide range of applications with stringent requirements that introduce new KPIs, which cannot be fulfilled with 5G systems and their updated releases. Achieving these KPIs demands new disruptive technologies, that come with technical challenges and limitations. In addition, research efforts are required to understand the limitations and develop solutions to overcome the challenges and push the performance to the limits. Although the concepts and related problems are well-known in many cases, achieving satisfactory solutions is challenging. 

In order to understand the technical and scientific challenges, we provide a high-level overview of the interrelation between different elements involved in the 6G system development. First, starting from the vision of connected intelligence applications and native AI as a technology at the top level and assuming that multiple smart entities are connected, the first question is how should these entities talk to each other in a smart and efficient way. This leads to the concept of semantic communication. Semantics and new languages can emerge from the interaction of multi-agents, that collect a massive amount of sensing data, exchange its representation with other agents, take actions, and receive feedback from the environment.
%In addition, as ML is involved, it is expected that a massive amount of training data are required from different nodes. It is true that it is possible to collect all the data in a central unit and process them, but this approach may not be efficient when considering the cost of communication and security. The foreseen solution is to develop multi-agent learning. This approach influences the way \ac{AI} technology is exploited and impacts the performance of learning the semantic language. 
Native AI and connected intelligence are developed on top of a large-scale distributed network that involves various communications technologies to interconnect end devices, and network infrastructure including access points, computation platforms, and switches, among others. Understanding the network model and optimizing the resources is the objective of large-scale communication theory, which is shown in the second level, and focuses on the network from an abstract level. 
The third level of the figure provides details related to the 6G wireless system. On the left side, several applications are listed and connected to the main relevant KPIs with a solid line arrow, whereas the dashed arrow indicates the relevance of the other KPIs. The 6G targeted values and ranges are indicated below the KPIs. Each KPI is connected to the most relevant technology, and for each technology, the foreseen scientific challenges are provided on the right side. The key technological challenges are noted below the technology components, and the main research topics of the scientific concepts are written below the scientific challenges. In particular:
\begin{itemize}
\item High data rate is required for \ac{XR}/\ac{VR}/\ac{AR} and holographic communication applications. The proposed technology to achieve Tbps is sub-THz and ultra-wideband. At this frequency, the response of the RF hardware component is non-linear, and the large bandwidth complicates the transceiver design. Thus, non-linear signal processing approaches are required, with the main challenges in obtaining the hardware models and developing low-complexity algorithms.
\item Extreme sensing performance, especially for localization, is important for applications of autonomous mobility with robots or vehicles, and to some extent is relevant to \ac{XR} applications. In addition to sensing, communication is needed with extreme KPIs in \ac{URLLC} and coverage. Therefore, \ac{ISAC} arises as an enabling technology with the challenges of designing hardware and sharing the spectrum in an optimized way such that the radio front-end is able to perform both sensing and communication efficiently. On top of that, the scientific challenge is in the design of joint waveform and exploiting super-resolution to guarantee achieving the required sensing and communication KPIs. Note that \ac{ISAC} technology is related to sub-THz, and the technologies of \ac{MIMO}, \ac{RIS}, \ac{OAM}, as they can be exploited to steer the sensing. Moreover, super-resolution parameter estimation is a non-linear problem, and non-linear processing approaches need to be explored. 
\item Capacity and coverage are connected to applications with a massive number of entities, such as digital twins, which can be exploited for applications such as smart cities, smart factories, and smart transportation. These applications require wide coverage and need to tackle the obstacles in the environment, as well as provide high spectral density. Therefore, technologies such as massive MIMO, RIS, and OAM, are explored. The technical challenge is on deriving the correct channel model considering a large number of antennas. In addition, estimating the channel requires a huge overhead. Thus, we need to understand the limitation of such approaches using electromagnetic information theory, and exploration of the potential of noncoherent communication.  
\item Extreme \ac{URLLC} is needed by time-sensitive applications in industrial automation, and for the tactile internet. Achieving the E2E requirements requires processing and AI technologies at the network edges. Besides the extreme requirements on the air interface, the impact of the upper layers queuing and code word length need to be jointly considered. Therefore, two scientific topics arise, namely combining queuing and Information theory and non-equilibrium information theory to investigate solutions for proper cross-layer design. 
\item High mobility is one of the applications that will be enabled by 6G to provide connectivity to flying objects such as \acp{UAV} in 3D space, and high-speed trains. The main challenge is associated with extreme Doppler shift. Thus, the signal design for the time-variant channel needs to be considered, to provide solutions with trade-offs between complexity and robustness, in terms of other KPIs such as URLLC. Note that, non-coherent communication is relevant to reduce the pilot overhead for receiver channel estimation, as in such scenarios, the channel can be outdated in a short interval.
In the fourth level of the figure, all the above techniques need to consider energy consumption as an important KPI for 6G sustainability. As a result, it is essential to study optimization approaches and understand the theoretical limits of energy consumption considering the cost of communication, and computation (including the processing of sensing and communication) based on the thermodynamics of computation and communication. Such limits will provide a reference for evaluating the performance of existing technologies and indicate the potential for advancement in the technologies.
\end{itemize} 
\section{Electromagnetic Information Theory} \label{sec:electromagnetic}
%\hl{Sami}\\
Research on electromagnetic information theory  is mainly related to the development of information theory principles and antenna engineering under limitations imposed by the laws of electromagnetism. It is noteworthy that this interplay between wave physics and information theory dates back decades ago \cite{Gabor}. %\newline
In this section, we shed light on the interaction between information theory and electromagnetism, which has led to the emergence of a new field known as \ac{EIT}. 

\begin{figure*}[ht]
  \includegraphics[width=0.9\textwidth,height=8cm]{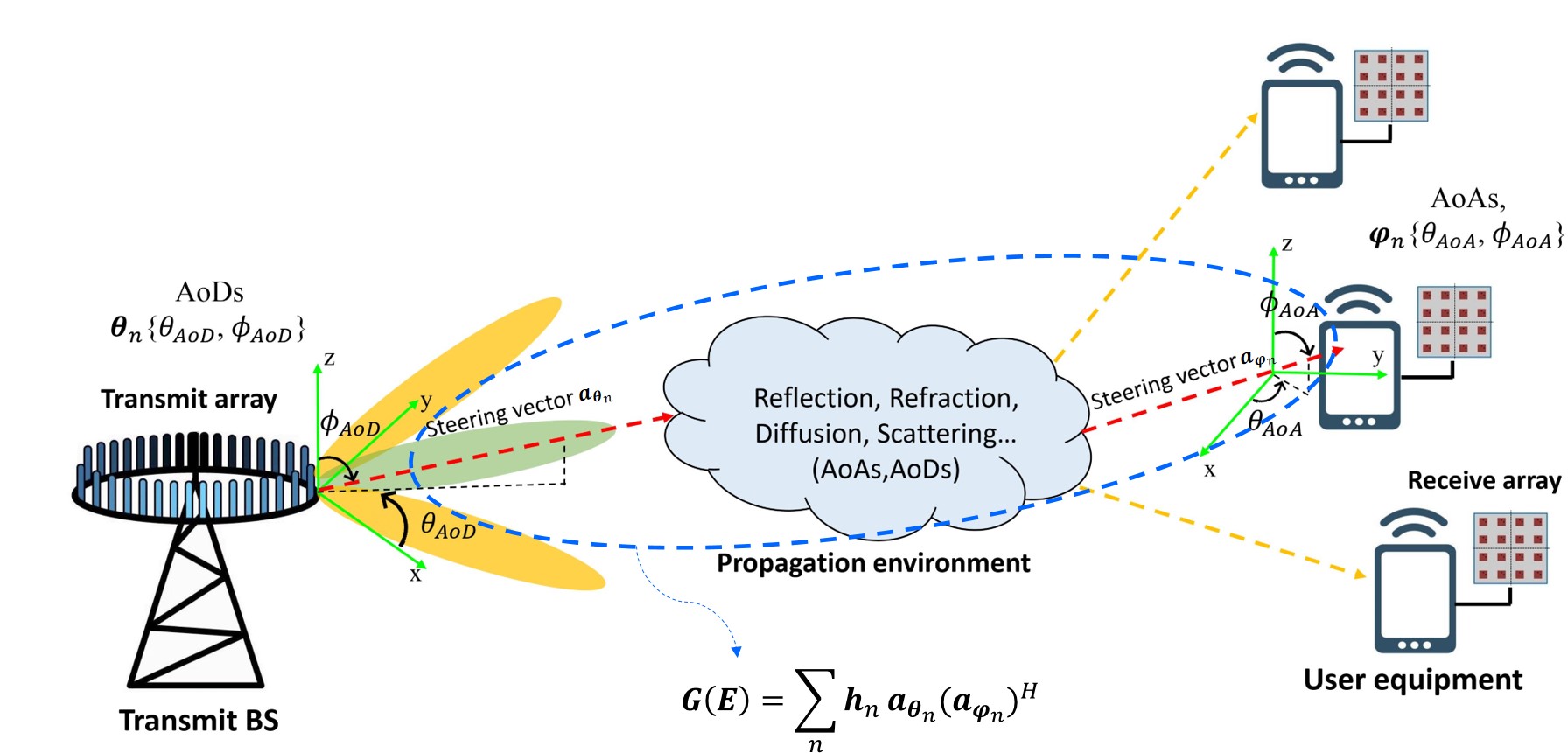}
  \caption{Generalized wireless channel.}
  \label{fig:channel}
\end{figure*}

In the following, we investigate the effect of  electromagnetism laws on the channel capacity, which is defined as the maximum mutual information and is given as 
%\begin{figure}[t]
%	\centering
%	\includegraphics[width=1\linewidth]{Figures/Channel.jpg}
%	\caption{Generalized wireless channel.}\label{fig:channel}
%\end{figure}
\begin{equation}
C=\max _{p(x)}\{I(x, y)\},
\label{Eq1}
\end{equation}
where $x$ and $y$ are the transmit (Tx) and receive (Rx) sequences, respectively, and $p(x)$ is the  distribution of $x$. The maximum is taken over all possible transmitted sequences subject to a power constraint $P_{c}=\left\langle x x^{+}\right\rangle \leq P_{T}$, where $P_T$ is the total power.

Considering the impact of electromagnetism laws, we define the spatial capacity $S$ as the maximum mutual information taken over both the Tx vector and the electromagnetic (EM) field distributions.  Consequently, $S$ can be written as

\begin{equation}
S=\max _{p(x), \mathbf{E}}\{I\{x,\{y, \mathbf{G}(\mathbf{E})\}\}\}.
\label{Eq2}
\end{equation}
Constraints: $P_c=\left\langle x^{+} x\right\rangle \leq P_{T},\\
\nabla^{2} \mathbf{E}-\frac{\mathbf{1}}{c^{2}} \frac{\partial^{2} \boldsymbol{E}}{\partial t^{2}}=\mathbf{0},
\quad \mathbf{E}=\mathbf{E}_{\mathbf{0}} \quad \forall\{\mathbf{r}, t\} \in B,$ 
\vspace{0.1cm}
\\ \noindent
where \textbf{E} is the electric field, $c$ is the speed of light, \textbf{r} is a position vector at time $t$. and B is the boundary condition.  $\mathbf{E_0}$ denotes the amplitude of the electric field, \textbf{G} represents the channel matrix, which is a function of the electric field, and the last constraint is due to boundary conditions.  Additionally, the first constraint is a classical total power constraint and the second is due to the wave equation. An exact solution to $S$ is intractable since the constraints contain a partial differential equation with arbitrary boundary conditions. However, by redefining $S$ as the maximum of conventional channel capacity under constraints dictated by Maxwell's equations, \eqref{Eq2}  can be written as \cite{loyka2004information} 
\begin{equation}
S=\max _{\mathbf{G(\mathbf{E})}}\{C(\mathbf{G(\mathbf{E})})\} \text {, const. : \textbf{G($\mathbf{E}$)} } \in \mathcal{M}\{\text { Maxwell }\},
\label{Eq3}
\end{equation}
where the constraint $\mathcal{M}\{\text { Maxwell }\}$ is due to the wave equations, and (3) is maximized
by varying \textbf{G($\mathbf{E}$)}, e.g., by changing antenna positions. Note that there is no known explicit form for the constraint $\mathcal{M}$.
\\

In Figure~\ref{fig:channel}, a generalized wireless channel, \textbf{G(E)}, which consists of a transmit antenna array, a wireless link, and a receive antenna array, is depicted. In the underlying scenario, \textbf{G(E)} is characterized by the convolution of the steering vectors and the multipath  channel, and is written as 
\begin{equation}
\mathbf{G}(\mathbf{E})=\sum_{n} h_{n} a_{\bm{\theta}_{n}} a_{\bm{\varphi}_{n}}^{H},
\label{Eq4}
\end{equation}
where $n$ is the number of multipath components in the physical propagation channel and $H$ is the Hermitian
transpose. The angle of arrival (AoA), $\bm{\varphi}_{n}$,  and the angle of departure (AoD), $\bm{\theta}_n$, are determined by the azimuth, $\theta_{AoA}$, $\theta_{AoD}$, and the elevation, $\phi_{AoA}$, $\phi_{AoD}$, angles. Furthermore,  $a_{\bm{\theta}_n}$ and $a_{\bm{\varphi}_n}$  
represent the steering vectors of the transmit and receive arrays, respectively,  and $h_n$ is  the fading gain of the $n$th multipath that accounts for reflection, refraction, and diffusion.\\

The concept of the number of degrees of freedom (DoF) was introduced  in \cite{pierri1998information} to represent the number of dimensions for effective communications dictated by the Maxwell equation. Several methodologies for determining the number of DoF were  introduced earlier in the literature \cite{slepian1961prolate}, \cite{landau1961prolate}. It is noteworthy that, in wireless environments, the number of DoF of the electromagnetic field must be explored to better understand their physical limitations. Within this context, there have been  a few sporadic results reported on DoF of wireless communications; see, e.g., \cite{linfoot1955information, bucci1989degrees, piestun2000electromagnetic}. In \cite{bjornson2019massive}, the effect of the laws of electromagnetism on the \ac{MIMO} channel capacity was investigated.  It was demonstrated that, for a given aperture size, the limit on the \ac{MIMO} channel capacity, dictated by Maxwell equations, is manifested in the form of minimum antenna spacing. 

Recently, massive MIMO has emerged as a disruptive technology and has been adopted as an integral part of the fifth-generation standard. Its advantages in terms of spectral and energy efficiency are well-defined and understood. Nonetheless, as the demand for data traffic continues to increase exponentially, the natural questions are: “Is it feasible to approach the limits of infinite antennas?” and “What can we expect next?” \cite{bjornson2019massive}.  The answer to these questions can be provided by EIT \cite{marzetta2018spatially}, \cite{pizzo2020degrees}.

More recently, new research directions have  emerged, which aim to identify the fundamental physical limitations of wireless communications systems and to offer new disruptive technologies to meet the demand of 6G  networks.  In the following subsections, we will focus on three novel technologies, which are expected to increase the number of DoF  of \textbf{G(E)}, as shown in Figure~\ref{fig:channel}, and discuss how the EIT's mathematical tool can  push further the limits of these technologies.

\subsection{Reconfigurable Intelligent Surfaces }
\Acp{RIS} is an emerging technology that is perceived as one of the  key enabling technologies for 6G. This is largely due to their unique capability of manipulating electromagnetic waves, e.g., steering, backscattering, and absorption, in a controllable manner. RISs are able to effectively control the amplitude, phase, and frequency of RF signals, without the need for complex signal processing techniques \cite{alghamdi2020intelligent}.

An RIS comprises an array of passive reflective elements, which can alter the phases of the incident signals \cite{alghamdi2020intelligent}. Specifically, by intelligently adjusting the phases of its reflective elements, the RIS can intelligently control the propagation environment to facilitate energy-efficient and reliable data transmissions. While recent investigations on RISs have mainly focused on the improvements of the signal-to-noise ratio (SNR) and beamforming, there have been limited results reported on theoretical capacity limits under particular constraints, e.g., feedback overhead, hardware resolution, etc. Within this context, EIT is expected to play a key role in demonstrating the impact of wave equations on the information-theoretic limits of RISs; see \eqref{Eq3}, \eqref{Eq4}.
\subsection{Time Reversal}
\Ac{TR} exploits rich multipath environments to create spatial-temporal focusing \cite{chen2016time}. It has shown great potential in underwater wireless environments \cite{fink1992time} and radar imaging \cite{chen2016time}.  It has been  shown that the resolution of the spatial-temporal focusing largely depends on the number of multipath components.  Thus, for extracting a high number of multipath components, a large bandwidth is required. Consequently, the integration of TR with ultrawideband (UWB) communications was proposed and further investigated with a focus on the error rate probability \cite{guo2007reduced}.  TR division multiple access was investigated in \cite{han2012time} in the context of multiuser scenarios.  

In TR systems, the SNR is maximized, if the symbol duration is larger or comparable to the channel delay spread. However, if it is smaller, then TR will experience \ac{ISI}, leading to degradation in the system’s performance. The problem becomes compounded in multiuser scenarios where inter-user interference is introduced \cite{han2012time}. To address these challenges, a signature waveform design was introduced in \cite{lerosey2007focusing}. The basic idea behind waveform design is to leverage channel information to carefully choose the amplitude and phase of each tap of the signature waveform to maximize the signal-to-interference-noise ratio.  

Finally, TR enjoys several attractive features, among which are the realization of virtual massive MIMO with a single antenna, realizing high energy efficiency and scalability \cite{chen2016time}. In TR systems, EIT is envisioned to play a major role in determining the performance limits, particularly in terms of the number of DoF of G(E),  which is given by \eqref{Eq4}.

\subsection{Orbital Angular Momentum}
EM waves contain angular momentum, which comprises \ac{SAM} and \ac{OAM}. SAM for a radio wave corresponds to polarization, which has already been used in radar systems. OAM, on the other hand, has not yet received fully the deserved attention. OAM consists of a wavefront with helical phase and a field strength, which can be used for information transmission. Theoretically, and in a single beam, OAM has an infinite number of modes, each of which forms an orthogonal basis, thus, creating the basic building block for new multiple access methods for future wireless networks.  
\\
By leveraging \eqref{Eq4}, it is envisaged that EIT theory will provide the necessary tools for unleashing the full potential of OAM and help uncover new wave characteristics. Likewise, it is expected that the EIT’s mathematical tools will guide designers to efficiently design antennas for \textit{large intelligent surfaces} and \textit{surface wave communication}.

\subsection{Summary}
%\hl{Sami}

To reap the full potential of 6G, novel communication paradigms have emerged, including RIS, \ac{TR}, OAM, surface wave communications, and ultra-massive MIMO.  Nonetheless, current information-theoretic tools are insufficient to fully characterize the performance limits of these technologies.  This is because current network design and performance analysis approaches consider scalar-quantity, far-field, planar-wavefront, and physically inconsistent assumptions, which may lead to discrepancies between system design and real propagation environments.  To address this issue, it is envisaged that EIT will play a key role in properly characterizing the fundamental limits of these technologies and potentially building better physically consistent communication system models. 
\section{Non-linear Signal Processing Theory} \label{sec:nonlinear}
%\hl{Marwa}
Conventional signal processing for communications assumes linear system model $\ma{y} = \ma{H}\ma{x} + \ma{v}$ with additive Gaussian noise in the development of signal processing algorithms such as channel estimation, equalization, and detection. Although a practical transceiver involves several non-linearities, the linear approximation still provides satisfactory results, where $\ma{v} = \ma{z} + \ma{\eta}$, consists of signal-dependent interference term $\ma{\eta}$ and independent noise $\ma{z}$ \cite{linearization}.
In current 5G  systems, with sufficient \ac{ADC} resolution, sub-6 frequencies, and bandwidth up to 100 MHz, such linearization and Gaussian noise assumptions are sufficient to achieve satisfactory performance.  However, for 6G considering mmWave and THz frequencies, with ultra-wideband $>$ 1 GHz, the non-linearity resulting from the \ac{RF} hardware including, \ac{PA}, \ac{LNA}, antenna coupling, as well as the limited resolution of \ac{ADC}, becomes significant. Accordingly, the linearization error $\ma{\eta}$ becomes dominant. For instance, under an impulsive noise environment: linear synchronization, detection, and equalization fail \cite{arce2005nonlinear}. Hence, there is a need for developing new approaches in designing signal processing algorithms beyond the linear assumption. 
This section provides an overview of non-linear systems, focusing in particular on the transceiver design. We show using a simple example the impact of wrong modeling on performance, where the theoretical evaluation based on linear assumptions can be misleading. Then, we introduce several techniques for non-linear signal processing and conclude the section with an overview of the relevant research challenges. 
\subsection{Non-linear Systems Overview}
Non-linear systems do not have a canonical representation, like impulse response for linear systems. There have been some efforts to characterize non-linear systems, such as Volterra and Wiener series using polynomial integrals as the use of those methods naturally extends the signal into multi-dimensions \cite{Volterra}. 
Typically, the behavior of a non-linear system can be described by a non-linear system of equations. The equations to be solved cannot be written as a linear combination of the unknown variables or functions that appear in them. Moreover, systems can be defined as non-linear, regardless of whether known linear functions appear in the equations. For example, a differential equation can be linear in terms of an unknown function and its derivatives, even though the functions are non-linear in terms of other variables. Another example is the \ac{DoA} estimation problem, which is linear in terms of steering vectors and non-linear in terms of the \acp{DoA}.

As non-linear dynamical equations are difficult to solve, non-linear systems are commonly approximated by linear equations. This works well up to some accuracy and some range for the input values, but some interesting phenomena such as solitons, chaos, and singularities are hidden by the linearization. For example, chaos may resemble random behaviour but it is actually not random, and thus, the linearization may affect the accuracy \cite{infeld2000nonlinear}. There are several sources of non-linearities, e.g., the system response is a non-linear function, the system does not fulfill the linear transform properties, or the involved random processes are not linear with sources of non-Gaussian noise such as impulsive noise or cross talk. Non-linear signal processing can then lead to improved performance.
\subsection {Non-linearities in Communications Systems}
A standard transceiver consists of a linear encoder in the Galois field, followed by digital mapping to a complex constellation set, e.g. M-QAM. This mapping is in fact a non-linear transform. Thereafter, the data symbols linearly modulate pulses to generate digital linear baseband waveforms, such as \ac{SC} and \ac{OFDM}, which is the dominant waveform in 4G and 5G systems \cite{RFtransceiver}. Then, another non-linear transform follows,  which is the \ac{DAC} with finite quantization. This conversion results in a quantization noise, that is typically uniformly distributed. The analog baseband  signal is upconverted to high frequency with a non-linear mixer. However, in an ideal design, the receiver mixer, which down-converts the signal back to the baseband, retrieves the linear relation between the transmitted and received analog baseband. The bandpass signal is then amplified prior to transmission with a \ac{PA}. \acp{PA} are practically non-linear, but linearization approaches, such as digital pre-distortion allows a linear operation within a limited range of the signal. This requires to control the signal amplitude variation by introducing an input back-off at the cost of decreasing the emitted signal power. Thus, signals with high \ac{PAPR}, like \ac{OFDM}, suffer more. One solution is to design constant envelop waveforms or rely on \ac{SC} with low QAM order. However, such waveforms achieve lower spectral efficiency, and thus,  \ac{MIMO} schemes are required for compensation. 

The communication  channel is modeled as a linear system, ignoring non-linearities. In fact, the channel can be non-linear as a sum of reflections from irregular surfaces or because of non-Gaussian interference in multi user scenarios. Moreover, surrounding rotating electric machinery generates impulses and atmospheric noise. Additionally, the receiver \ac{LNA} may also impose non-linearity to the passband signal. The \ac{ADC} with finite resolution introduces another quantization noise, in addition to non-linear clipping when non-ideal \ac{AGC} is considered. Finally, IQ imbalance is also another source of nonlinearity, which manifests due to the imperfect orthogonality between the inphase and quadrature components at the local oscillator \cite{rfImpairment}. Note that a $90^\circ$ phase shift is very hard to achieve in analog domain. Besides the IQ phase imbalance, there will also be gain mismatches due to the analog low pass filters at the inphase and quadrature, which are also difficult to match in the analog domain. In general, for wideband scenarios and even for 802.11ax where bandwidths exceed 100MHz, the IQ imbalance starts to depend on frequency, in which case nonlinearity would enforce itself as a natural phenomenon \cite{RFmmWave}.
\begin{figure}[t]
	\centering
	\includegraphics[width=.98\linewidth]{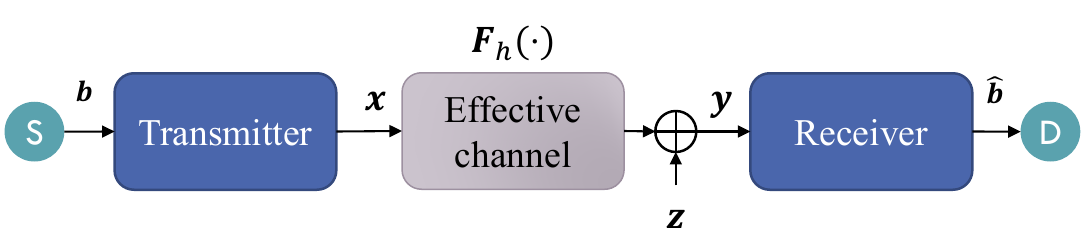}
	\caption{Transceiver model.}\label{fig:transceiver}
\end{figure}

In general, as illustrated in \figref{fig:transceiver}, the transceiver can be modeled by a transmitter block that consumes bits and generates samples, an effective channel that summarizes the hardware and communication channel response, and a receiver block that extracts the information bits from the received sequence. Considering all the non-linear effects,  the relation between the digital transmitted and received sequences  $\ma{x}, \ma{y} \in \compl^{N\times 1}$ at the digital processing interface is defined by a non-linear transform $\ma{F}_h: \compl^N\rightarrow \compl^N $ such that
\begin{equation}
\ma{y} = \ma{F}_h (\ma{x}) + \ma{z}, \label{eq:nonlinear model}
\end{equation}
where $\ma{z}$ is additive noise. In linear systems, $\ma{F}_h $ is a linear transform, defined by a channel matrix $\ma{H}\in \compl^{N\times N}$, with
\begin{equation}
\ma{y} = \ma{H}\ma{x} + \ma{v}, \label{eq:linear model}
\end{equation}
and $\ma{v} = \ma{z}$ is \ac{AWGN}. Such assumption is considered in linearization using the generalization of Bussgang decomposition \cite{non-linear-mimo}, given by $\ma{F}_h (\ma{x}) = \ma{H} \ma{x} + \ma{\eta}$, where $\ma{H}$ is obtained by \ac{LMMSE} estimation and $\ma{\eta}$ is a random error vector independent of $\ma{x}$. In the case of Gaussian vector $\ma{x}$,   $\ma{\eta}$ is also Gaussian. Therefore, if the additive noise $ \ma{z}$ is also Gaussian, then we obtain the typical linear of \eqref{eq:linear model}, with $\ma{v} = \ma{z} + \ma{\eta}$. 

The linear channel matrix $\ma{H}$ is statistically modeled based on measurements to constrain its structure to facilitate practical channel estimation. For example, in a multipath channel and by employing a \ac{CP}, it becomes a circular matrix with a few unknown parameters. In fact, the performance is limited by the covariance of $\ma{v}$, which depends on the additive noise $\ma{z}$, and  the linearization error $\ma{\eta}$. 
\subsection{MLD Problem in non-Gaussian Noise}
\begin{figure}[t]
	\centering
	\includegraphics[width=.98\linewidth]{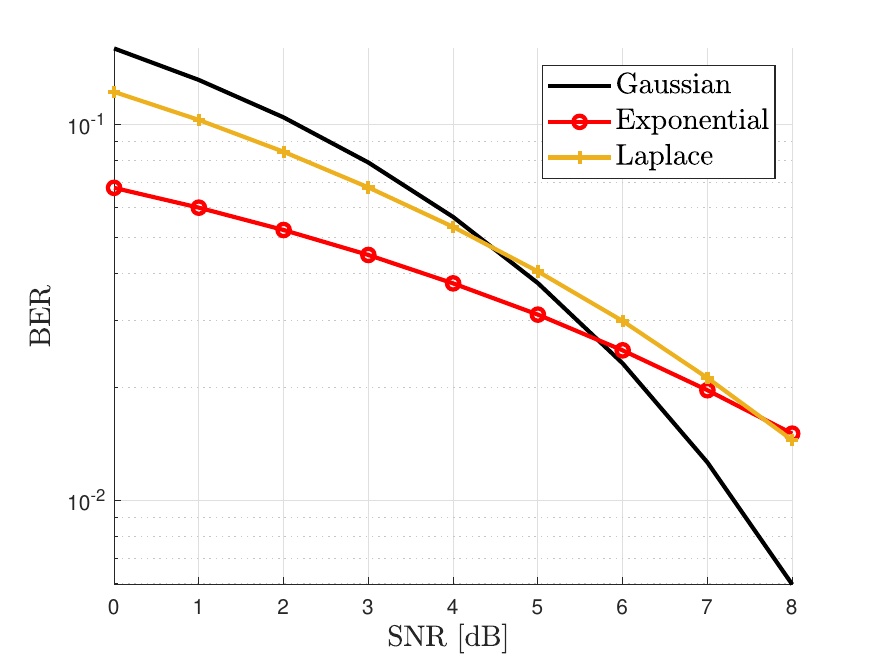}
	\caption{Example of applying minimum distance for different distributions using BPSK for different additive noise.}\label{fig:MLD_example}
\end{figure}

When considering a Gaussian noise, the \ac{MLD} solution is a well-defined problem given by the minimum distance 
\begin{equation}
\hat{\ma{x}} = \arg \min_{\ma{x} \in \set{X}}\Norm{\ma{y}- \ma{H} \ma{x}}^2, \label{eq:MLDD}
\end{equation}
where $\set{X}$ is the set of possible vectors corresponding to the input data vector $\ma{b}$. When the noise is arbitrarily distributed, such as impulsive noise, using the solution  \eqref{eq:MLDD} does not lead to the best detection performance, and the exact \ac{MLD} is hard to be solved \cite{arce2005nonlinear}. Additionally, the solution \label{eq:MLD} under non-Gaussian noise leads to non-optimal solutions. For example, assume the simple additive noise channel, $y = d + v$, where $d \in \{1,-1\}$ is a BPSK data symbol. The \ac{MLD} solution for Gaussian noise is given by the solution of $\min_{d \in {1,-1}} |y-d|$, but using the same solution for other types of noise with the same power leads to a non-optimal performance, as shown in \figref{fig:MLD_example}.
Similarly, linear \ac{MMSE} (LMMSE) estimation and prediction considering second-order statistics are commonly used. However, using exact distributions other than Gaussian deviates from the exact non-linear \ac{MMSE}. Note that criterion \eqref{eq:MLDD} is dedicated to hard-input-hard-output decoding, in which an example is where forward error correcting codes are not utilized. Another version of \eqref{eq:MLDD} is to integrate soft-input-soft-output scheme through the maximum aposteriori principle, i.e. in cases where \ac{LDPC}/\ac{BCC} codes are utilized. This is where log-likelihood distance metrics are computed to generate soft-bit outputs, rather than taking a mere hard decision. It is also anticipated that for soft-input-soft-output, the nature of the noise would also perturb the optimality of the decision that has to be taken.

\subsection{Non-linear Signal Processing Approaches}

One possible approach relies on the modeling of the non-linearities for systematic design and rigorous analysis based on mathematical models. The scope of non-linear signal processing includes analysis of signals, non-linear filtering, and modeling of the system to find the best approach. At the receiver, given an observation waveform, one goal is to extract information that is embedded within the signal such as modulated data and environmental parameters.
The main challenge is to provide accurate models, which require extensive measurements and considering the dependency between many parameters. Another challenge arises from the lack of a unified and universal set of tools for analysis and design, in addition to the computational complexity.

As an example of non-linear transceiver design, we present an example related to fiber optical communication. The propagation is governed by \ac{NLSE}, where solving for different noise realization is impractical. As a processing technique,\ac{NFT} is proposed, where it decomposes a wave into non-linear degrees of freedom, and encodes information on the non-linear spectrum \cite{non-linear-FT1}. A periodic \ac{NFT} version  (PNFT) for finite length signal, can be used as an alternative to linear \ac{FT}.  However, the mathematical complexity of the inverse of periodic PNFT is high. To overcome that, designing solutions based on algebra-geometric approaches  with non-linear frequency amplitude modulation is proposed in \cite{non-linear-FT2}.

Another possible approach for transceiver design  follows data-driven modeling using machine learning, which relies on \ac{E2E} training of neural networks at the transmitter and receiver blocks as depicted in \figref{fig:transceiver}. In these techniques, the combined effect of hardware response and communication channel can be learned and handled by the trained models\cite{DL_overair}. One main challenge of this approach is that the performance of the learned models  highly depends on the specifications of the used hardware and  channel conditions, and it may fail to work when the channel changes. The generalization issue of deep learning techniques is still an open problem. Moreover, since deep learning lacks theoretical foundations, there are in general no performance guarantees or bounds of the developed algorithms.
  %Moreover, as the wireless channel is subject to changes, the challenges is on adapting the models to the dynamical channel in a similar way, conventional channel estimation and adaptive modulation is exploited.
  %Moreover, as a common challenge in \ac{AI}, how to define the model parameters, such as the input data packet length, number of layers, type of activation functions, etc. 
%In general, with the lack of developed information theory for general non-linear channel, it is challenging to judge the performance of the developed non-linear algorithms. 

\subsection{Summary}
In order to satisfy the extreme data rate requirement of certain {6G} applications in the range of 100 Gbps-1 Tbps, exploiting ultra-wideband at the sub-THz band is one potential enabling technology, which is associated with technical challenges in the transceiver design. This arises from the non-linear behavior of the \ac{RF} components, and other impairments caused by the hardware design constraints. Relying on conventional linearization approaches and assuming Gaussian noise in the signal processing results in poor performance. Thus, considering non-linear signal processing methods is essential, which requires addressing several challenges concerning proper modeling of the transceiver components, developing low-complexity algorithms, in addition to the need for tools to evaluate the information theoretical limits as an alternative to the conventional \ac{AWGN} model.

\section{Multi-Agent Learning Systems}\label{sec:multiagent}
%\hl{Lina}
The uniqueness of 6G networks, compared to previous wireless network generations, lies in the realization of ubiquitous intelligence, in which native \ac{AI} will be the key to orchestrating wireless networks from the core to the edge, and to the cloud. Although \ac{ML} algorithms are anticipated to be an indispensable tool in future 6G networks, which operate on the data collected from all network segments in order to enable smart resource management, access control, and smart decision-making, the speculative vision of 6G networks goes beyond leveraging ML to replace particular modules in the network. Rather, it is envisioned that each network node will enjoy a level of intelligence that enables it to continuously learn from the environment, and therefore, adapt to the network changes, and realize self-optimization. The inherent heterogeneous characteristics/requirements of different nodes in heterogeneous dynamic environments pose extreme challenges to the implementation of \ac{ML} algorithms, owing to the resulting heterogeneous data. Additionally, in conventional \ac{ML} algorithms, raw data generated and stored at local devices should be sent to centralized servers for processing, training, and aggregation, yielding compromised users’ privacy and security, and increased network overhead. Furthermore, centralized \ac{ML} suffers from long propagation delay, rendering it unsuitable for real-time applications. Motivated by the increasing demand for secure \ac{AI} tools, and the enhanced on-board computing and storage capabilities of wireless devices, the research has started to shift from centralized to distributed learning approaches. In this respect, \ac{DML}, including federated learning, has been recently identified as an enabling technology that is capable of training wireless networks without leaking private information or consuming network resources \cite{wahab2021federated}. In particular, \ac{DML} allows a set of local devices to locally and collaboratively participate in the training process of a global model without having to upload their raw local data to centralized servers. Among others, multi-agent systems have stood out as a promising sub-field of distributed \ac{AI}, allowing the realization of intelligent, systematic, and self-coordinating networks. 

On the other hand, current and past wireless networks have relied primarily on feedback-based communications, where several mechanisms, e.g., \ac{FDD} and \ac{TDD}, are adopted for pilot transmission and CSI feedback. With the advent of massive MIMO systems and large-scale connectivity, such mechanisms fail to meet the throughput requirement imposed by the available resources in wireless networks. In this regard, multi-agent systems constitute a promising solution for the feedback problem, and for alleviating the pilot transmission overhead, by allowing multiple agents to interact with the environment and reach a consensus in performing a particular task (e.g., resource allocation, trajectory design, etc) without consuming the spectrum resources for feedback.

\begin{figure}[t]
	\centering
	\includegraphics[width=1\linewidth]{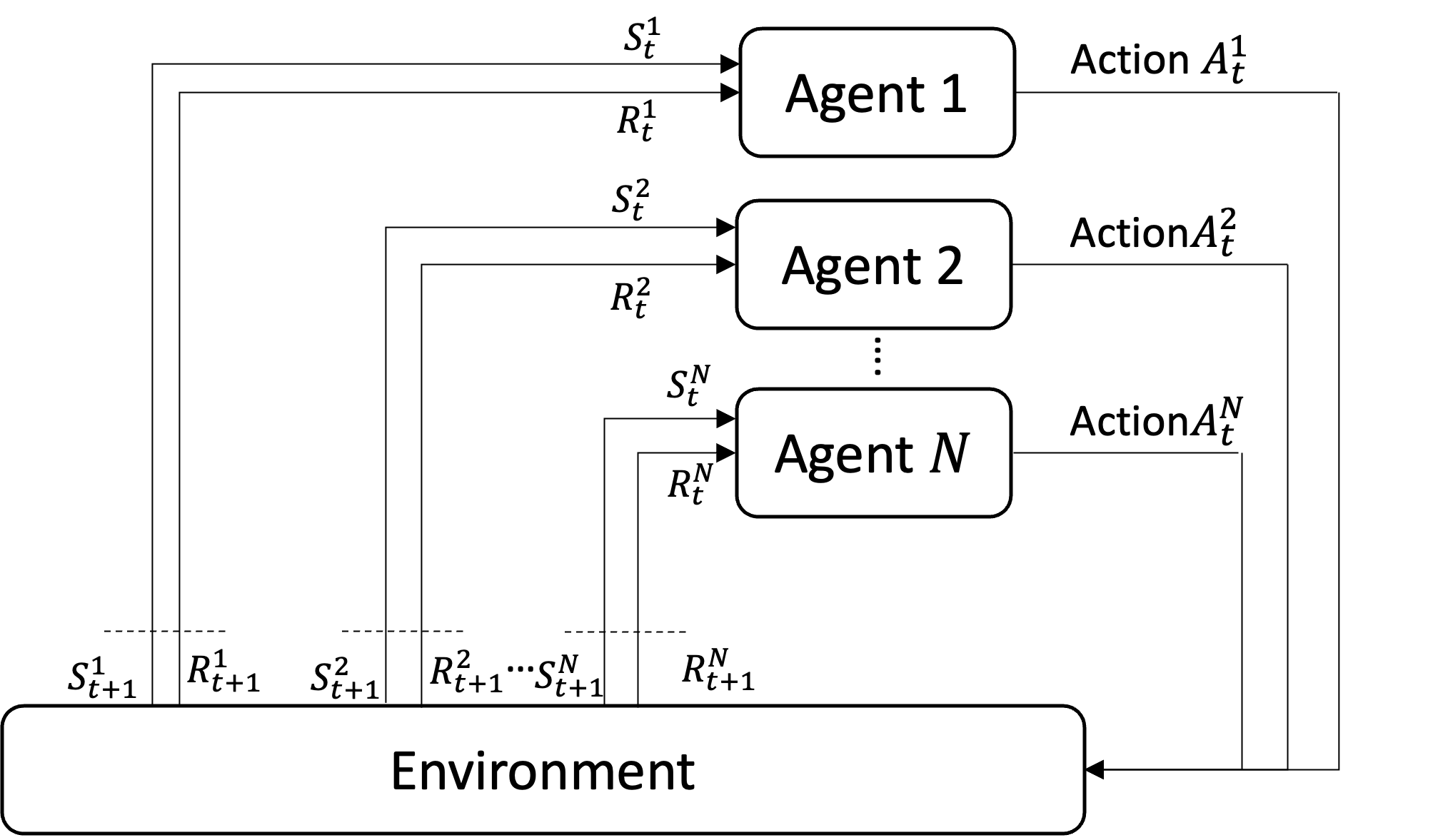}
	\caption{Architecture of multi-agent systems.}\label{fig:MA}
\end{figure}

As its name indicates, multi-agent learning systems comprise a group of autonomous computing entities (agents) that leverage \ac{ML} algorithms and the flexibility feature offered to them by the network, in order to learn, solve complex tasks in a collaborative manner, and autonomously perform smart decision-making process. The key principle of multi-agent systems relies on the orchestration of multiple network agents and facilitating their interaction with each other and with the surrounding environment, with the aim to enable them to develop new knowledge pertaining to related contexts and actions, and hence, to efficiently solve their allocated sub-tasks \cite{dorri2018multi}. In this sense, the required processing and energy resources can be significantly scaled down. From a different perspective, when an agent faces a failure due to a power outage, compromised communication, etc, the sub-task allocated to this agent can be easily re-allocated to another available agent, overcoming the single point of failure experienced in single-agent systems. As demonstrated in Fig. \ref{fig:MA}, each agent in the network is considered as an active goal-oriented node, and hence, for agent $n$, at each observation period, it observes part of the environment to investigate its state, and accordingly, execute a corresponding action. In the case of collaborative agents, where agents are generally performing a single task, the agents may need to observe the whole environment, in addition to observing what other agents are doing, in order to perform the needed action for the allocated sub-task. In this scenario, an agent manager is required for orchestrating the actions taken by different agents, in order to obtain the global action of a particular task. 

\subsection{Challenges} Although the merit of multi-agent learning systems is harnessed through the massive, ubiquitous deployment of agents throughout the network, large-scale network architecture imposes several challenges, pertaining to agents formation and the dynamic nature of wireless networks. For the former, it is deemed that the organization of agents in the network needs to be maintained for a particular period of time, in order to ensure the successful delivery of the tasks. A fixed agent formation is considered a highly challenging problem, owing to three main factors, namely, the selection of optimum agent formation to perform a particular task, organizing the agents based on the optimum formation, and maintaining this structure. While the full potential of multi-agent learning systems depends on such a fixed formation, it is a difficult step to achieve, due to the highly dynamic nature of wireless networks when implemented on a large scale. Furthermore, fixing the agents formation renders such systems unscalable, and hence, limiting the advantages of multi-agent systems. From a similar perspective, such a dynamic behavior exacerbates the difficulty of the learning process at the agents, which will necessitate frequent environment sensing, re-connection with new neighboring agents, and a re-training process. This yields an increased processing and communication overhead, and hence, significantly drains the agents' onboard resources.

\subsection{Emergent Communications in Multi-agent Learning Systems}

Wireless communication is a key enabler for information sharing between machine learning agents working together to complete a complex task. However, transmitting raw data between sensors and agents running in a centralized data center will unlikely be sustainable on large scale, given the growth of the massive deployment of devices with connectivity and \ac{AI} abilities in our environment. To address these future needs, investigations are ongoing on pushing the performance boundaries of communication and exploring high frequencies. A complementary approach shall be investigated by efficiently sharing information between agents by learning to communicate in order to coordinate towards the execution of complex tasks. 
Communication is learned or emergent in the sense that at the beginning of a simulation, the messages the agents transmit do not have any semantics. Meaning and syntax emerge through interaction \cite{lazaridou2020emergent}. Therefore, communication messages are not hard-coded but learned. Agents are able to learn both policies and communication protocols to compensate for the partially observable environments in order to perform a complex task \cite{simoes2019multi}.
There are two main areas of research around this problem. One direction investigates decoding the emergent protocol in order to understand its generality and similarity to human language, while the second direction investigates employing powerful \ac{AI} tools in order to learn an efficient communication protocol able to perform a collaborative complex task. Several challenges need to be solved; it is actually difficult to measure the degree of effectiveness of emergent communications and understand to what extent the received message impacts the agent's action. New deep reinforcement learning algorithms that include a communication learning network need to be developed. Most of the previous work on this topic ignores the effect of noisy communications where messages can be lost or received with distortions due to the wireless channel. Constrained and unreliable communication should be considered within this framework. For each use case, several choices need to be made such as the size and the nature of the channel (i.e. discrete or continuous), how many rounds of communications should be considered before taking an action, how we perform the training and execution (i.e. centralized or distributed), how do we decide that there is a piece of relevant information to share, how do we define and represent the context of communication. 

\subsection{Summary}
As demonstrated earlier, 6G networks are envisioned to be shaped to cater for novel applications, that require extremely high data rates, ultra-low latency, and high quality of service. One of the critically identified challenges is pertinent to the feedback theory, which represents a bottleneck in the realization of the 6G vision, due to the severely degraded spectral efficiency and increased delay. As a solution, multi-agent theory represents a prominent approach to improve the throughput in future wireless networks, and to eliminate the feedback overhead experienced in massive-scale systems, particularly with the undergoing spectrum crisis. However, efficient mechanisms to coordinate agents formation and stability in the network, while ensuring the needed adaptivity and scalability requirements are maintained, are yet to be explored, particularly in highly dynamic networks. Within this regard, emergent communication in multi-agent systems can potentially reduce agents' communication costs, in addition to reducing the processing overhead at the multiple agents, which is a result of network dynamic, by allowing the agents to learn the messages instead of coding/decoding them, i.e., capture the semantics within the messages.
\section{Super-Resolution Theory} \label{sec:resolution}
%\hl{Marwa}

\begin{figure}[t]
	\centering
	\includegraphics[width=.98\linewidth]{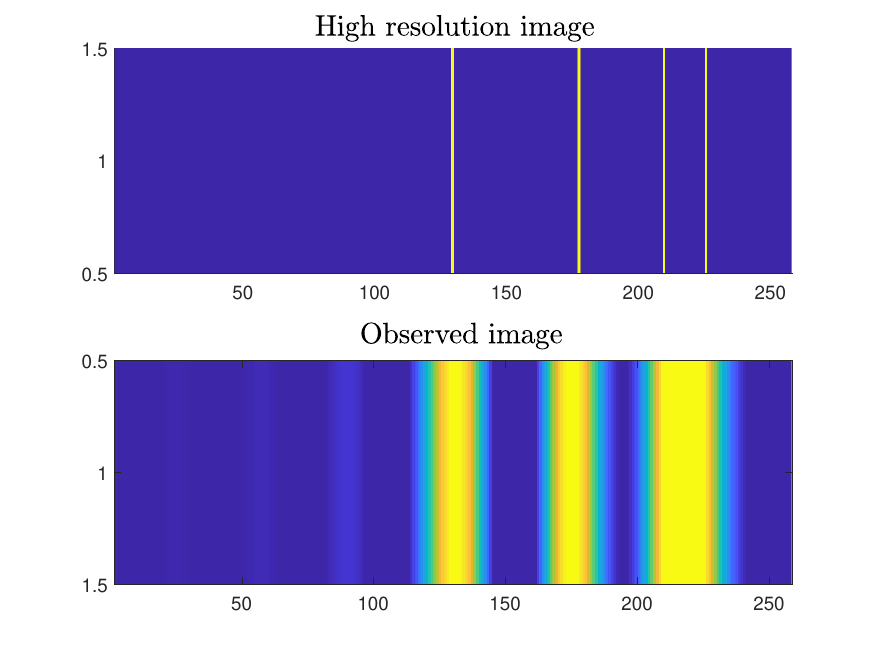}
	\caption{Comparison of high-resolution image and observed one.}\label{fig:SR_image}
\end{figure}
\begin{figure}[t]
	\centering
	\includegraphics[width=.98\linewidth]{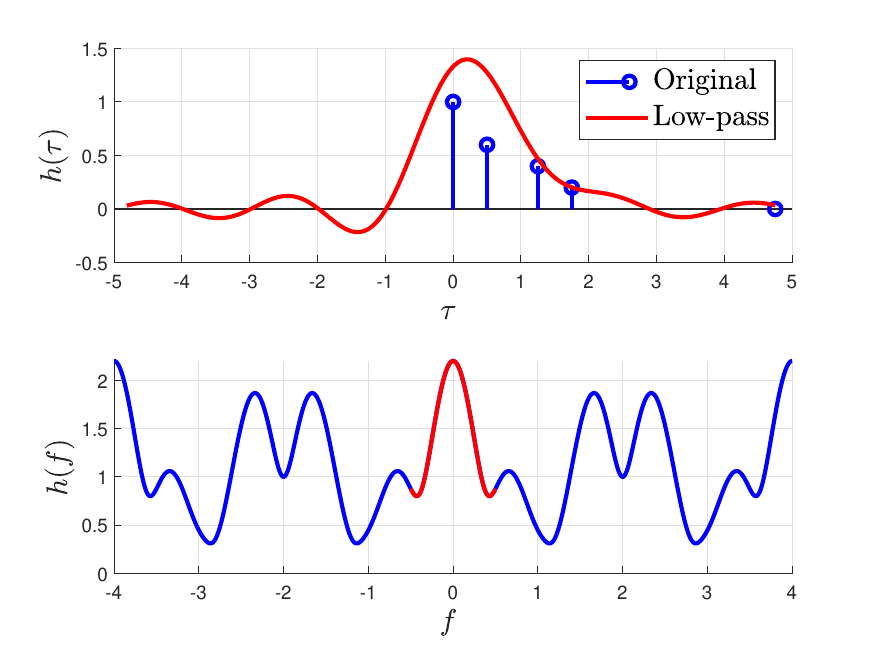}
	\caption{Multipath channel and the observed baseband response.}\label{fig:SR_channel}
\end{figure}
Several emerging {6G} applications require  communication and sensing, which can be enabled by \ac{ISAC} technology for sharing the spectrum, hardware, and waveforms \cite{arnold2022maxray}. To achieve the extreme sensing requirements, e.g., for 1 cm positioning accuracy, without degrading the performance of communication, and at the same time optimizing the utilization of hardware and spectrum resources, super-resolution techniques need to be considered. Besides the conventional super-resolution approaches, heterogeneous 6G technologies will provide multiple sources of signals, that can be fused in the processing to enhance the sensing accuracy. However, there is a trade-off between performance accuracy and processing complexity, especially because of the non-linear nature of the problem. Therefore, research is needed to address this trade-off and to provide tools to assess the theoretical limits of performance. In this section, we present an overview of the super-resolution concept, and highlight the 6G-related challenges.
\subsection{Super-resolution Concept}
Super-resolution theory deals with the problem of recovering the fine details of an object from coarse-scale information \cite{candes2014towards}. For example, an image with sparse spots appears blurred because of the hardware response of the imaging system, as shown in \figref{fig:SR_image}. This is because of the low-pass filter response. The task of super-resolution is to reconstruct the sparse image from the observed data. A similar problem arises when measuring sparse signals, such as multipath channels, using band limited system, as shown in \figref{fig:SR_channel}. Here, only the  low-end spectrum is observed at the receiver, and to find information about the high-end spectrum, an extrapolation problem needs to be solved using super-resolution techniques  for estimating the sparse signal components. This problem corresponds to a general sparse  representation of a function $f(x)$ as a sum of  basis functions $\varPhi_m(x)$, such that
\begin{equation}
f(x) = \sum_{m=0}^{M-1} a_m \varPhi_m(x).
\end{equation} 
Super-resolution algorithms aim at finding the parameters $a_m$ and $\varPhi_m(t)$ with the smallest number of components $M$ \cite{sauer2018prony}. 

In many practical signals, the basis functions are exponential and characterized by a variable $b_m$, where  $\varPhi_m(x) = b_m^x$. And therefore, from $K$ equally-spaced  discrete  observations $f(x_k)$, where $x_k = k \Delta_x$, the goal is to find the parameters $\{a_m\}$, $\{b_m\}$ in addition to the number $M$, by solving the set of equations
 \begin{equation}
 	f(x_k) = \sum_{m=0}^{M-1} a_m {\left(b_m^{\Delta_x}\right)}^k.
 \end{equation} 
This is a classical problem that has been addressed in different approaches relying on structuring the  measurement data.  The first solution from 1795 known as Prony's method  is based on restructuring the data in a matrix. The solution  is performed in two steps; the first involves solving a linear system of equations, and in the second, the zeros of a polynomial are computed.  However, Prony's method  suffers from numerical  instability under measurement perturbation and high computational complexity. Newer techniques have been proposed to tackle the stability such as approximate Prony method \cite{potts2010parameter}, and to allow one step low-complexity alternative computation such as the matrix pencil method \cite{matrixPencil}, which employs eigenvalue decomposition. 
\subsection{Super-resolution Techniques in Wireless Systems}
A common example is the frequency response of a multipath channel, which is estimated at discrete frequencies such that
  \begin{equation}
 	\tilde{h}[k] = \sum_{l=0}^{L-1} h_l e^{j2\pi k\Delta_f \tau_l },
 \end{equation} 
and it is required to estimate $\tau_l$ and $h_l$, e,g.,  for localization applications \cite{Li202204} .  
A similar problem arises in estimating \ac{DoA} using a linear array. The measurements are obtained from antenna elements and the goal is to estimate the number and directions of multiple signal sources. This  problem has been solved differently using  signal subspace techniques such as MUSIC and ESPRIT \cite{MUSIC}  to tackle the impact of noise. However, these  approaches require some knowledge of the number of parameters. There are methods like \cite{bazzi2016jaded} that do parameter estimation (\ac{DoA}/\ac{ToA}) without the knowledge of the number of sources and without resorting to compressed sensing type methods. Its limitation is that it relies on the geometry being used, i.e. only works for uniform linear arrays, even though there are some transformations between array geometries like \cite{wax1994direction}.
To overcome this limitation, optimization techniques based on norm $L_1$ are proposed in \cite{sparse}. Optimization techniques are also suitable for  multidimensional cases, where the exponential functions are defined by multiple parameters. These are encountered in 2D problems such as estimating the delay-Doppler, azimuth and elevation angles, or 3D problems for estimating delay-Doppler-angle in \ac{MIMO} radar \cite{heckel2016super}.  The sparse problem formulation follows atomic norm minimization, 
\begin{equation}
\min_{\ma{x}} \|\ma{x}\|_{\alpha} \mbox{ subject to } \ma{y} = \ma{A}\ma{x}, \label{eq:sparse}
\end{equation} 
where, $\ma{y}$ is the vector of observations, $\ma{A}$ is the system matrix, and $\ma{x}$ is the sparse vector under estimation. The atomic norm is used to measure the sparsity and can be replaced by $L_1$ norm. However, this formulation ignores the impact of the noise in the solution. %$\min \Vert \ma{x} \Vert_\alpha \Vert y - \ma{A} \ma{x} \Vert< \epsilon$ generalizes \eqref{eq:sparse} to include some margin on the fitness of the affine constraint, through $\epsilon$. This is known as the basis-pursuit denoising by \cite{}.

Based on that, the same sequence of measured data can be processed differently to provide various solutions with different capabilities. This raises the question about the best structuring of data, especially in high-dimensional parameter estimations. On the other hand, most of  the sparse-based solutions use  norm $L_1$, which  achieves satisfactory  results, especially in noise-free data. In practice,  the noise characteristics should be considered in the algorithm design  to maintain robust performance at different noise levels. It should be also noted that the noise correlation may change depending on the structuring of the data. Moreover, most of the techniques assume linear systems, and only consider the impact of low-pass filtering. Nevertheless, the non-linearity effects of the hardware need to be also studied. Even under the noise-free assumption, the available approaches show limited resolution and impose requirements on the measured signals, such as the relation  between the  bandwidth and time resolution, and the relation between array geometry and angular resolution. 

Thus, one of the research challenges is to investigate the limits of estimation performance for given measurements in the presence of noise and nonlinearity. Another research direction is to study approaches for  improving parameter estimation by fusing different measurements at different bands and from different sensors.  Furthermore, for applications that rely on parameter estimation, such as localization, it is important to provide the estimation error distribution to evaluate the performance and derive the requirements for resolution and estimation accuracy.  Finally, super-resolution approaches are, in general, non-linear problems, and deriving an optimal solution is a hard problem. Thus, investigating new techniques based on machine learning is a natural step. 
\subsection{Summary}
Super-resolution techniques have the potential to infer information from ubiquitous radio signals in 6G, such as delay, Doppler, and angles, which can be used to achieve extreme performance metrics such as extreme positioning accuracy. High position precision is needed in several 6G scenarios such as robotics, industrial automation, healthcare, smart cities, as well as augmented and virtual reality applications.
The complexity scales with the number of signal sources, which will be available from different technologies (e.g. mmWave, sub-THz) and distributed deployment of radio units (e.g. distributed MIMO and RIS).
 Moreover, considering non-ideal effects such as hardware impairments, and interference, the conventional approaches may not work properly. Therefore, research is required to develop new methods and investigate the theoretical limits to benchmark the performance and provide insights into the complexity-performance trade-off. 

\section{Thermodynamics of Computation and Communication} \label{sec:thermo}
In previous generations of mobile networks, the main source of energy consumption is the radio unit, and in particular, the energy consumed by the analog \ac{RF} frontend and the transmit power. Moving towards 6G and with the increased demand for high data rate, the energy consumption of the baseband processing and \ac{DSP} becomes more significant \cite{6g_power}. Moreover, the native integration of \ac{AI} in 6G networks either to replace conventional network functions or as a service requires exploiting extra computation resources. Accordingly, 6G systems will act as a connect-compute platform to support varieties of distributed applications, especially, the AI-enabled use cases \cite{hexa_x_vision}.
 This section introduces a high-level view of the connect-compute platform with different types of wireless and wired communication and processing resources. Then, the joint design of communication and computation is discussed, showing an abstracted framework for the evaluation of latency and energy consumption. Moreover, we shed light on the theoretical limits of computation and communication from the perspective of quantum physics and thermodynamics. 
\subsection{Connect-Compute Platform}
\begin{figure}[t]
	\centering
	\includegraphics[width=.98\linewidth]{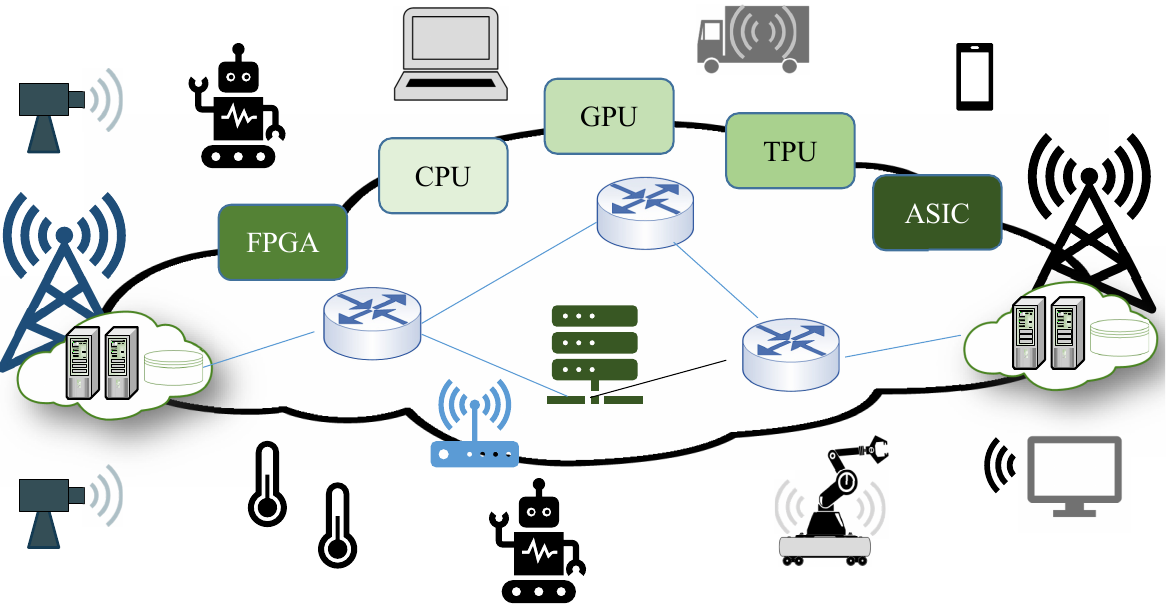}
	\caption{6G system as a connect-compute platform. Communication types involve heterogeneous wireless and wired technologies for connecting devices, computation platform, and network infrastructure. }\label{fig:network_platform}
\end{figure}

As depicted in \figref{fig:network_platform}, a 6G system encompasses devices of different communication and computation capabilities, ranging from simple \ac{IoT} sensors to mobile phones and network infrastructure equipment \cite{6g_platform}. Available computation hardware  includes on-the-shelf general purpose \acp{CPU}, special \acp{GPU} for image and signal processing,  \acp{TPU} for \ac{AI} applications, in addition to \acp{FPGA} and \ac{ASIC} as hardware accelerators. On top of this infrastructure, different computation platforms  provide  services in form of functions, virtual machines, or even allow access to the physical hardware.

\begin{figure}[t]
	\centering
	\includegraphics[width=.98\linewidth]{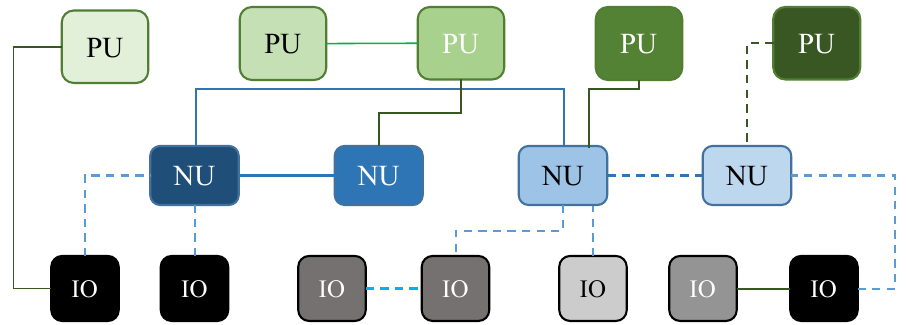}
	\caption{Abstracted model of connect-compute. The solid lines present wired connections such as a serial bus, twisted pair, or optical fiber. The dash lines express wireless connection.  }\label{fig:network_platform_model}
\end{figure}

In an abstracted view, as illustrated in \figref{fig:network_platform_model}, the connect-compute platform consists of \ac{I/O} terminals, \ac{PU}, and \ac{NU}.
An \ac{NU} can be a wireless access point, gateway, router, or switch. The \acp{I/O} represents a device connected  to an \ac{NU} or directly to another \ac{I/O} terminal. 
Some \acs{PU} are  available at the devices, and thus directly connected to an \ac{I/O} terminal, or co-located with an \ac{NU},  connected to other \acs{PU} to form a computation platform, and other \acs{PU} are deployed and connected on demands.  Diverse applications can be developed using this platform, such as AI-based 6G applications, where  data are collected from \ac{I/O} terminals and forwarded via \acp{NU} to train  models using the processing power of different \acs{PU}. 
\subsection{Joint Communication and Computation Co-design}
 In the context of distributed network applications,  joint communication and computation co-design plays an important role in fulfilling application requirements such as data rate, \ac{E2E}  latency, security,  and performance accuracy while minimizing energy consumption and reducing the cost \cite{joint_com_comp}. 
Such co-design considers, on the first hand, the network infrastructure planning and deployment, and on the other hand,  the optimization of runtime resource allocation depending on the running applications and the engaged devices. This implies the joint allocation of the communication and computation resources, e.g., to reduce the \ac{E2E} latency  \cite{joint_com_comp_latency}, or reduce the energy consumption 
\cite{joint_com_comp_latency_energy}, unlike the conventional approaches, where the network resource allocation is decoupled from the computation resources.  Namely, network resource allocation focuses on assigning and scheduling the  bandwidth and power, at different network segments, whereas computation resource allocation is performed by the corresponding operating system.  However, by joint design, the optimization problem considers the availability of both types of resources and aims to  exploit them considering communication channel conditions and other hardware constraints.

\subsubsection{Communication-computation runtime abstracted model}
\begin{figure}[t]
	\centering
	\includegraphics[width=.98\linewidth]{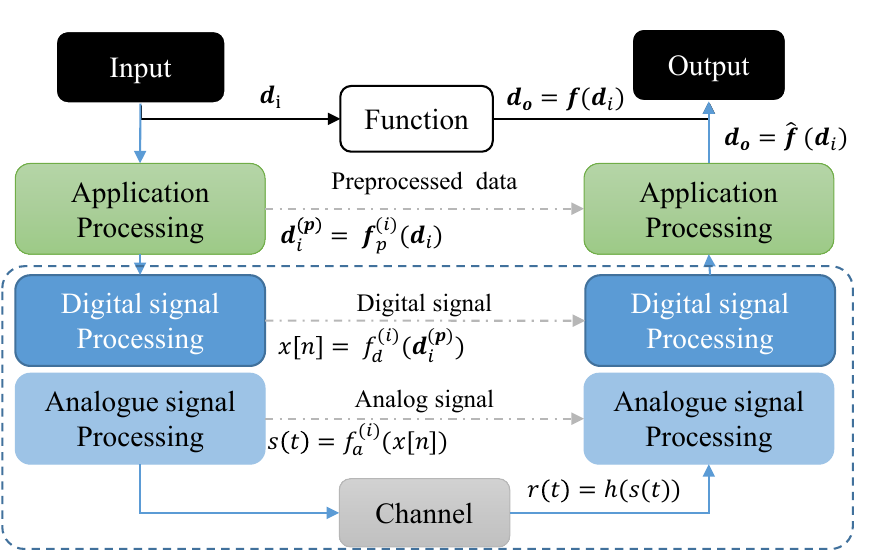}
	\caption{Communication and computation link model.}\label{fig:comunication_computation}
\end{figure}

An application can be modeled by a \ac{MIMO} function $\ma{f}(\cdot)$ that maps an input data vector $\ma{d}_i$ to an output vector $\ma{d}_o$ such that $\ma{d}_o = \ma{f} (\ma{d}_i)$. For example, in \ac{AI} model training,  $\ma{d}_i$ represents the training data, $\ma{d}_o$ the model parameters, and $\ma{f}(\cdot)$ the training algorithm. The implementation of this function depends on the studied scenario. Consider a single link, as shown in  \figref{fig:comunication_computation}, where $\ma{d}_i$ is generated by an input terminal, and the output terminal expects the output $\ma{d}_o$. When the input and the output are collocated with the processing platform that executes $\ma{f}(\cdot)$, the impact of communication cost can be neglected in comparison with the computation cost in terms of energy consumption and latency. In the other case, where the input and the output are remotely connected, communication needs to be involved. In fact, different possibilities can be examined; such as executing the function completely at the input terminal and communicating the results to the output terminal, or splitting the processing load over the input and output resources, as illustrated in \figref{fig:comunication_computation}. Accordingly, the main function can be decomposition into two functions $ \ma{f}^{(i)}_p$ and $\ma{f}^{(o)}_p$ at the input and output processing terminals, respectively such that 
\begin{equation}
\ma{f} (\ma{d}_i) =  \ma{f}^{(o)}_p\left(\ma{f}^{(i)}_p(\ma{d}_i)\right) = \left(\ma{f}^{(o)}_p \circ \ma{f}^{(i)}_p\right) (\ma{d}_i) .
\end{equation}
This corresponds to an exact decomposition, but such a case might not be necessary, and therefore, $\ma{f}^{(o)}_p \circ \ma{f}^{(i)}_p = \hat{\ma{f}}$.
The communication complexity and cost depend on the transmission technology, the distance, and the required data rate.  For instance,  wired serial communication technologies, such as  USB3, allow high data rates for short distances with relatively low cost and energy consumption for communication processing. To achieve a high data rate at longer distances, optical fibers are exploited at the cost of additional hardware resources for signal processing. Accordingly, the modulation produces a discrete signal from the reprocessed data $\ma{d}^{(p)}_i =  \ma{f}^{(i)}_p(\ma{d}_i)$, and this can be expressed by a function 
\begin{equation}
x[n] = 	{f}_d^{(i)} \left(\ma{d}^{(p)}_i\right).
\end{equation}
To overcome the limitations of wired networks for mobility and to reduce the material cost of cables, wireless technologies provide alternative solutions. However, in addition to the \ac{DSP}, analog frontend hardware is employed to convert the electrical signal to a medium  signal such as electromagnetic waves, which requires more energy to provide sufficient power for conveying the signal at the target range. The signal at the output of the fronted function can be expressed as 
 \begin{equation}
 s(t)=  	{f}_a^{(i)}(x[n]) =  	\left({f}_a^{(i)} \circ {f}_d^{(i)}\right) \left(\ma{d}^{(p)}_i\right).
 \end{equation}
After propagating through the channel, the received signal is given by 
 \begin{equation}
	r(t)=  	h(s(t)) =  	\left(h\circ {f}_a^{(i)} \circ {f}_d^{(i)}\right) \left(\ma{d}^{(p)}_i\right).
\end{equation}
The receiver functions aim at inverting the transmitter functions to deliver the processed data $\ma{d}^{(p)}_i$ such that 
 \begin{equation}
	\ma{d}^{(p)}_i =  \ma{f}^{(i)}_p(\ma{d}_i) =  	{f}_a^{(i)}(x[n]) =  	\left({f}_d^{(o)} \circ {f}_a^{(o)}\right) (r(t)).
\end{equation}
Achieving this requires reliable communication, where the requirements depend on the channel function. This impacts the design of the analog functions  ${f}_a^{(o)}$ and  ${f}_a^{(i)}$, which are implemented on analog circuits, and the \ac{DSP} functions  ${f}_d^{(o)}$ and  ${f}_d^{(i)}$. The complexity and energy consumption of such a function depend on the data rate of $\ma{d}^{(p)}_i$. Thus, when the preprocessing compresses the input data, the communication requirements can be relaxed. 
\subsubsection{Digital video transmission example}
The concept can be clarified by the application of video transmission, where the input terminal is a camera and the output is a display. 
The camera sensors convert the optical scene to electrical signals, and these signals are sampled and quantized, which results in a raw data bit stream. At the display,  the  bit stream  is converted to an analog signal to drive the display elements such as the emitting diodes in an \ac{OLED}. When both the camera and display are of the same resolution and refresh rate, the corresponding function is identity $\ma{I}$, i.e  $\ma{d}_o = \ma{I}(\ma{d}_i) = \ma{d}_i$. 

The raw data rate $R_d$ depends on the number of pixels per frame $S$, the quantization resolution $Q$, and the frame refresh rate $F_r$. Considering three colors per pixel, then 
\begin{equation}
	R_d = 3 F_r S Q. 
\end{equation}
For instance, in full HD video, $S = 1920 \times 1080 $ pixels, $F_r = 60$ Hz, $Q = 8$, then $R_d = 2.986 \times 10^9$ bps.
By using a video encoder such as H.264, where the preprocessing function $\ma{f}^{(i)}_p$ performs compression, the data rate of the compressed data $\ma{d}^{(p)}_i$ using a compression ratio $1/216$  becomes only $R_c = 13.815\times 10^6 $. This compression is essential to reduce the video size for storage and to reduce the bandwidth requirements for transmission.  The video decoder at the output terminal, which is represented by  $\ma{f}^{(i)}_p$, aims at reconstructing the raw data, such that $\ma{f}^{(o)}_p\circ \ma{f}^{(i)}_p = \ma{I}$. Nevertheless, typical video compression schemes are lossy and irreversible, which means that the original raw data cannot be perfectly reconstructed. However, the human perceived quality may not be degraded by a lower compression ratio, but rather by the techniques that require significant computing power to achieve high fidelity. Therefore, when the video transmission is over a link that supports the raw data rate with low-cost communication such as over HDMI and USB3 cables, compression should not be used. In the case of a wireless link, the decision is based on the available radio and computation resources. 
\subsection{E2E Latency and Energy Consumption}
The \ac{E2E} latency: as defined by 3GPP TS 22.261 is  "the time that takes to transfer a given piece of information from a source to a destination, measured at the communication interface, from the moment it is transmitted by the source to the moment it is successfully
received at the destination". Based on that, it can be formally defined by the time difference between consuming the first bit of the input data $\ma{\ma{d}}_i$ and producing the last bit of the output data $\ma{\ma{d}}_o$. In the context of tactile internet, \cite{itu2014tactile}, the \ac{E2E} latency consists of air interface latency (modem at the user device and base station, and propagating delay), computing latency, and server latency. From an implementation perspective, \cite{saifullah2014end}, this latency can depend on the data acquisition, buffering, communication, and processing architecture. For instance, the process of sampling a signal is sequential and the delay of completing a frame acquisition is larger or equal to the frame duration. Reading the data frame from memory for processing induces a delay that is related to the speed of memory access. The communication latency is governed by the propagation delay, in addition to the delay induced by the analog and digital processing, as well as the retransmission delay when applied. The processing delay depends on the architecture, where the sequential execution like in \ac{CPU} requires times proportional to the number of operations, whereas the parallel processing with combinatory logic used in hardware accelerators is impacted by the propagation delay of the circuits. Referring to  \figref{fig:comunication_computation}, the \ac{E2E} latency of a single segment assuming reliable communication without retransmission is the sum of all delays between the input and output, this can be expressed as 
\begin{equation}
	\tau_{\text{E2E}} = \set{T}\left(\ma{f}^{(i)}_p\right)+\set{T}\left(\ma{f}^{(o)}_p\right)+\set{T}\left(\ma{f}^{(i)}_d\right) +\set{T}\left(\ma{f}^{(o)}_d\right) + \tau_{\text{ch}},
\end{equation}
where $\tau_{\text{ch}}$ represents the channel latency, including the propagation delay and the analog hardware latency. The other terms correspond to the latency of the computation and communication functions.  In the application requirements, $\tau_{\text{E2E}}$ needs to be smaller than a threshold, $\tau_{\text{ch}}$ can be estimated from the deployment scenario. The other terms can be tuned according to the communication-computation co-design considering other constraints on the hardware processing architecture, which impacts energy efficiency.

The main source of energy consumption in wireless communications systems is the analog fronted, where sufficient transmit power is required according to the range needed for achieving reliable communications \cite{energyconsumption}.  The \ac{PA} efficiency is constrained by the electronics and depends on several factors such as the frequency, bandwidth, and the input signal envelope. This indirectly impacts the design of the waveform at the \ac{DSP} functions. Accordingly, current works only consider the energy consumption of the analog functions in addition to the application computation in the co-design \cite{joint_com_comp_latency_energy}. Nevertheless, the \ac{DSP} needs also to be considered in the evaluation of energy consumption. Therefore, the total energy consumption can be defined as 
\begin{equation}
	E_{\text{E2E}} = \set{E}\left(\ma{f}^{(i)}_p\right)+\set{E}\left(\ma{f}^{(o)}_p\right)+\set{E}\left(\ma{f}^{(i)}_d\right) +\set{E}\left(\ma{f}^{(o)}_d\right) + E_{ch},
\end{equation}
where $E_{ch}$ is the energy consumption at the frontend, which depends on the transmission parameters (bandwidth, range,  frequency, and antenna configuration), in addition to the data rate. This term can be estimated from the analog components models and the link budget required to achieve reliable communication. As discussed in this section, the data rate also  impacts the energy consumption of the application computation and signal processing. Based on that, the energy optimization problem depends on the data rate, which is related to the compression ratio on one side, and on the other side, it determines the required resources. 
 \subsubsection{Energy consumption and latency optimization}
 Combining both latency and energy, the main optimization problem for an application can be formulated as 
\begin{equation}
	\min~ E_{\text{E2E}},~~\text{s.t.} ~\tau_{\text{E2E}} \le \tau_{\text{app}},
\end{equation}
where $\tau_{\text{app}}$ is a threshold defined by the application requirements.  From this basic problem, various problems can be derived depending on other practical constraints related to the scenario,  availability of  resources, processing architecture, and hardware technical  limitations.  In pure computation scenario,  where the application function $\ma{f}$ is to be computed, the problem can be translated to find a computation architecture that achieves the latency requirements with minimum energy consumption. This can also be expanded with technical constraints on the memory size and speed, area of the chip,  clock rate, etc. On the other corner, when the problem is pure communication, the aim is to minimize the energy per bit by designing the digital and analog functions constrained by the available technology. 

Optimization problems for certain applications may not have feasible solutions using certain technologies. For instance, achieving low-latency  software-based video coding using an old computer is not feasible, but it becomes affordable with new generations. High-speed communication was not available as it is today. This is indicated by Moore's law as a human law on how technology develops and gains more capabilities over time. However, the question that arises is what are the physical limitations?
\subsubsection{Physical limits of communication and computing}
 The question can be reformulated as follows given no constraints on the technology, what are the physical limits of communication and computing? Answering these questions is the subject of the study of quantum physics and thermodynamics. In particular, to determine the  physical limitations of the energy required to transmit a bit over a certain distance and what is the minimum delay for that, and what is the energy required to perform a certain operation in a given time. As discussed  in \cite{limitofcomput}, to perform an elementary logical operation in time $\Delta t$ the required energy is given by 
\begin{equation}
	E\ge \frac{\pi}{2} \hbar \Delta t, \label{eq:energy of compute}
\end{equation}
where $\hbar = 1.0545 \times 10^{-34}$ joule/second. 
The physical limit of communication data rate, employing power $P$, and using a channel modeled as a region in space with cross-section $A$ is discussed in  \cite{limitofcomp}, and it is given by 
\begin{equation}
	R\le \sqrt{\frac{AP}{l_p^2 \hbar}},
\end{equation}
where $l_p = 1.6 \times 10^{-35}$ m, and $P = \frac{E}{\tau}$, $\tau$ is the delay. 
By evaluating the number of operations required by an algorithm within a latency constraint,  and the  involved communication in terms of data rate and range,  it is possible to estimate the physical limit of the energy requirements.
\subsubsection{Computational complexity} 
The computational complexity of an algorithm can be measured by the number of primitive  arithmetic operations such as multiplications/divisions, summations, and subtractions. When considering digital computation, each preemptive operation is implemented by a number of universal logical operations that depend on the word length of the digital representation. For example, the basic logical functions {AND, XOR} are common in implementing arithmetic operations. Accordingly, the number of elementary bit operations can be estimated,  and by giving a time constraint, from \eqref{eq:energy of compute}, the energy limit can be estimated. 
Such estimation can be used to determine the energy requirements of processing hungry applications, such as \ac{DL} \cite{DL_limit}. Such estimation provides a reference, which can be very optimistic in comparison to the current technologies. In addition, when employing distributed learning to speed up the training, there can be extra processing cost and communication that needs to be considered when designing the training algorithms.
\subsection{Summary}
 In summary, 6G will bring new distributed applications with extensive computation enabled by extreme communication in terms of high data rate and low latency, and the integration of computation platform. To achieve a sustainable system, the implementation of new applications, which are mostly \ac{AI}-based ones, should consider joint optimization of communication and computation. As a methodology that is independent of the technology, relying on  the physical limits of communication and computation to evaluate the implementation will give an insight into the energy cost. However, in practice, different constraints should be considered. The main challenge in this context concerns finding solutions to varieties of optimization problems, that can be adapted to different constraints, rather than solving hundreds of individual problems. Thus, a joint communication and computation optimization framework is an interesting challenging topic for 6G research.

\section{Signals for Time Varying Systems Theory} \label{sec:varying}
%\hl{Marwa}
6G is expected to support mobility of terminals in a wide range of use cases including V2X communications, \acp{UAV}, wireless connectivity for high-speed trains among others.
In such scenarios, the wireless channel becomes doubly selective, meaning that the channel is selective in both time and frequency domains. Estimating such channels or designing robust waveforms against the selectivity of the time-varying channel is an active research area that raises several research questions. In this section, we will first overview the channel properties in fast fading, then highlight the recent advances of robust signal waveform design against time-varying channels, and provide open research questions.
\subsection{Doubly-selective Channels}
\begin{figure*}[ht]
	\begin{subfigure}[b]{0.5\linewidth}
		\centering
		\includegraphics[width=\textwidth]{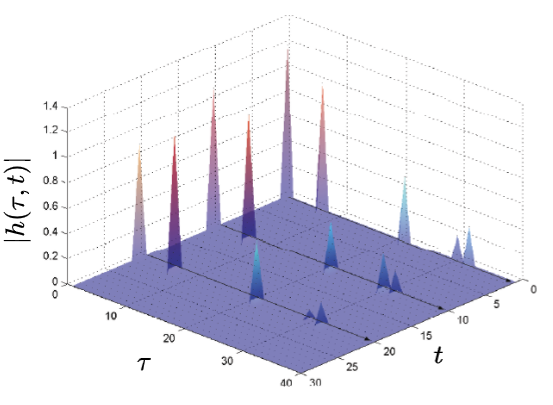}
	    \caption{\label{CIR} Channel impulse response ${\ma{h}}(\tau,t)$}
	\end{subfigure}\hfill
	\begin{subfigure}[b]{0.5\linewidth}
		\centering
		\includegraphics[width=\textwidth]{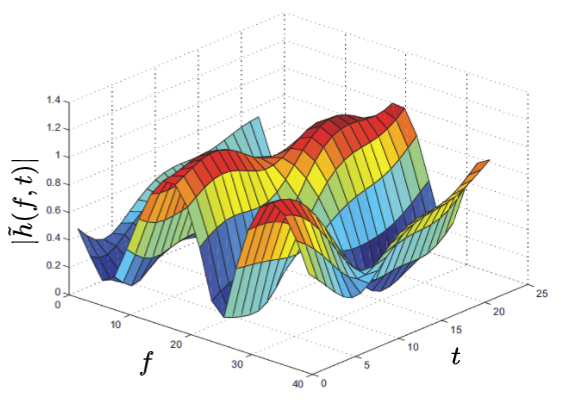}
		\caption{\label{CFR} Channel frequency response $\tilde{\ma{h}}(f,t)$}
	\end{subfigure}
	\caption{Channel impulse and frequency responses}.
	\label{fig:ChannelIFR}
\end{figure*}

 Due to multi-path propagation resulting from the signal traveling along different paths, and therefore received at slightly different times, the channel becomes frequency selective. Moreover, relative motion between the transmitter and receiver (or surrounding environment) causes random frequency modulation. Since each multi-path component has a different Doppler shift, the channel then experiences fast fading, and becomes doubly selective i.e. varies in time and frequency.
The doubly-selective wireless channel can be modeled as a \ac{LTV} channel that can be represented by a two-dimensional  impulse response
\begin{equation}
    \small
    \ma{h}(\tau, t) = \sum_{l = 0}^{L-1} \alpha_{l}(t)   \delta (\tau - \tau_{l}(t)),
    \label{eq:DT_R}
\end{equation}
where $\alpha_{l}(t)$ and $\tau_{l} (t)$ denote the channel propagation gain (known as attenuation) and delay of the $l$-th propagation path as functions of time, respectively. The channel frequency-time response, i.e the time-variant transfer function is the Fourier transform of the impulse response with respect to the delay variable $\tau$:
\begin{equation}
    \small
    \tilde{\ma{h}}(f,t) = F_{\tau} \{\ma{h}(\tau, t)\} 
    %= \int_{-\infty}^{\infty} {\ma{h}}(t, \tau) e^{-j2\pi f \tau} d\tau 
    = 
    \sum_{l=0}^{L-1} { \alpha_{l}(t) e^{-j2\pi f \tau_{l}(t)}} .
    \label{eq:FT_R}
\end{equation}
Figure~\ref{fig:ChannelIFR} shows the \ac{LTV} impulse and frequency responses.
%

%%%%%%%%%%%%%%
%\begin{figure*}[t]
%\begin{subfigure}{.5\textwidth}
%	\centering
%	\includegraphics[width=2\columnwidth]{Figures/ImpulseA.png}
 %\caption{Channel impulse response ${\ma{h}}(\tau,t)$}
%\label{CIR}
%\end{subfigure}
 %\begin{subfigure}{.5\textwidth}
%	\centering
%	\includegraphics[width=2\columnwidth]{Figures/ImpulseB.png}
%\caption{Channel frequency response $\tilde{\ma{h}}(f,t)$ }
%  \label{CFR}
%\end{subfigure}
% \setlength{\abovecaptionskip}{6pt plus 3pt minus 2pt}
% \includegraphics[width=2\columnwidth]{Figures/Impluse_Frequency_Response.png}
% \centering
% \caption{Channel Impulse ${\ma{h}}(\tau,t)$ and frequency $\tilde{\ma{h}}(f,t)$ responses.}
% \end{figure*}
%%%%%%%%%%%%%%%%%%%%%%%%%%%
%%%%%%%%%%%%%%%%%%%%%%%%%%%%%%%%

%	\subfloat[\label{CIR} Channel impulse response ${\ma{h}}(\tau,t)$]{\hspace{.5\linewidth}}
%	\subfloat[\label{CFR} Channel frequency response $\tilde{\ma{h}}(f,t)$ ]{\hspace{.5\linewidth}}

The reliability of communication in such time-varying systems highly depends on how robust is the employed communication waveform against the channel variations and how accurate are the channel estimation and tracking \cite{molisch2012wireless}.
\paragraph{Channel Coherence Time} is defined as
\begin{equation}
    \small
    T_{c} = \sqrt{\frac{9}{16 \pi}} \frac{1}{f_{d}} = \frac{0.423}{f_{d}},
\end{equation}
where $f_d$ is the maximum Doppler shift. A slowly changing channel has a large coherence time with respect to the symbol duration, while a fast-varying channel has a small coherence time.
\paragraph{Coherence Bandwidth} is defined as the reciprocal of the multipath spread as follows:
\begin{equation}
    \small
    B_{C} = \frac{1}{ \beta~\tau_{max}}, 
\end{equation}
where $\beta$ is a variable factor related to the definition of the coherence bandwidth. If $B_{C}$ is small with respect to the bandwidth of the transmitted signal, the channel is said to be frequency selective, in this case, the transmitted signal is distorted by the channel. On the other hand, if $B_{C}$ is large in comparison with the bandwidth of the transmitted signal, then the channel is said to be frequency flat.
\subsection{Spreading Waveforms}
\paragraph{Equal Gain Criterion}
In \cite{bomfin2018theoretical, bomfin2021robust}, it has been demonstrated that, under the assumption of \ac{CSI} only available at the receiver side, and when a perfect feedback equalizer \cite{studer2011asic, benvenuto2009single} is employed, an optimal waveform in terms of coded modulation capacity, needs to spread the data symbols over the domain of selectivity of the channel, such that all data symbols experience the same gain. This criterion has been referred to in the literature as the equal gain criterion. Hence, in this sense, \ac{OFDM} and \ac{SC} are not optimal in doubly-dispersive channels, while spreading waveforms such as \ac{OTFS} \cite{nimr2018extended}, Walsh Hadamard modulation \cite{bomfin2020robust}, \ac{BM} \ac{OCDM} \cite{bomfin2019performance} are optimal in doubly-selective channels.

\paragraph{Block Multiplexing for Time Selective Channels}

\begin{table*}[t!]
	\centering
	\renewcommand{\arraystretch}{1.3}
	\caption{Expression and complexity of the waveform matrix ${\ma{A}_{{\rm F}}}$.}
	\label{tab:spreading}
	\begin{tabular}{|l|c|l|l|}
		\hline
	Spreading waveform	&  ${\ma{A}_{{\rm F}}} = \kronProd{ \mathbf{B}}{ \F_{M} ^{\text{H}} \mathbf{A}'}$   & Multiplications & {Additions}\\ 
		%		\hline 
		%		OFDM& $\ma{I}_N$ &  $N\log_2 (K)$ & $2N\log_2 (K)$\\ 
		\hline 
		SC& $\kronProd{\ma{I}_M}{\F_{K}}$   & $N/2 \log_2 (K)$ & $ N \log_2 (K)$ \\ 
		\hline
		OCDM& $\kronProd{\ma{I}_M}{\ma{\Gamma}_K \F_{K}}$   & $N/2 \log_2 (K) + N$  & $ N \log_2 (K)$ \\ 
		\hline 
		OTFS (BM-SC)&  $\kronProd{\F_{M} ^{\text{H}}}{\F_{K}}$  & $N/2\log_2(N)$ & $ N\log_2(N)$\\ 		
		\hline 
		BM-OCDM &  $\kronProd{\F_{M} ^{\text{H}}}{\ma{\Gamma}_K \F_{K}}$  & $N/2\log_2(N) + N$  & $ N\log_2(N)$\\ 	
		\hline
		SWH &  $ \left(\kronProd{\ma{W}_{Q'}}{\ma{I}_{P'}}\right)\otimes \left(\kronProd{\ma{W}_{Q}}{\ma{I}_{P}}\right)$  & $N$, {\small $\sqrt{QQ'}$ not radix-2}&  $  N \log_2(Q Q')$  \\ % &{\tiny ($\sqrt{N}$ not integer)} \\ 	\cline{3-4}
		%		 &   & $N\log_2(K)$ & {\tiny ($\sqrt{N}$ integer)}\\ 	
		&  & 0, {\small $\sqrt{QQ'}$ radix-2} &\\	           
		\hline 	  	
	\end{tabular} 
\end{table*}

Spreading waveforms have been studied in a general framework of \ac{BM} modulation \cite{bomfin2021robust}, which multiplexes the data vector in sub-blocks so that the data symbols reach the receiver with the same gain. Although the unpredictable time-varying nature of the channel prevents the design of a perfect equal gain modulation scheme, \ac{BM} spreading waveforms satisfy the equal gain criterion under the assumption that the channel is practically static during one sub-block transmission.

Let $N=MK$ such that $K$ is the number of subcarriers and $M$ the number of sub-blocks. $\ma{I}_M$ stands for the identity matrix of size $M$, and $\F_{K}$ refers to the unitary Fourier transform matrix of size $K$.
Let $\mathbf{A}$ be the modulation matrix in the time domain, it is defined by the \ac{BM} structure as follows:
\begin{equation}\label{eq:A_conv}
\mathbf{A} =  \mathbf{B} \otimes \mathbf{A}', %= \begin{bmatrix} \mathbf{A}' & & \\   & \ddots & \\ & & \mathbf{A}' \end{bmatrix},
\end{equation}
where $\otimes$ stands for the Kronecker product, $\mathbf{A}' \in \mathbb{C}^{K \times K}$ is the modulation matrix per sub-block, and $\mathbf{B} \in \mathbb{C}^{M \times M}$ is the \ac{BM} precoding matrix.
In the frequency domain, the modulation matrix is expressed as:
\begin{equation}\label{eq:A_F2}
\begin{split}
\mathbf{A}_{\rm F}
& = \mathbf{B} \otimes \F \mathbf{A}'\\
& = \mathbf{B} \otimes \mathbf{A}_{\rm F}'.
\end{split}
\end{equation}
\begin{figure*}[ht]
  \includegraphics[width=\textwidth]{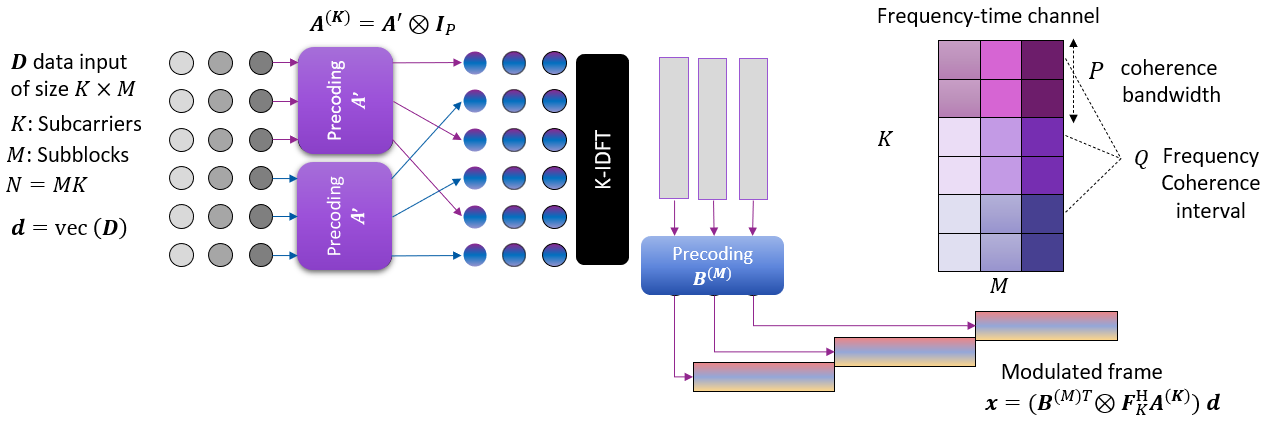}
  \caption{Implementing time and frequency spreading waveforms.}
  \label{fig:spreading}
\end{figure*}

Figure~\ref{fig:spreading} illustrates the steps for designing full \cite{bomfin2021robust} and sparse \cite{bomfin2020robust} spreading waveforms. By sparse spreading, we mean that the energy of the symbols is spread only over the coherence bands and intervals instead of fully spreading the symbols over all the subcarriers and subblocks. In this figure, the input circles represent input data symbols e.g. \ac{QAM} symbols. We first apply precoding to spread the data in the frequency domain, and then we apply the \ac{IFFT}. At the output of the IFFT, our waveform can be regarded as precoded \ac{OFDM}. After that, another precoding is applied in order to spread the data over time. The resulting signal has its symbols spread over the bandwidth and the time frame.

\paragraph{Conditions on Designing Spreading Matrices}
In order to spread the data symbols over all sub-blocks with no coupling at the transmitter,  $\mathbf{A}_{\rm F}$ should be orthogonal, i.e., $\mathbf{A}_{\rm F}\Ht\mathbf{A}_{\rm F} = \mathbf{I}_N$, otherwise the data symbols will be coupled by the transmitter in addition to the channel, causing an extra distortion of the signal which decreases the performance with respect to the orthogonal case.

In order to spread the data symbols equally over all sub-blocks, we constraint $| \mathbf{B}[m,m']|^2 = 1/M \, \forall \, _{m,m'}$. 
Another consideration is the low-complexity implementation of the transmitted signal, which is achieved by considering \ac{FFT} or Fast \ac{WHT}-based transforms.
 
Note that setting $\mathbf{B} = \mathbf{F}_M$ and $\mathbf{A}_{\rm F}' = \ma{F}_K$ leads to the OTFS modulation, which can be interpreted as single-carrier with BM, namely, BM-SC, because $\mathbf{A}_{\rm F}' = \ma{F}_K$ is the SC waveform in the frequency domain.
Similarly, we can also have the BM-OCDM by setting $\mathbf{B} = \mathbf{F}_M$ and $\mathbf{A}_{\rm F}' = \ma{\Gamma}_K\ma{F}_K$, where $\mathbf{\Gamma}_K[n,n]$ is a diagonal matrix defined as
\begin{equation}
\mathbf{\Gamma}_K[n,n] = \exp(j \pi n^2/K).
\end{equation}

WH waveform is achieved by setting $\mathbf{B} = \ma{W}_M$ and $\mathbf{A}_{\rm F}' = \ma{W}_K$. The \ac{WHT} is recursively defined by
\begin{equation}
\begin{split}
\ma{W}_N &= \kronProd{\ma{W}_2}{\ma{W}_{N/2}},~
\ma{W}_2 = \frac{1}{\sqrt{2}}\left[\begin{array} {lr}
1 & 1\\
1 & -1
\end{array}\right].\label{eq:WH}
\end{split}
\end{equation}
One clear advantage of the \ac{WHT} in comparison to the FFT-based precoding matrices is that its construction uses only 1 and -1, i.e., the transform does not require multiplications besides normalization. 

The choice of spreading precoding matrices also depends on the application requirements. For example, Walsh Hadamard modulation and sparse spreading waveforms achieve the same \ac{FER} performance as \ac{OTFS} while significantly decreasing the complexity of the modem. Another example is the \ac{CP}-free transmissions, where \ac{BM}-\ac{OCDM} has been shown to perform better than \ac{OTFS} \cite{bomfin2021robust}.
In Table~\ref{tab:spreading}, we summarize the expression and the complexity of the waveform matrix for different spreading waveforms. 
 
\subsection{Challenges}
One of the main challenges of spreading waveforms is that they require an iterative receiver in order to resolve \ac{ISI} e.g. \ac{LMMSE-PIC}. Iterative receivers are in general highly complex, which prevents deploying spreading waveforms in practice with iterative equalization. Researchers have been working on reducing the receiver's complexity of the spreading waveforms  \cite{bomfin2019low, thaj2020low}, by making approximation of the LMMSE matrix. While these methods reduce the computational complexity of the original iterative receiver, they are still difficult to implement in practice. Some recent works \cite{enku2021two,naikoti2021low} propose using ML-based receivers in order to approximate the non-linear iterative receivers with low complexity, while others \cite{9813719,gizzini2021cnn,gizzini2020deep,gizzini2021temporal} propose improving channel estimation in doubly dispersive channels using deep learning.
However, supervised machine learning methods need a large amount of training data, and they do not generalize well when the environment changes. Therefore, the current methods investigated in the literature suffer from practical problems, and investigating a low-complexity receiver for time-varying systems is an open challenge. %\textcolor{blue}{add here something like: spreading waveforms promise to improve the FER reliability by XXX order of magnitude, hence solving the receiver complexity challenge would make 6G XXX times more reliable in high mobility applications}

In addition to the receiver complexity, another challenge is to define a clear tradeoff between the accuracy of the channel estimation and the required overhead. For well-defined system parameters, we should be able to know how much control overhead is needed for the required reliability. This is directly related to the frame design problem, where we should define the best pilot allocation scheme for channel estimation. 
In \cite{ehsanfar2020uw}, it has been shown that unique word-based frame design achieves nearly one order of magnitude higher channel estimation than conventional pilot-aided \ac{CP} based design, which is explained by the fact that training sequences in the first scheme are concentrated in a short time slot compared to the coherence time, and then experience less number of channel realizations. However, this scheme suffers from high complexity at the receiver. A practical frame design for a time-varying system is still an open research problem.

Future 6G communications should rethink the whole radio design by considering a flexible waveform that adapts to the channel selectivity while utilizing the available resources and minimizing energy consumption. A similar concept has been referred to in \cite{fettweis20216g} as Gearbox PHY,  which provides different gears i.e. radio architectures options. Each gear refers to different spectrum bands and transmission schemes with adaptability in the modulation and coding schemes. The flexibility of switching between gears should be decided based on the channel conditions and the available communication resources such that the most energy-efficient hardware implementation is selected. For instance, if a considerably large contiguous bandwidth is available for only one user, it should employ techniques with low spectral efficiency that can be implemented with 1-bit ADC resolution \cite{neuhaus2021enabling} which consumes less energy in the terminal.
Investigating this concept and implementing it in practice is an ongoing active research problem.

%Additionally, numerical evaluations comparing the performance gain of OTFS in relation to SC under different channel configurations have been conducted. In particular, OTFS reduces the FER of SC to half under 350km/hr and block duration of 128μs, which corresponds to 1.367 times the coherence time. For a FER decrease of 10 times, the transmission time should be approximately 9.373 times higher than the coherence time, which is a considerable high number and typically is not met in sub-6GHz systems.

\subsection{Summary}
One of the main KPIs of 6G is to support high mobility for different applications. In these scenarios, the communication channel becomes time-varying and hence challenging to estimate. Communication signals can be designed properly to be robust against channel variations. Spreading waveforms, which spread the symbols over time and frequency, have been demonstrated to be resilient to channel time variations and frequency selectivity. Enhancing channel estimation improves the reliability of communications. However, spreading waveforms require iterative receivers to mitigate \ac{ISI}. Such receivers are highly complex which prevents spreading waveforms from being deployed in practice. The most important research direction for communication signals in time-varying channels is designing low-complex receivers for spreading waveforms. Rethinking the frame design, pilot allocation, and adaptive waveform design are also active research problems in this area.

\section{Semantic Communication Theory} \label{sec:semantic}
%\hl{Lina}
Over the last couple of decades, the development of wireless technologies has fundamentally relied on Shannon’s information theory, which mainly focuses on how accurately and reliably symbols can be transmitted. Accordingly, all network segments, from the core to the edge, have been optimally designed with the aim to ensure that wirelessly transmitted messages over noisy channels are correctly received and decoded, while the meaning of the conveyed messages has been considered less important, and hence, it has been ignored during the design and optimization process \cite{shi2021semantic}. Although this approach has been an appropriate fit for current communication paradigms and has fulfilled the requirements of wireless networks in the earlier generations, it is debatable whether it will be sufficient in future 6G networks and beyond. With the strong emergence of machine-to-machine communication paradigms and the recent progression in solid-state technology, it is envisaged to witness revolutionary changes in the roles that machines will play in future wireless networks, which is foreseen to be much more than end devices that are leveraged to blindly transmit and receive. The key questions to be answered are: in the envisioned era of native \ac{AI}, where each network node is anticipated to enjoy a sort of intelligence, will machines be able to develop a level of common understanding that allows them to communicate meaningfully, without the human intervention? Doing so, how does incorporating the semantic aspect of communication enable the realization of pervasive service-based networks, with the aim to facilitate the delivery of knowledge and experience-based applications, including haptics, tactile internet, and holographic presence? 

\begin{figure*}[ht]
	\centering
	\includegraphics[width=1\linewidth]{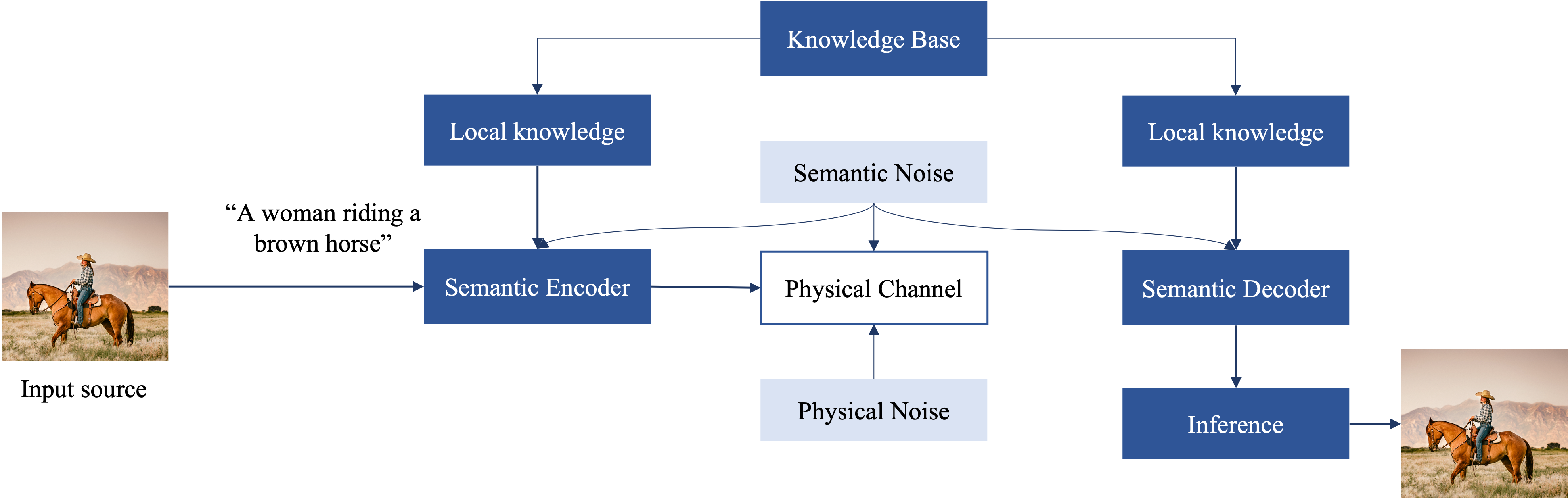}
	\caption{Architecture of semantic communication systems.}\label{fig:SC}
\end{figure*}

In this respect, \textit{semantic communication} paradigm has recently emerged as a consequence for the rise of machine intelligence, where the tangible meaning of received messages can be understood/sensed and leveraged by machines, to autonomously perform human-centric tasks without the involvement of humans (see Fig. \ref{fig:SC}) \cite{lan2021semantic}. In other words, semantic communication conceptualizes the principle of humans’ languages, i.e., it enables machines to communicate meaningfully and to understand each others’ behavior, and hence, develop a common language-like type of communication.  It is envisioned that semantic communications will bring several unique advantages, pertaining to reliability, communication efficiency, and \ac{QoE}. In particular, in conventional data-oriented communication, continuous signals (e.g., speech) need to be discretized due to limitations imposed on channels capacity, yielding a tolerable level of information loss. Meanwhile, in semantic communications, it is anticipated to realize lossless information transmission, while relieving the demand on communication resources. This can be achieved by transmitting the meaning of the message conveyed over a continuous voice signal (the main idea in a compact form), instead of fully transmitting the whole speech (which can be hours-long), potentially reducing the communication overhead. From a reliability perspective, leveraging the semantic aspect of communication enables efficient prediction for missing information from a corrupted signal, by understanding the overall concept transferred over the wireless signal, and filling the gaps within the message context. Under the same concept, delivering the context, intentions, interpretations, or physical content (in other words all human-based semantic information) of a message allows the realization of improved \ac{QoE}.     
\subsection{Challenges}
Albeit the bright speculations of such a communication paradigm, several challenges have been identified when semantic communication is incorporated to develop the new generation of wireless applications. For example, in holographic-type communication, it is essential to create a close-to-real presence of a remote object, in order to enhance the virtual interaction capabilities and to deliver a fully-immersive experience. In order to achieve high-fidelity holograms, orders of terahertz data rates and sub-millimeter latency communication is required. In addition, to realize a realistic interactive and high-resolution virtualization, sophisticated encoding schemes of haptic information, including the object's/human's shape, details, position, and movements, need to be designed. Furthermore, it is equally important to direct the attention toward developing a delicate human-machine interface for improved semantic information transfer, and hence, enhanced real-time holographic interaction. From a wider angle, an all-sense five-dimensional communication, incorporating the sight, smell, taste, touch, and hearing senses, constitutes one of the 6G pillars. It should be highlighted that, in such a tactile communication framework, sensed haptic information are characterized in multi-dimensional representations, severely exaggerating the implementation complexity of semantic communication. 

From a different perspective, the speed and accuracy of acquiring the semantic information constitute key performance indicators in quantifying the efficiency of semantic communication frameworks. In this regard, it is articulated that data clustering, identification, classification, and recognition are conventionally performed through the exploitation of deep neural networks, which demand heavy computational complexity and the availability of a sufficient amount of labeled data. This calls for a need for energy-efficient and simple mechanisms for fast, efficient semantic information acquisition and identification. Such a problem is still an open research direction. Finally, from a mathematical point of view, the lack of efficient mathematical tools poses extra challenge to the implementation of the semantic theory. Owing to the unexplored mathematical foundation of the semantic theory, the quantitative metrics and the solid foundations that are used to quantify the performance of such a theory when implemented in different scenarios are yet to be understood. Despite the limited initiatives aimed to re-derive Shannon's theorem or propose non-Shannon-based semantic theorem \cite{baudot2015homological, vigneaux2019topology}, perspective studies on the entropy, as well as the algebraic, probabilistic, combinatorial, and dynamic behavior of the semantic communication, are not touched in the literature, rendering it an attractive open research problem. 

\subsection{Summary}
With the advancements of semantic communication technology, which conceptualizes the principle of humans’ languages, efforts should be directed towards investigating the integration of semantic communication in various use-cases of wireless networks, identifying performance bottlenecks, and tackling several challenges pertaining to relations between multiple objects, multi-user semantic communication, and semantics generalization. Moreover, it is essential to further explore the potential of extending the semantic concept into communication protocol learning. Different than classical medium access control (MAC) protocols, which are rather generic, uninterpretable, and with unadaptable control signaling messages, semantic communication protocols are anticipated to enjoy task-specific control signaling messages, with an acceptable communication, computing, energy consumption, and memory usage overhead.

\section{Signals for Integrated Sensing and Communications} \label{sec:sensing}

\begin{figure*}[!t]
\centering
\subfloat[]{\includegraphics[width=3.5in]{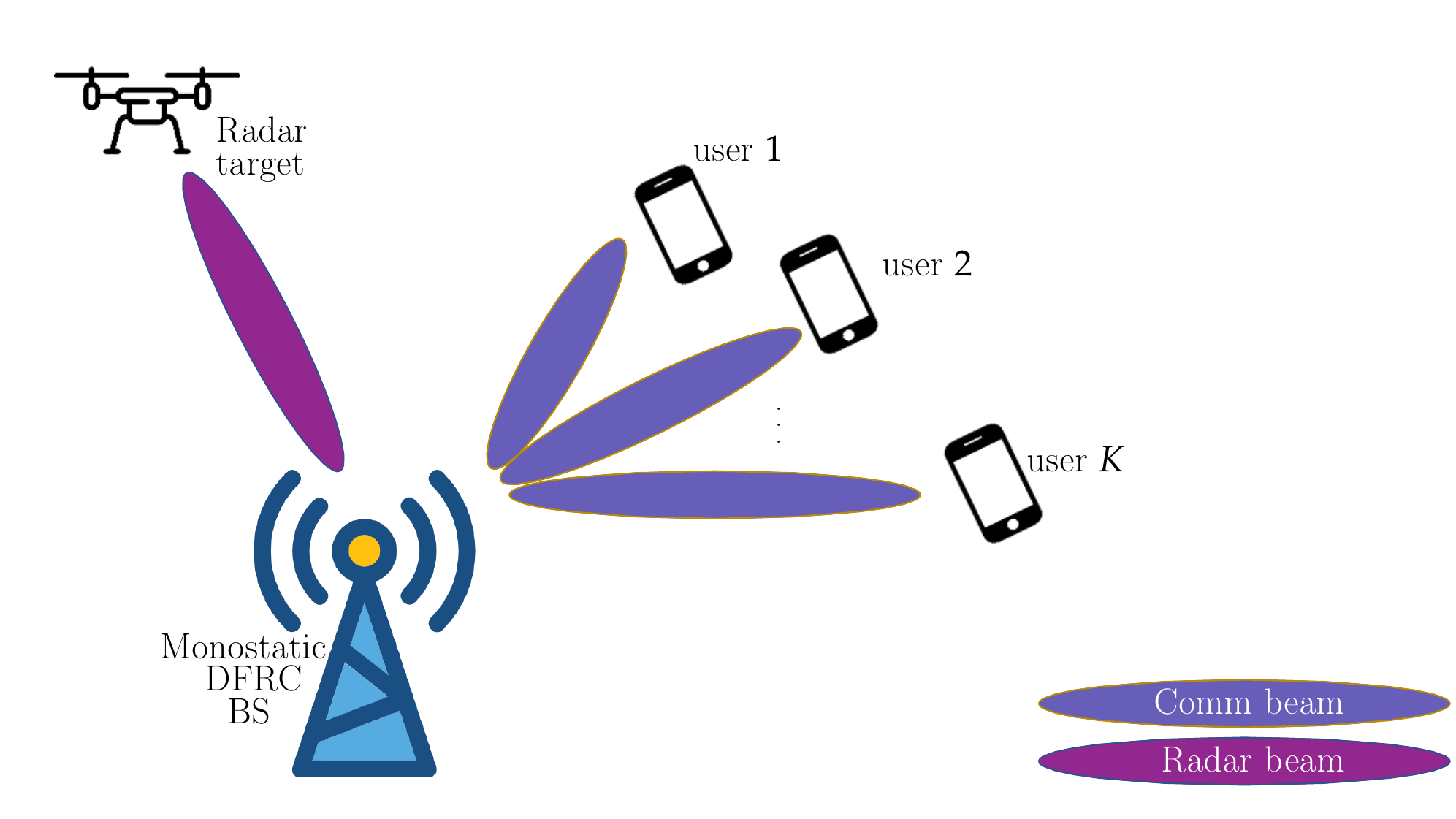}%
\label{fig_first_case}}
\hfil
\subfloat[]{\includegraphics[width=3.5in]{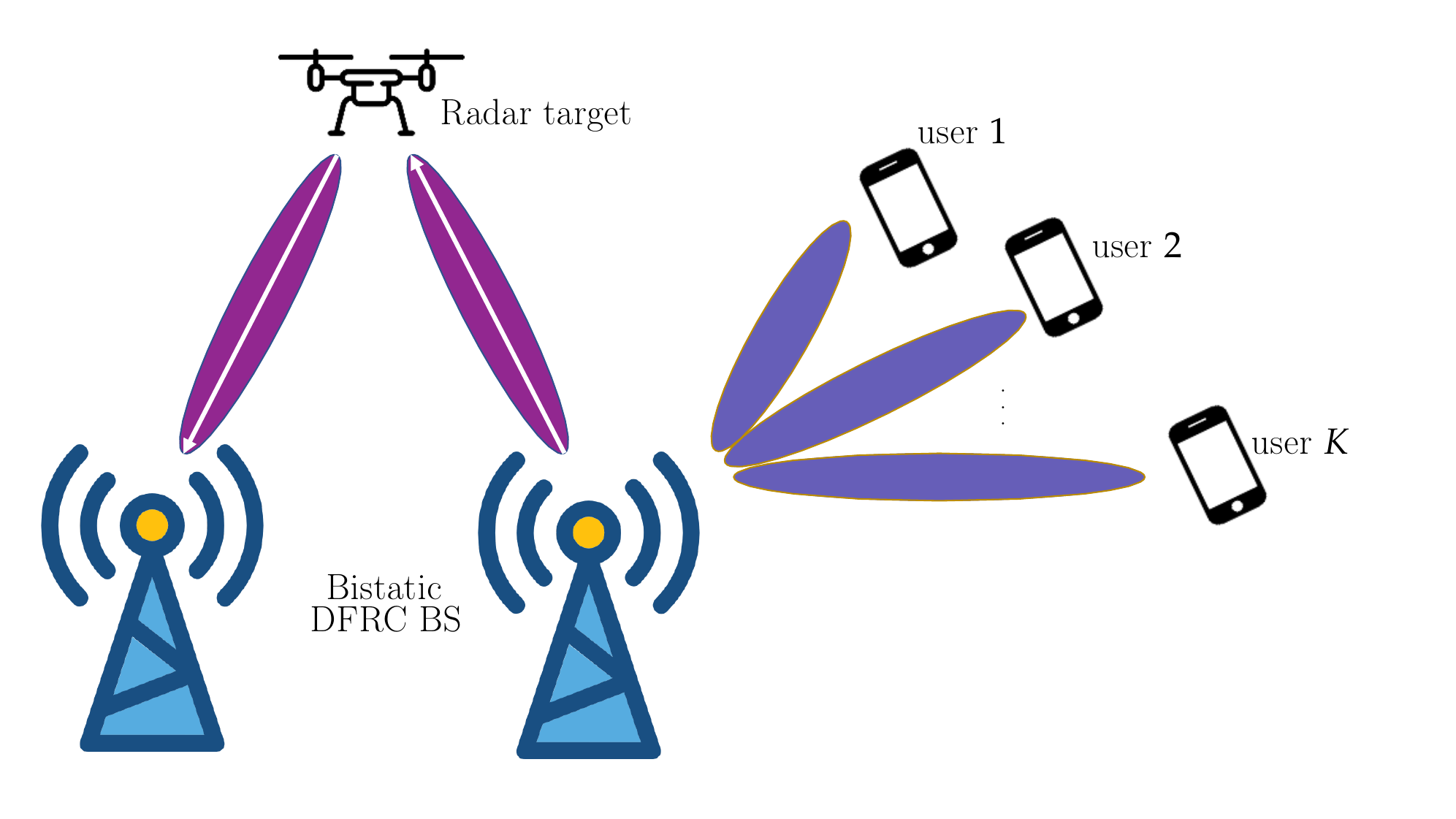}%
\label{fig_second_case}}
\hfil
%\subfloat[]{\includegraphics[width=3.5in]{Figures/fig3.pdf}%
%\label{fig_third_case}}
\caption{Integrated sensing and communication in (a) Dual-functional radar and communication base station operating in a monostatic fashion, where the task is to illuminate a radar beam towards an intended target and serve $K$ communication users, simultaneously (b) Dual-functional radar and communication base station operating in a bistatic fashion, where the task of both radar components is to steer the beam towards the radar and serve the users in the scene. %and (c) Monostatic DFRC operating in full duplex mode.
}
\label{fig_sim}
\end{figure*}

%\hl{Marwa}
Modern applications such as automatically controlled vehicles, \acp{UAV} in industrial environments and autonomous driving on roads require wireless communication and radar sensing. Both radar sensing and wireless communications use electromagnetic waves and have a very similar system architecture. This inspires the idea of combining radar and communications at different levels: first, by using a single reconfigurable radio front-end to save hardware resources; second, by sharing the same spectrum to improve frequency utilization; and third, by using a joint signal for sensing and communications.
This integration brings several scientific challenges, which are reviewed in this section.
%we mainly focus on the waveform design challenges in Section~\ref{subsec:waveformISAC}, and give an overview of other general challenges in Section~\ref{subsec:generalISAC}.

\subsection{Joint Waveform Design Challenge} \label{subsec:waveformISAC}
From a signal processing perspective, both communications and sensing require transmitted signals to carry the communicated information or to probe the medium. Thus, waveform design is a common processing block of both radar and communication transmitters. At the receiver side, the communications system needs to know the state of the channel to decode the data from the received signal, whereas the radar exploits the knowledge of the probing signal to determine the environment. The environment information is embedded in the channel state information. The communications require the knowledge of the composite gain while for the radar getting high-resolution separable parameters is necessary. In particular, the high-resolution delay and doppler. 
%Thus, after channel estimation, the radar employs high resolution parameter estimating algorithms. Thereafter, the estimated parameters undergo further interpretation, such as determination of the range and velocity. The channel estimation accuracy requirements for communications mainly depends on the modulation, whereas for radar, it directly impacts the accuracy, resolution, and false alarm rate. As a result, there are three research areas to be considered at the radar receiver  

The joint design should consider both radar scenarios as illustrated in Figure~\ref{fig_sim}, each one bringing different challenges.
\paragraph{Monostatic Active Radar} 
The transmitter is equipped with a radar receiver to listen to the echoes of its own transmitted signal. The communications transmitter is used as a radar illuminator, and the transmitted signal, which is known at the transmitter, is used as the probing signal for radar. As this scenario is deployed in a full duplex, the main challenge here is to be able to cancel the self-interference from the transmitter \cite{5776640}. In conventional radar systems, such interference is eliminated by using a short pulse and switching off the radar receiver while the transmitter is active. However, this approach is not suitable for high data rate communications. In high frequencies, where large antenna arrays can be deployed, isolating the transmitter and the receiver may be possible with beamforming. The challenge here is to design proper beams for self-interference cancellation.

\paragraph{Bistatic Radar}
The receiver in this scenario is not collocated with the transmitter, and the interference is reduced by isolation. Two modes can be considered here, first the passive mode between the communications transmitter and receiver. The passive radar exploits the reference signal for channel estimation. As a special case, the WiFi signal can be exploited for RF sensing based on the received signal strength, or based on the channel state information that can be obtained from the preamble \cite{9497736}. In this mode, the challenge lies in the appropriate design of the reference signal for both communications and sensing capabilities. For communications, it can be sufficient to use short training signals to estimate the channel gains with adequate accuracy. However, the radar may require more training signals to achieve high accuracy estimation of the radar parameters. The joint design of the waveform here should mainly focus on the frame design in terms of allocation schemes of the payload and training signals. Moreover, the successfully decoded signal can be used to reconstruct the original communication signal and exploit it in refining the channel estimation for the radar. The challenge here is to design a waveform so that the reconstructed signal is beneficial for radar. The impact of latency on radar estimation should be considered. The second mode of bistatic radar is the active mode where the communications signal is conveyed to the radar receiver. The full communications signal is then used as a probing signal for the radar, which is similar to the monostatic case, but with a relaxed hardware design to cope with the cross-talk. An additional challenge here lies in how the communications signal is conveyed to the radar while both are isolated from each other. A general challenge of the bistatic scenario is the synchronization problem where both the transmitter and the radar receiver should be synchronized for accurate parameters estimation.

%\paragraph{Hybrid Radar} 
%This scenario arises especially in frequency division duplex \ac{FDD}, where a device can implement the active monostatic radar while transmitting (downlink), and implement the passive radar approach while receiving the communications signal (uplink). 
%The research opportunity and challenge here is how to fuse both radar information from different bands (downlink and uplink) to improve the radar resolution. 

%Moreover, by considering multi-antenna system, there can be more scenarios such as when certain beams are used for communications, while others for sensing.  Multistate radar is an available scenario, especially considering multiuser in the uplink, where the receiver can obtain information from different perspectives. In most of the radar scenarios, the locations of the nodes are required to create a map. The location can be estimated from the fusion of information collected by different nodes, and this results in simultaneous localization and mapping (SLAM).
%subsection{Other Challenges} \label{subsec:generalISAC}
%There are several challenges beyond the waveform design that need to be addressed to make the integration between communications and radar sensing happen \cite{tong20226g}, in particular:
\subsection{Performance Indicators}
Performance indicators need to be defined according to different use cases. The sensing accuracy measures how close the estimated sensing parameters are relative to the true ones. In localization use cases, one would need to include parameters, such as propagation delays, angle of arrival, time difference of arrival, to position the intended nodes. For instance, a one-meter level of accuracy is attained for vehicular use cases, as reported in \cite{9154212}. In some use cases including finger-level precision, the angle and delay estimates of the multipath components are required at a cm-level precision. In other applications, however, this sensing accuracy will not be sufficient to enable services empowered by semantic localization, such as drones and robotic waiters in future restaurants, where drone-like waiters are expected to locate and serve guests at a millimeter-level \cite{tong20226g}.  Other key performance indicators include sensing resolutions. For example, in applications such as gesture recognition, the sensing resolution is a common indicator reporting the quality of greater resolution and accuracy in apprehending finer actions. Moreover, the probability of detection/false alarm is another vital indicator in delineating the presence of a target of interest, such as objects in the scene. In addition to the mentioned performance indicators, hybrid indicators will also be considered. One example of hybrid indicators can be reflected in situations where multipath can be regarded as a friend instead of a foe. In particular, the amount of multipath included in certain contributions of the propagation environment can be leveraged to boost the probability of detection, which has a direct impact on improving localization accuracy. Another instance is to formulate sophisticated waveform optimization frameworks, such as maximizing radar power in certain directions (in the Bartlett or MVDR senses) while guaranteeing a minimal achievable sum rate with outage considerations \cite{bazzi2022outage} or minimizing the Cramer-Rao Bound of the involved sensing parameters, while targeting a minimum SINR \cite{liu2021cramer}.

\subsection{Theoretical Performance Limits}
Theoretical performance limitations need to be defined for radar sensing in the same way information theory defines the limitations for communications. The Cramer-Rao bound being a metric for estimation accuracy and Shannon's information-theoretic capacity being a bound for the communication rate, have to be combined in a sophisticated way to report the overall achievable performances of the resulting sensing and communications system. To this extent, the inner bounds derived in \cite{chiriyath2015inner} allow us to assess the performance of radar and communication co-existence, i.e. when one sub-system is a source of interference with the other. The estimation and communication rates are combined by developing a radar estimation rate (measured in bits-per-second) is used to measure the quality of time-delay estimation and is further combined with the communication data rate to study several inner bounds depending on the resource allocation strategy being adopted, such as the Fisher information bound, water-filling bound, and the successive interference cancellation (SIC) bound.

In\cite{he2018performance}, a hybrid active-passive MIMO radar network and a distributed MIMO communication system are considered within a radar communication co-existence scenario. The system makes use of the back-scattered communication signals from the target, which are then exploited at the radar receiver. To this end, the CRB and the communication of mutual information are derived. Operating in a cooperative manner, it is demonstrated that the radar target can assist the communication system. In addition, tradeoffs between communication and radar are discussed for both cooperative and uncooperative scenarios. Furthermore, the paper in \cite{cui2018perspective} transposes the concept of \textit{Rényi entropy} to exploit the structure of the degrees of freedom (DoF) for radar-communications coexistence scenarios. Furthermore, connections are made between DoFs and diversity order and gain. For example, if a Neyman-Pearson detection strategy is adopted under a target exponential distribution, then the DoF is the same as the diversity gain. Generalizations towards interference multiple user radar communication interference channels are also introduced and discussed. Moreover, \cite{fortunati2020massive} provides a shrewd analysis for co-located MIMO radar target detection in a single-snapshot massive MIMO setting, i.e. when the product of the number of transmit antennas and receive antennas go to infinity. In particular, the paper derives closed-form expressions for the probability of detection and false alarm in the massive MIMO regime, and in the presence of random measurement disturbances, caused by the so-called clutter and white Gaussian noise measurements. The entire analysis assumes no prior knowledge of the disturbance probability density function. A simple integrated sensing and communication model is explored in \cite{kobayashi2018joint} from an information-theoretic approach. The system is composed of a transmitter, who exploits a strictly causal channel output with generalized feedback for channel state sensing with perfect state knowledge. The optimal capacity-distortion trade-off is characterized as the Pareto frontier between the minimum distortion and the achievable communication rate. This idea is further built up in \cite{ahmadipour2022information}, where both inner and outer bounds on the capacity-distortion region are derived for general broadcast channels. It is worth noting that the assumptions made here assume an independent and identically distributed process on the channel state, i.e. varying from one symbol to another. Within the same context, the paper in \cite{joudeh2022joint} proceeds with different assumptions, i.e. the channel state is assumed to remain fixed throughout the transmission block and two main channels were studied, the binary symmetric channel (BSC) and the Gaussian one. On the one hand, communication-detection trade-offs were established in the BSC case. On the other hand, the authors in \cite{joudeh2022joint} conclude that there  exists no trade-off between the communication rate and the state detection error exponent in the Gaussian case.

\subsection{Stochastic or Deterministic Channels}
According to \cite{han2018propagation}, the communication channel model could be divided into three categories: stochastic, deterministic and hybrid channel models. From a joint sensing and communications perspective, communication channel modeling differs from sensing channel modeling, which typically calls for high-resolution environmental data. Having that said, the communications nongeometrical stochastic model, such as Saleh-Valenzuela \cite{saleh1987statistical}, Zwick \cite{zwick2002stochastic} and the like,  may not be appropriate for radar, which needs a deterministic model based on ray tracing and measurements. Note that ray tracing is a high-frequency approximation of the well-known Maxwell’s equations, which allows us to assume geometrical wave propagation rays. The beams are traced and assessed using geometric optics rules such as geometric optic, the geometric theory of diffraction, and uniform theory of diffraction \cite{li2021integrated}. In the context of integrated sensing and communications, for example, \textit{MaxRay} \cite{arnold2022maxray}, is a versatile tool, that is developed to simulate realistic scenarios, making use of ray-tracing to reproduce ISAC channel responses. The framework is based on a popular open-source 3D computer graphics software, Blender \cite{blender2018blender}. Also, different metrics (detection probability, SINR, prominence and isolation) were considered to measure the overall system performance, with particular emphasis on the performance of clutter removal. Note that for sensing use cases that involve tracking and localization, ray tracing is a viable contender for channel modeling since it does not need the definition of intricate contours and EM properties. Furthermore, measurement-type models allow for a more deterministic characterization for sensing applications. In addition, computational electromagnetics (CEM) is also another potential candidate for deterministic channel modeling for sensing applications. CEM, based on Maxwell Green’s function and empowered through known numerical methods such as the finite element method and the method of moments, is a cross-disciplinary subject that may address challenging electromagnetic theory and engineering issues and is interconnected with electromagnetic field theory and engineering and pervades every area of electromagnetics. CEM modeling seems to be well suited for applications that include imaging. On the other hand, there have been geometry-based stochastic channel models (GSCM) that could as well tailor themselves as potential candidates from \ac{ISAC} channel models due to their inherent characteristic nature of randomness, which could serve  communication channel modeling and, at the same time, leverage the geometric information of the environment, since it is directly related to physical reality. GSCM selects scatterer positions in a stochastic manner based on a probability distribution, then the channel impulse response is found by a ray tracing process \cite{almers2007survey}. Indeed, GSCM is an interesting channel modeling that regards the channel in a stochastic way, thus in favor of both communications and sensing applications.

\subsection{Impact of Integrated Sensing on the Communications performance}
The impact of integrated sensing on communications performance should be investigated. In this context, communication-centric approaches focus on integrating sensing capabilities in a system that is already performing communications. Therefore, in such a class of integrated sensing and communication schemes, the radar functionality is built on top of the communication system, where the main focus is on communication. It is then natural to optimize the algorithms and hardware to support the added task. This approach may offer less tunable trade-offs, and may leave us with very volatile sensing performance. On the other hand, a joint design of sensing and communication would tailor itself more flexible due to its versatile nature.

\subsection{Radar Fusion}
Radar fusion refers to a framework that will leverage information, such as communication bounces off targets, which is to be fused along with radar probing echoes, thus radar fusion from different sensors (radar, lidar, etc). Indeed, this will allow the base station, to perform sensing parameter estimation, with higher confidence and better control of the environment. Let a consider a monostatic base station that transmits a probing signal, $x(t)$, which is then received as
\begin{equation}
	\pmb{y}_{\tt{probe}}(t) = \alpha_0 \pmb{a}(\theta_0) x(t- 2\tau_0),
\end{equation}
where $\alpha_0$ is a two-way channel coefficient, accounting for path-loss and reflection of the intended target and $\theta_0$ is the angle-of-departure (in this case also angle-of-arrival) between the base station and the intended target. The steering vector is denoted as $\pmb{a}(\theta)$, which depends on the antenna geometry configuration and the propagation delay is denoted as $\tau_0$. The radar sub-system at the base station would then process $\pmb{y}(t)$  to give an estimate on $\theta_0,\tau_0$ so as to localize the intended target, thanks to favorable auto-correlation properties of the signal $x(t)$.  Compared to this classical approach, the base station could jointly exploit the uplink signal components, as well:
\begin{equation}
    \pmb{y}(t) = \pmb{y}_{\tt{UL}}(t) + \pmb{y}_{\tt{probe}}(t),
\end{equation}
where $y_{\tt{UL}}$ contains the received uplink signal from all the communication users, as well as the bounces off the targets. Indeed, for the single-user case, we have that
\begin{equation}
\label{radar-fusion-SU}
    y_{\tt{UL}}(t) = \pmb{h} s(t-\delta) + \alpha \pmb{a}(\theta_0) s(t-\tau_0 -\tau),
\end{equation}
where the channel $\pmb{h}$ is the communication channel between the communication user and the base station and $\delta$ is the propagation delay between the communication user and the base station. The other contribution is a single-bounce component that hits the target and goes straight toward the base station. The channel gain of the single-bounce component is $\alpha$ and the propagation delay between the communication user and the target is $\tau$. The base station would then sample the observed signal in equation \eqref{radar-fusion-SU} at the Nyquist rate, then leverage the signal $s(t)$, which could be a training sequence to fuse the radar, as well as the communication information, in a sophisticated way.

%\paragraph{How to represent and share sensing data as well as how to define the trade-off between computation resources and communication resources}

\subsection{PHY Layer Security for \ac{ISAC}}
Due to the broadcast nature of wireless signals, security prevails itself as a primary concern. In fact, security has been regarded as an issue that is addressed by the higher layers rather than the PHY layer. It seems very natural that security will even be a concern with higher priority in the context of integrated sensing and communications, due to the dual information held by the ISAC waveform. Indeed, a sophisticated eavesdropper\footnote{By a sophisticated eavesdropper, we mean that the eavesdropper is aware of the communication standardization used.} could definitely acquire communication, as well as sensing data, which leads to colossal consequences from a security standpoint. Cryptographic methods were built on the assumption that the PHY layer produces error-free bits of data. Higher-layered security management, on the other hand, appears to be difficult to implement and vulnerable to assaults in some contexts, such as relay and ad-hoc networks. As a result, there has been a lot of recent focus on creating protected PHY-layered modems with sophisticated privacy-aware beamforming methods that take security precautions into consideration. Such beamforming methods could build on top of higher-layer cryptographic methods to form a well-secured system from both PHY, MAC and application perspectives. In \cite{su2020secure}, a \ac{DFRC} model has been adopted, where the DFRC base station focuses on designing waveforms that are deemed “suitable” for both radar and communication tasks. The design part focuses on a well-structured probing waveform that contains communication information, while also having desired beampatterns that are helpful for radar. Furthermore, the target in \cite{su2020secure} is assumed to be the Eavesdropper, hence the following dilemma: The waveform design aiming at illuminating radar power in the direction of Eve, should as well reveal minimal communication information towards the direction of Eve, while maintaining acceptable communication rate towards the communication users. In Wi-Fi applications, the demo presented in \cite{jiao2021openwifi} allows communication between users under obfuscated physical channel. For example, imagine a Basic Service Set (BSS) with $2$ users and $1$ access point – then users could carry out communications, but require authorization to perform sensing. This is achieved by fabricating an artificial channel that cascades the physical (environmental) channel, where the latter obviously contains the sensing parameters. Therefore, the key here to access the latter is to obtain the former, which could be seen as an exchange key to access sensing parameters.

\subsection{Imperfect Channel State Information}
Imperfect \ac{CSI} refers to the state where the base station has a limited channel estimate. The same challenge will prevail in the context of integrated sensing and communication: The CSI acquired by the \ac{DFRC} base station and the communication users cannot be perfect, due to hardware imperfections, such as non-linearities and saturations in the RF front-end. The paper in \cite{bazzi2022outage} focuses on a scenario, where the DFRC base station communicates with downlink communication users, with imperfect CSI knowledge, and performs target detection, all via the same transmit waveform. A suitable optimization framework has been adopted taking into account all system requirements. Assuming the communication model reads
\begin{equation}
	\pmb{y}_c = \pmb{H}\pmb{x} + \pmb{z}_c ,
\end{equation}
where $\pmb{H}$ is the multi-user downlink channel matrix and $\pmb{y}_c$ is the received vector over all users. From the radar side, assuming a colocated mono-static MIMO radar setting
\begin{equation}
	\pmb{y}_r = \gamma_0 \pmb{a}(\theta_0)\pmb{a}^T(\theta_0)\pmb{x} + \pmb{z}_r,
\end{equation}
where the angle $\theta_0$ is the angle between the base station and the intended target. Notice how the signal $\pmb{x}$ is shared between the communication and radar models. In \cite{bazzi2022outage}, the joint waveform design assumes a precoded $\pmb{x}$ as 
\begin{equation}
    \pmb{x} = \pmb{W} \pmb{s}
\end{equation}
where $\pmb{s}$  contains the QAM symbols intended for all users, and assumed to be independent from one another with unit variance, i.e. $\mathbb{E}(\pmb{s}\pmb{s}^H) = \pmb{I}$. The following optimization problem is thus formulated 
\begin{equation}
 \label{eq:problem1}
\begin{aligned}
(\mathcal{P}):
\begin{cases}
\max\limits_{\lbrace \pmb{w}_k \rbrace}&   P(\theta_0)\\
\textrm{s.t.}
 &  \Pr( \SINR_k \leq \gamma_k ) \leq p_k , \quad \forall k \\
 & \sum\limits_{k=1}^K \Tr(\pmb{W}_k) \leq 1, \\ 
 & \pmb{W}_k = \pmb{w}_k\pmb{w}_k^H  , \quad \forall k , \\
\end{cases}
\end{aligned}
\end{equation}
where $P(\theta)$ is the output power of the Bartlett beamformer. The parameters $\gamma_k, p_k$ are outage-related parameters. The trace constraint is a power constraint over all the $K$ communication users. A series of relaxations and approximations are discussed to arrive at a convex optimization problem, which is proved to give rank-1 solutions. Furthermore, closed-form solutions for the single-user case are derived and discussed. It is also shown that various trade-offs are achieved as a function of the outage parameters \cite{bazzi2022outage}. Another interesting result is that when the target approaches one of the communication users, the efforts spent on maximizing the $\SINR$ are obtained via the radar metric. How close? This depends on a correlation threshold, which is a function of the outage parameters. 

\subsection{Summary}
Integrated sensing and communication can be implemented along with various radar architectures, such as monostatic and bistatic. Each of these architectures presents its own challenge. For instance, due to the colocation of transmitting and receiving units at the monostatic radar, signal leakage due to the full-duplex operation of the monostatic active radar causes natural self-interference. If not dealt with, the analog-to-digital converter at the receiving unit will be driven towards saturation, i.e. the self-interference component will contribute to an added signal operating outside the dynamic range of the \ac{ADC}. The most straightforward solution is antenna separation, which aims at creating some sort of isolation between the transmitting and receiving arrays. However, this isolation will increase the area occupied by the radar, which is not favorable in certain applications. Meanwhile, the literature involves methods that address the self-interference problem. In particular, analog cancellation, involving analog operations, can be leveraged by taking full advantage of directional diversity so that the notches of the transmitting and receiving antenna arrays are steered towards each other \cite{haneda2010measurement}. 

On the other hand, digital beamforming \cite{ahmed2015all} can further aid in reducing residual self-interference. However, it becomes nonessential if preceded by a strong analog canceller. A combination of the aforementioned solutions can provide additional advantages. For example, the work in \cite{snow2011transmit} considers a combination of antenna isolation, as well as analog and digital suppression, which leads to a self-interference suppression by more than 70 dB. For bi-static radar, the first challenge related to waveform design can be addressed by carefully constructing waveforms with proper ambiguity functions, as discussed in \cite{griffiths2009passive, bazzi2022integrated}. Also, an \ac{RIS} aided architecture can help boost the sensing capabilities of the radar subsystem, as shown in \cite{bazzi2022ris}. As for the second challenge, information can be transported through the means of fiber optics, as alluded by \cite{futatsumori2016design,shin2016distributed}. We have also presented a framework dedicated to fusing radar information to enhance sensing capabilities. Furthermore, various performance indicators will aid in designing integrated sensing and communication waveforms oriented toward application needs, and based on different channel models. This depends on whether the application is communication-centric, where the radar sub-system is an add-on, or whether the application is radar-centric, i.e. the communication capability comes as an extra feature.  For example, outage \ac{SINR} probabilities per user are reasonable metrics, while maximizing the output radar power, in situations when imperfect \ac{CSI} knowledge is available. 

We have also stressed the importance of physical layer security for ISAC due to the presence of malicious eavesdroppers and the potential harm that can be caused. For this reason, \ac{DFRC} design for \ac{ISAC} systems should also account for eavesdroppers and carefully perform communication beamforming, under the constraint that useful information should not be leaked toward the eavesdroppers. Future \ac{DFRC} designs should also address the multi-cellular case and provide insights when coordination between different DFRC base stations is possible.

% \begin{figure*}[t]
% 	\setlength{\abovecaptionskip}{6pt plus 3pt minus 2pt}
% 	\centering
% 	\includegraphics[width=1\columnwidth]{Figures/ISAC01/fig1.pdf}
% %	\subfloat[\label{CIR} Channel impulse response ${\ma{h}}(\tau,t)$]{\hspace{.5\linewidth}}
% 	%\subfloat[\label{CFR} Channel frequency response $\tilde{\ma{h}}(f,t)$ ]{\hspace{.5\linewidth}}
% 	\caption{test}
% 	\label{fig:test00001}
% \end{figure*}

\section{Large Scale Communication Theory} \label{sec:large}
%\hl{Lina}
With the emergence of novel applications, and the evolution of telecommunication paradigms, including human-based and machine-based communications, it has become evident that future wireless generations will be designed vertically, incorporating the underwater, ground, air, and space domains. Such an architecture exacerbates the communication management and coordination difficulty, owing to three main factors, namely, i) the need for a scalable massive deployment in order to cope with the vertical and horizontal growth of wireless networks sizes, in which a massive number of components in ultra-dense multi-layer heterogeneous environments needs to be connected/configured, ii) the urgency for rapidly adaptive mechanisms in order to manage continuous variations in dynamically evolving environments, and iii) the necessity of decentralized processing, by realizing nodes automation, intelligence, and adaptivity, with the aim to enable self-optimizing, self-organizing, and self-healing networks. Till recent days, the modeling and optimization of large-scale wireless networks was accomplished by leveraging the recent development of sophisticated computing services and the advancements in \ac{ANN}. The former, which solely relies on heavy computer simulations, lacks the real time adaptation to network continuous variations and the on-demand scalability requirements imposed by future wireless generations. Furthermore, such heavy simulations drain a considerable amount of resources, that may not be affordable by resources-limited devices. On the other hand, although heuristic solutions, such as \ac{ANN} algorithms, are identified as an alternative method in order to solve more complex multi-objective multi-attribute optimization problems, the absence of a solid mathematical interpretation limits their applicability in scenarios where a solid mathematical foundation is a necessity, for system improvement purposes. In addition, the deployment of such algorithms may be very challenging in realistic scenarios, due to the statistical heterogeneity experienced in large-scale networks, which directly affects the accuracy and convergence rate of these algorithms. This issue is particularly pronounced in highly dynamic environments. 

Accordingly, it is essential to develop robust, flexible, and systematic mechanisms that enjoy a solid mathematical foundation, and are versatile in a way that allows them to fit in multiple network scenarios, covering a wide range of use-cases at a large-scale. Among others, three major tools have manifested themselves as promising, generalized, yet resilient methods for modeling, analyzing, and optimizing large-scale dynamic wireless networks. This includes random matrix theory, decentralized stochastic optimization, and tensor algebra and low-rank tensor decomposition.
\subsection{Random Matrix Theory}
Over the last couple of years, \ac{RMT} has found its applications in the field of wireless communications, and it has been regarded as a powerful tool that can be leveraged in order to model and optimize wireless systems, as well as quantify their performance \cite{couillet2011a}. Recalling that \ac{RMT} relies on characterizing the eigenvalues/eigenspaces of considerably large matrices with random entries, \ac{RMT} has shown to be a robust method for analyzing large-scale multi-dimensional dynamic wireless systems, encompassing the antenna \cite{ dupuy2011, 8320821}, spectrum \cite{bianchi2009}, precoder \cite{ geraci2013}, users \cite{couillet2011}, and cells dimensions \cite{ sifaou2016}, as well as channel time and frequency variations. Additionally, RMT has demonstrated an efficient performance when employed to develop innovative detection and estimation schemes in MIMO systems \cite{mestre2008 }. 

\paragraph*{Will \ac{RMT} and Deep Learning Handshake?} Although the merit of \ac{ML} is shown by the complex deep neural networks that constitute a massive number of layers/parameters, the burden of understanding such complex networks represents a major challenge, particularly when deploying \ac{ML} in large-scale environments \cite{adlam2019}. In fact, in extremely large-scale networks with heterogeneous datasets, a large-dimensional \ac{ANN} is required, and therefore, deep learning in such scenarios is treated as a black box, in which inputs and outputs can be measured, while the complex operations in between are not easy to comprehend and analyze. Given the fact that large complex \acp{ANN}, with randomized initialization, can be modeled as random variables, it is anticipated that \ac{RMT} will be an effective tool in understanding \acp{ANN}, when implemented in large-scale networks. It is worthy to mention that understanding such complex networks paves the way for further large-scale system optimization and improvement. In particular, as a result of exploiting \ac{RMT} in studying the characteristics of neural networks, the depth and weights initialization can be optimized for improved learning process in \acp{ANN}. Moreover, \ac{RMT} can be utilized to analyze the spectra of the data covariance matrices, and subsequently, to explore and design a number of efficient activation functions that are capable of ensuring stable eigenvalues at neural networks. Note that the advantages of such an approach can be more understood when data distributions are highly skewed, and hence, the features space experiences strong anisotropy, resulting in an increased training overhead \cite{ge2021}.

\subsection{Decentralized Stochastic Optimization}
With the aim to enable self-optimizing large-scale networks, decentralized stochastic optimization is proposed as an efficient method to realize network-wide optimization, with leveraging a large number of resources-constrained devices which communicate with each other in a decentralized ad-hoc manner \cite{xin2020}. This is initially motivated by the fact that, in large-scale networks, raw data samples are available at multiple distributed devices, with limited communication capabilities. In order to perform a centralized network optimization, raw data generated and stored at local devices should be sent to a centralized location for storing and processing, yielding compromised privacy and security, and increased network overhead. Furthermore, due to the massive number of connected homogeneous nodes, centralized network optimization in large-scale environments suffers from long propagation delay, rendering it unsuitable for real-time applications. In decentralized stochastic optimization, the enhanced on-board computing and storage capabilities of wireless devices are leveraged in order to allow multiple distributed devices across a dense network to cooperatively solve a network-wide problem, through integrating stochastic approximation and gossip algorithms, where the latter represents an information infusion method in ad-hoc networks \cite{bianchi2011}. This can be achieved by dividing the problem into several dependent sub-problems, which are then optimized in a distributed manner, by allocating each node in the network a sub-problem to solve. Each node then shares the obtained optimized parameters with the neighboring nodes, until an agreement on the optimum global parameters is achieved \cite{xin2019}.

\subsection{Tensor Theory}
With conventional optimization algorithms being unscalable to accommodate larger datasets with diverse practical scenarios, tensor algebra, which enjoys a solid mathematical foundation, has been recently identified as a promising mathematical tool that is capable of handling the linear/sub-linear increase in datasets sizes. In particular, low-rank tensor network approximations can be utilized to break down a large-scale mutli-objective optimization problem into multiple small-size less-complex sub-problems \cite{cichocki2016}. It should be noted that tensors can be modeled as a $D$-dimensional data structure, and therefore, can be efficiently used to manage datasets, and allow a compressed and distributed representation of large-scale data. Tensor network, also known as tensor decomposition, converts high-order tensors into multiple sparsely interconnected low-rank tensors, enabling efficient computation of eigenvalues/eigenvectors of high-dimensional matrices. It was further shown that tensor networks demonstrate a robust performance when dealing with missing values and noisy data, rendering it suitable for a wide range of large-scale network scenarios \cite{cichocki2014tensor}. Moreover, blind sources separation, for the case of a large number of sources, represents a direct application of tensor networks, i.e., non-coherent non-orthogonal signals detection, supporting unsourced massive random access use cases \cite{decurninge2020}.

\subsection{Summary}
Large-scale communication theory has imposed several challenges that need to be tackled by future wireless generations, including on-demand scalability for ultra-dense massive access support, adaptivity to accommodate the need for dynamic networks, as well as edge intelligence for network automation. While \ac{AI} has been deemed as a magical tool to deliver the identified \acp{KPI} of large-scale networks, it is yet essential to develop solid mathematical underpinnings to allow a better understanding of how \ac{AI} operates and to facilitate the reasoning of decisions made by \ac{AI}. Several tools have been deemed as potential approaches to offer a flexible, reliable understanding of large-scale networks, with the incorporation of an \ac{AI} element, including, \ac{RMT}, decentralized stochastic optimization, and tensor algebra and low-rank tensor decomposition.

\section{Non-equilibrium Information Theory} \label{sec:nonequilibirum}

Since the early development of wireless generations, 2G, 3G, and up to 5G networks, channel coding has always been an essential component in the transmission process, and therefore, several efforts were devoted for developing resilient channel coding mechanisms, spanning from the basic algebraic codes, which aims at maximizing the error correction probability, such as Hamming \cite{HC} and Reed–Muller codes \cite{paterson2000generalized} (with constrained code rates and lengths) and \ac{BCH} \cite{chien1964cyclic} and Reed–Solomon codes \cite{wicker1999reed } (with more relaxed constraints and enhanced performance), to more advanced coding schemes including convolutional \cite{viterbi1971convolutional}, turbo \cite{vucetic2012turbo}, polar \cite{ tal2013construct}, and \ac{LDPC} codes \cite{gallager1962low}. This evolution in channel coding was fueled by the urge to achieve the theoretical channel capacity limit, which was not possible with length-constrained codes, with the aim to offer extended coverage, enhanced \ac{QoS}, and improved data rates and energy efficiency, which are the main verticals of all wireless generations. The basic principle of coding is to superimpose multiple packets with the aim to achieve higher coding gain, in addition to performing better interference management. The successful implementation of channel coding relies on its performance, power consumption, and cost. This was demonstrated through the development of near-Shannon limit codes with acceptable power consumption \cite{arora2020survey}. Recall that the main purpose of channel coding is to maintain the needed reliability and \ac{QoS} constraints, even under severe channel conditions. Meanwhile, although 6G networks are generally characterized by their highly dynamic topologies and heterogeneous nature, they are expected to serve a wide range of use cases with distinct requirements, in terms of ultra-high reliability, extremely high data rates, ultra-low latency, and significantly improved energy efficiency. This calls for a compelling need to revisit current coding methodologies, explore their limitations, and accordingly construct innovative channel coding schemes that are capable of delivering the required \ac{QoS}, under various severe channel conditions. 

\begin{figure*}[t]
	\centering
	\includegraphics[width=0.9\linewidth]{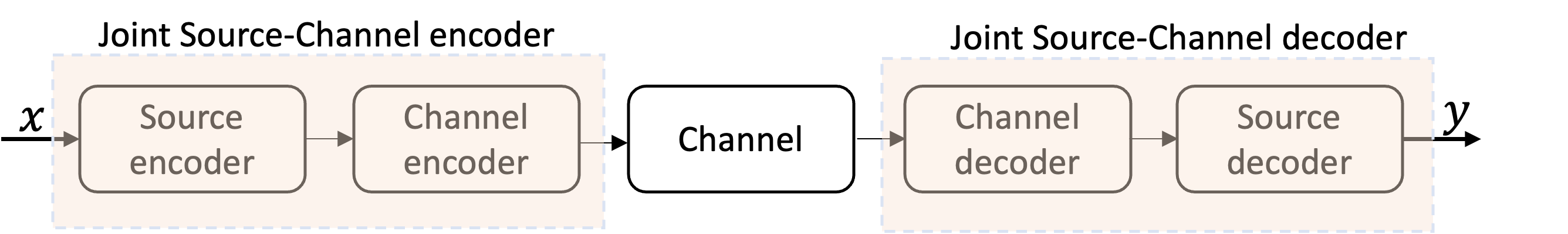}
	\caption{Joint source and channel coding.}\label{fig:JSC}
\end{figure*}

In conventional methods, an idealistic assumption of achieving Shannon channel capacity, through separated channel and source coding schemes, is considered. While this assumption is valid for large information block size, it fails to deliver the promised complexity and latency requirements of channel codes in current and future wireless networks. Accordingly, it is envisioned that coding in 6G networks and beyond will leverage joint designs of source and channel codes. Motivated by the emergence of several applications with critical requirements in 6G, including autonomous driving, VR/XR, \ac{mMTC}, and \ac{URLLC}, the key requirements of future channel codes will incorporate a peak data rate of Tbps, less than 200 bits code-lengths with simplified decoding process, and adaptive latency-complexity-reliability tradeoff (see Fig. \ref{fig:KPI-codes}) \cite{lu20206g}. Accordingly, the following design principles will shape the vision of channel codes in future 6G networks:
\begin{figure}[t]
	\centering
	\includegraphics[width=1.2\linewidth]{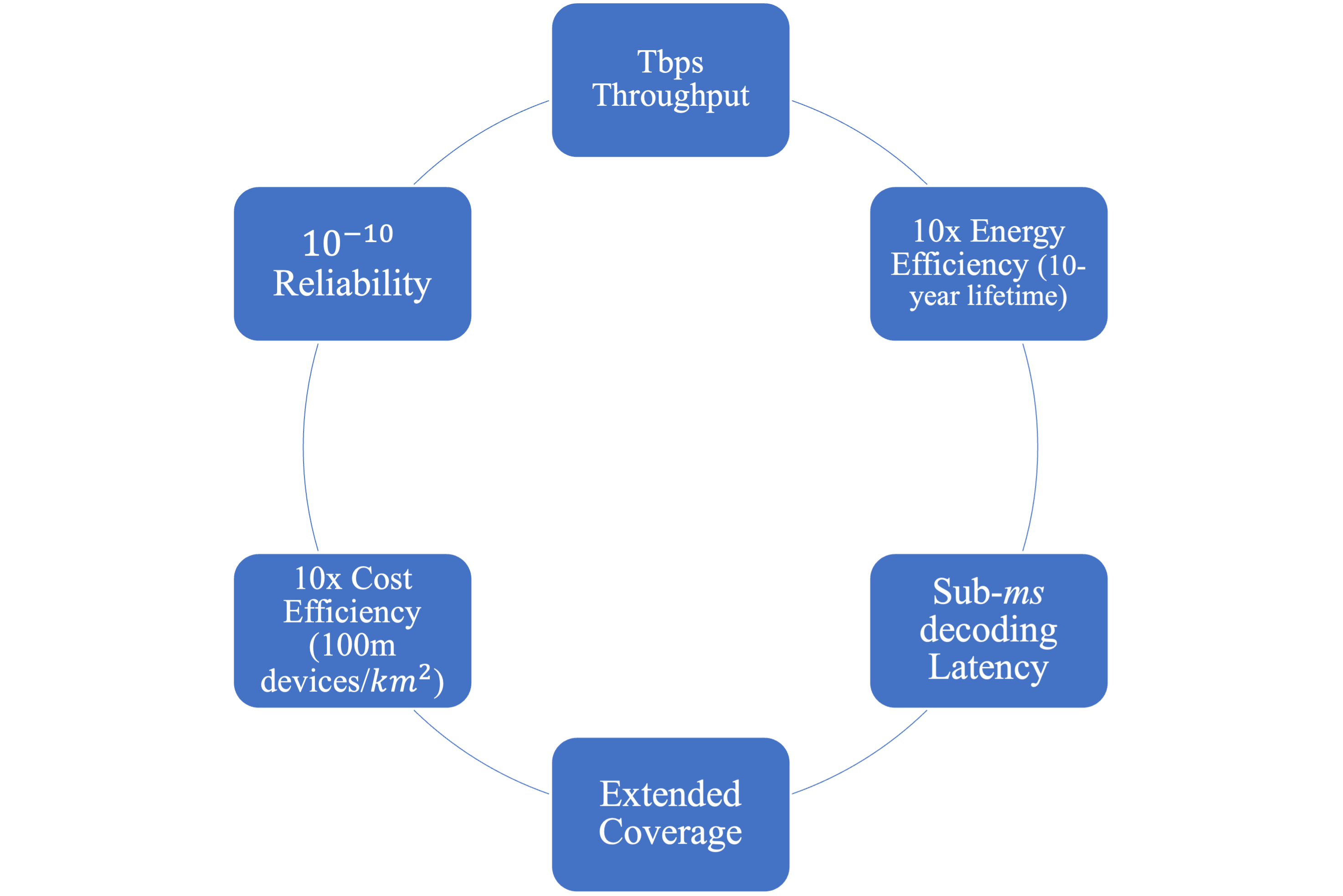}
	\caption{6G channel codes KPIs and design principles.}\label{fig:KPI-codes}
\end{figure}
\begin{enumerate}
    \item \textbf{One-size-fits-all channel codes:} While current coding schemes have successfully met the needs of existing use-cases, they cannot be readily applied to different use-cases, with diverse requirements, envisioned in 6G networks. This is stimulated by the fact that current coding schemes are not adaptive enough to meet diverse \acp{KPI}, and therefore, the process of selecting the proper coding scheme for each scenario in the heterogeneous 6G exacerbates the design complexity in a way that cannot be tolerated by the network capabilities. Hence, future channel coding schemes are anticipated to enjoy a level of flexibility and adaptivity that enables them to fit within a wide range of use cases with the minimum complexity. 
    \item \textbf{Tbps throughput channel codes:} Achieving extremely high throughput channel coding, while maintaining an acceptable complexity and power efficiency is constrained by the limited resources in current wireless networks.  Furthermore, albeit the witnessed advancements in precoder hardware design, none of the available coding schemes can realize an area efficiency of 1Tbps/mm$^2$. For example, the silicon-based routing in \ac{LDPC} decoder and the successive decoding architecture in the polar decoder represent a performance limiting factor in achieving the needed area efficiency. This is attributed to the reduced die area and increased level of regularity associated with simpler routing. Accordingly, Tbps-throughput coding designs, that take into consideration complexity, degree of parallelism, as well as hardware implementation and limitations, necessitate revisiting current architectures in order to implement the needed modifications that enable the earlier discussed \acp{KPI}.  
    \item \textbf{Short-length coding:} It is noted that, at an acceptable implementation complexity, current channel coding schemes exhibit around 1 dB performance difference with the theoretical finite-length performance limit. This has motivated the adoption of basic short algebraic channel codes, such as Reed–Muller, \ac{BCH}, and Reed–Solomon, as well as polar codes, in order to fill this gap. In particular, the latter enjoy an optimal reliability-complexity trade-off in short-length events, and hence, they represent a good candidate for 6G networks. In addition to that, it was demonstrated that the random coding union bound can be achieved by increasing the list size in polar codes with short block length \cite{ cocskun2019efficient}. This opens the door for exploring the merit of ML to construct efficient codes with optimized list size to strike a balance between performance, decoding threshold, reliability, complexity, and coding gain \cite{huang2019ai}. On the other hand, further investigations need to be carried out with the aim to construct efficient channel coding schemes, not only with short block length, but also with an optimized minimum coding distance, due to the high impact of the latter on the performance. This lead as well to the need for channel codes that are adaptive enough to accommodate various rate-length scenarios.
    \item \textbf{Joint source and channel coding:} The separation theory indicates that the characteristics of the channel encoder/decoder and source encoder/decoder are independent, and have no effect on each other. This theory, however, does not hold when the block length is finite. This has fueled the research over the last couple of decades with the aim to develop joint source-channel coding (see Fig. \ref{fig:JSC}), which later showed, through theoretical performance evaluation, robust performance and introduced several advantages, in terms of complexity and latency, over conventional separated encoders/decoders \cite{dong2022joint}. This can be achieved by utilizing the source residual redundancy as a mean to realize improved error rate performance, which readily impacts the spectral and energy efficiency of the system. Accordingly, several variants of the joint coding design were proposed in the literature. For example, in \cite{guyader2001joint, pu2007ldpc, grangetto2005joint}, the advantages of integrating sparse-graph-based channel decoding with high-level or a posteriori information of source codes are investigated. Furthermore, in \cite{garcia2003ldpc,bhattad2006decision}, novel approaches for tackling the distributive compression of multi-terminal sources, by exploiting the Slepian–Wolf theorem, were proposed. Despite their promising potential, such joint codes rely heavily on the adopted compression scheme at the source and are generally amenable to generalization. This means that the decoder should be completely redesigned if the compression scheme at the source is changed, introducing a new level of complexity that should be addressed in order to allow such schemes to fit within the vision of 6G networks.   
\end{enumerate}

\subsection{Summary}
The evolution of 6G networks, with the associated novel requirements, necessitates revisiting current coding schemes, in order to ensure their flexibility to serve various use cases, with diverse \ac{QoS} requirements. Within this context, joint designs for channel and source coding are envisioned to replace conventional separated coding schemes, with the aim to achieve Tbps data rates, code lengths that are shorter than 200 bits, and low-complex decoding mechanisms. However, this comes at the expense of increased overhead in terms of compression complexity, minimum coding distance optimization for rate-length tradeoff, and sophisticated hardware design for Tbps communication support.

\section{Combining  Queuing Theory and Information Theory: A Cross-Layer Approach} \label{sec:queuing}
%\hl{Sami} \\
Information theory has made a significant impact in the area of communications theory.  However, its impact on communication networks is less pronounced. Specifically, apart from sporadic results; see, e.g., \cite{telatar1995combining}, there have not been noticeable attempts to explore the synergy between information theory and communication networks. This is predominantly attributed to the fact that information theory has overlooked the bursty natures of data sources. Another key issue is that information theory has ignored the effect of delay on the trade-off between rate and error rate.  Interestingly, the majority of recent works continue to ignore the effect delay and burstiness of communication sources.  

In this section, we illustrate how cross-layer optimization addresses these issues, focusing on the joint design of the Medium Access Control (MAC) and the physical (PHY) layers. We are motivated by the fact that although scheduling algorithms prioritize requests based on quality-of-service (QoS) requirements, the wireless link can indeed realize efficient resource allocation, and thus, it must be considered. Depending on specific requirements, the optimization problem can be formulated to address particular quality of service requirements, e.g., delay minimization, queue status, and throughput maximization. 
\begin{figure*}[ht]
  \includegraphics[width=\textwidth,height=8cm]{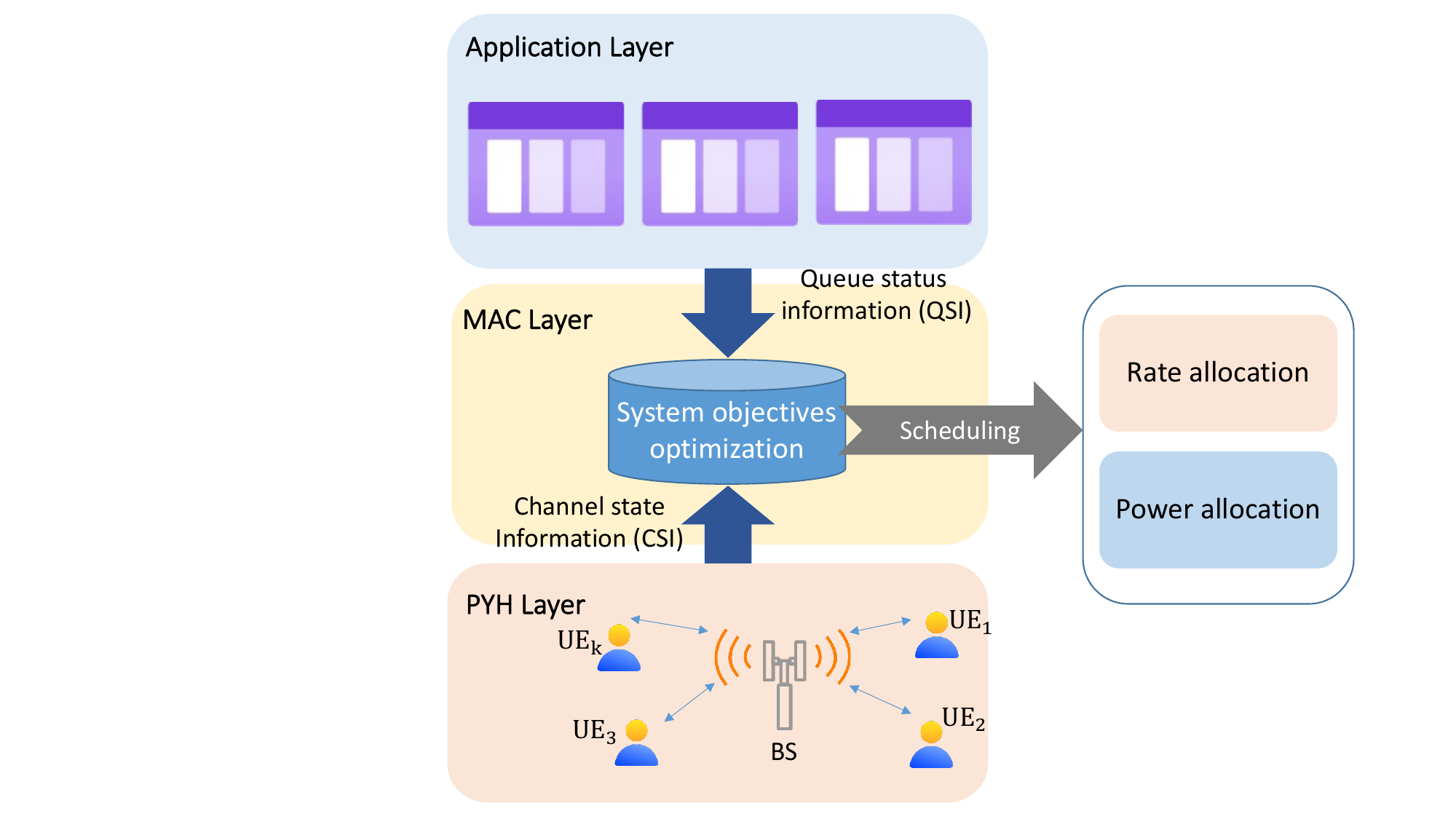}
  \caption{Cross-layer scheduling model \cite{lau2006channel}.}
  \label{fig:queueing1}
\end{figure*}

\begin{figure*}[ht]
  \includegraphics[width=\textwidth,height=8cm]{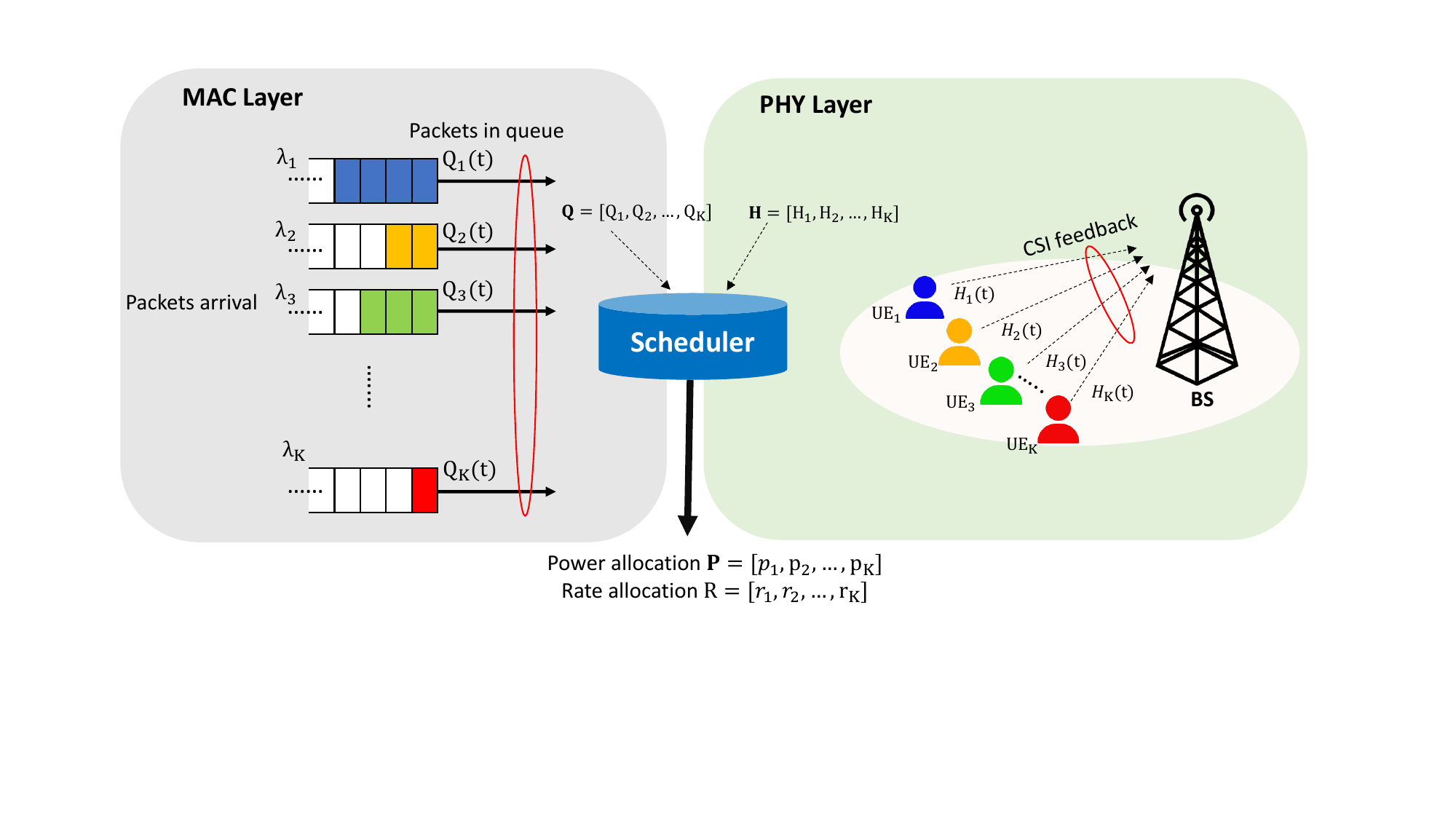}
  \caption{Cross-layer scheduling for delay minimization \cite{lau2006channel}.}
  \label{fig:queueing2}
\end{figure*}
In the context of 6G networks, in which ultra-reliable and low-latency communications (URLLC) is envisioned to support various new use cases that require high network reliability, and extremely low latency, cross-layer  optimization is clearly of paramount importance. It is noteworthy that the current state-of-the-art does not overall fulfill these requirements. Most of the existing literature treats the MAC layer and the physical layer independently \cite{Berry}. However, due to the dynamic nature of wireless channels, current approaches are considered suboptimal, which motivates the cross-layer design approach. A key challenge, however, is that  MAC scheduling algorithms have to adapt to both  channel conditions and source statistics.  

Recently, there has been a growing interest in cross-layer optimization for 6G networks. In \cite{she}, the authors discussed how the domain knowledge, i.e., the analytical and cross-layer optimization frameworks,  can be integrated with deep learning  to enable URLLC. \cite{zhang2021} proposed a cross-layer optimization framework to achieve a trade-off between spectral and energy efficiency.

As reported in the literature, MAC layer scheduling design is realized through either queuing theory or information theory. In the queuing theory approach, MAC scheduling adapts to the source dynamics. On the other hand, in the information theoretic approach,  users are considered to have an infinite queue size and are assumed to be delay-insensitive. The adoption of these approaches independently leads to suboptimal solutions. Thus, the development of a cross-layer framework that can adapt to source and PHY-layer dynamics is of great interest. This ultimately requires the design of a framework that combines both queuing theory and information theory \cite{ephremides1998, gallager1985}. 

\subsection{Cross-Layer Scheduling Model}
Figure~\ref{fig:queueing1} depicts a cross-layer scheduling model that comprises  an optimization problem subject to particular PHY layer parameters and application QoS constraints. In particular, at time slot $t$, the scheduling algorithm  produces the rate  allocation vector  $ \mathbf{r}=\left(r_{1}, \ldots, r_{K}\right)$ and the power allocation vector $ \mathbf{p}=\left(p_{1}, \ldots, p_{K}\right)$, for all $K$ users,  based on the current channel state information (CSI)  and  the queue state information (QSI).  

In a multiuser scenario, the scheduler produces the rate and power allocations, $\mathbf{r}$ and $\mathbf{p}$, respectively, at every time slot $t$, based on the system's state $(\tilde{\mathbf{H}}(t), \mathbf{Q}(t))$. The set of rate allocations over  the time slot $t$  is called the rate allocation policy $\mathcal{R}$ while the set of power allocations is called the power allocation policy $\mathcal{P}$. 
\\
The underling cross layer framework is to find the optimal policies  $\mathcal{R}$ and $\mathcal{P}$ that maximize/minimize  the following utility function
\begin{equation}
U\left(\bar{R}_{1}, \ldots, \bar{R}_{K} ; \bar{Q}_{1}, \ldots, \bar{Q}_{K}\right),
\label{eq1}
\end{equation}
where $\bar{R}_{k}=\mathbb{E}\left[r_{k}\right]$ is the average data rate of user $k$. Note that the optimization problem may  be subject to PHY layer  as well as application layer QoS constrains. For instance, the instantaneous data rate vector must lie in the feasible capacity region, $\mathbf{r}=\left(r_{1}, \ldots, r_{K}\right) \in C(\tilde{\mathbf{H}})$, where $C(\tilde{\mathbf{H}})$ is the instantaneous feasible capacity
region. Additionally, the power allocation vector must  satisfy an average power constraint.

In some applications, where the system's buffer size is rather large, i.e., in delay-insensitive applications, the cross-layer optimization framework is  described as a function of  $\tilde{\mathbf{H}}(t)$. On the other hand, in delay-sensitive applications, the cross-layer optimization objective  is given as a function of the  channel conditions and the buffer status, i.e.,  $S(t)=(\tilde{\mathbf{H}}(t),\mathbf{Q}(t))$.

The system performance depends largely on the choice of the utility function (also called here the scheduling objective) in (\ref{eq1}).  In particular, for delay-insensitive applications,   a natural choice of the  system utility function  is expressed as \cite{fattah2002overview} 
\begin{equation}
U\left(\bar{R}_{1}, \ldots, \bar{R}_{K}\right)=\sum_{k=1}^{K} \bar{R}_{k}=\mathbb{E}\left[\sum_{k} r_{k}\right],
\label{eq2}
\end{equation}
where $\bar{R}_{k}=\mathbb{E}\left[r_{k}\right]$ is the average throughput of user $k$ and $r_k$ is the instantaneous
throughput of user $k$ at any scheduling slot. The optimization of (\ref{eq2}) will lead to the highest achievable  system's capacity.  Note that the objective function in the cross-layer scheduling optimization of delay-sensitive applications is represented in terms of  \emph{system stability} or \emph{delay minimization}.   

\subsection{Multiuser Physical Layer Model}

Consider a system with $K$ users, where user $i$ transmits at data rate $r_i$. The
 data rates of all users can be expressed as $\mathbf{r} = (r_1, . . . , r_K)$, where the rate tuple $\mathbf{r}$ is called feasible when a reliable transmission of all data rates is achieved.  The set of all feasible rate tuples, $\mathbf{r}$, is known as the feasible region, whereas the capacity region is defined as the closure of the union of all achievable rate regions.  In the downlink,  multiuser capacity region can be described by the broadcast region. Note that broadcast
channels can be classified as degraded broadcast channels and the non-degraded broadcast channels. 

 For the degraded broadcast channels and denoting $n_T$ as the number of transmit antennas, the instantaneous
capacity region $\mathrm{C}_{n_{T}=1}\left(H_{1}, \ldots, H_{K}\right)$, conditioned on the channel gains, can be written as  $\forall k \in[1, K]$
\begin{eqnarray}
r_{k} \leq \log _{2}\left(1+\frac{\left|H_{k}\right|^{2} p_{k}}{\sum_{j}^{K} p_{j}\left|H_{k}\right|^{2} \mathbf{1}\left(\left|H_{k}\right|<\left|H_{j}\right|\right)+\sigma_{z}^{2}}\right),
\label{Eq.11.11}
\end{eqnarray}
where $P_{0}=\sum_{j=1}^{K} p_{j}$  is the total transmit power constraint. For  non-degraded broadcast channels, i.e., $n_T > 1$, the capacity region can be determined by dirty paper coding \cite{amraoui2003coding}.  
\\
On the other hand, the  capacity  region of the uplink  multiple access channel, conditioned on the channel gains, is given by 
\begin{equation}
\sum_{k \in S} r_{k} \leq \log _{2}\left|\mathbf{I}_{n_{R}}+\frac{\sum_{k \in S} p_{k} \mathbf{H}_{k} \mathbf{H}_{k}^{*}}{\sigma_{z}^{2}}\right|,
\label{Eq.11.12}
\end{equation}
for all $S \subset[1, K]$. In this case, the capacity region is achieved by successive interference cancellation.

%Lau VK, Kwok YK. Channel-adaptive technologies and cross-layer %designs for wireless systems with multiple antennas: theory and %applications. John Wiley & Sons; 2006 Feb 10.

\subsection{Cross-Layer Scheduling in Multiuser  Networks: Delay Minimization}
The following conclusions
illustrate the advantages of cross-layer scheduling \cite{lau2006channel}:

\begin{itemize}
\item \textit{CSI-aware scheduler outperforms a regular scheduler which does not exploit the channel knowledge}. This highlights the fact that CSI is needed in scheduling  to exploit multiuser diversity. 
\item
\textit{A buffer-aware scheduler gives a better average delay performance compared to a scheduler that does not have buffer information}. This conclusion illustrates the advantage of buffer status information, particularly when the objective is to minimize the average delay. 

\item
\textit{Information-theoretic water filling is not optimal}.
Under the assumption of perfect CSI knowledge, it has been shown that optimum scheduling, which gives the maximum rate, is water filling. However, this is built on the assumption of large buffer size. When the buffer size is finite, the optimum cross-layer scheduler should  adapt to both the CSI and the buffer status. 
\end{itemize}

\noindent 
Figure~\ref{fig:queueing2} shows how cross-layer scheduling can be used to minimize the average delay, where the packet arrival of the  $k\mathrm{th}$ user is modeled by Poisson random variable, with average arrival rate $\lambda_k$.  The delay minimization problem can be cast into the following optimization problem.
%\noindet \\

\begin{Delay Minimization}{}
\textit{Find the rate allocation policy $\mathcal{R}$ and the power allocation policy $\mathcal{P}$ such that the average delay $\lim _{t \rightarrow \infty} \mathbb{E}\left[\Sigma_{k} Q_{k}(t)\right]$ is minimized.}
\end{Delay Minimization}

\noindent 
Broadly speaking, although cross-layer optimization has shown great potential, particularly for 6G networks, its implementation brings key challenges that must be addressed before the successful deployment:
\begin{itemize}
\item Communication overhead:  Due to the dynamic nature of wireless channels, the scheduler needs to update its status and adjusts resource allocations frequently. 
\item Although the above delay minimization problem could be solved in a closed form, cross-layer optimization problems are usually NP-hard which are difficult to handle in practical scenarios due to their high complexity. Various approaches have been proposed in the literature to solve complex cross-layer optimization problems \cite{wang2021deepnetqoe, chiang2007layering}.
\end{itemize}
\noindent

In summary, there is a pressing need for the development of new  cross-layer solutions for 6G networks which will address key challenges in 6G  and give insights into the performance limits in terms of different requirements such as reliability and  latency. Furthermore, it is envisaged that these solutions will provide guidelines for the design of near-optimal solutions that will lend themselves to low complex implementation.    
\subsection{Summary}
%\hl{Sami}
The interaction between queuing theory and information theory is typically overlooked in the literature; because generally, information theory ignores the bursty natures of data traffic and considers delay-insensitive sources. On the other hand,  queueing theory focuses on delay analysis from queuing perspective.  As pointed out earlier, the adoption of these approaches independently addresses the scheduling algorithm design problem partially and eventually results in suboptimal solutions. Therefore, there is a need for the development of a cross-layer framework that can  jointly adapt to source and PHY layer dynamics, which ultimately requires the design of a framework that combines both queuing theory and information theory.

\section{Non-Coherent Communication Theory} \label{sec:noncoherent}
%\hl{Lina}
Since the evolution of data communication and wireless systems, \ac{CSI} acquisition, generally through the assistance of a training sequence, has been always a concerning problem for the research community. This is primarily due to the increased overhead and degraded spectral efficiency experienced in coherent systems. Such overhead is further manifested in pilot contamination scenarios, which necessitate pilot re-transmission to ensure reliable coherent signals detection, as well as imperfect \ac{CSI}, which might lead to a noticeable degradation on the system performance \cite{chowdhury2015}. As an alternative method, blind signal detection, also called non-coherent detection, has demonstrated a reduced overhead. Yet, this has come at the expense of a severely degraded performance and increased complexity.  

As it is now agreed that large-scale networks will be one of the 6G verticals, a rapid and massive growth of the number of antenna arrays/users/nodes will be witnessed in the near future, especially with the emergence of novel applications and wireless services for humans and machines. While the research continues to deliver promising solutions at a large-scale, simultaneous \ac{CSI} acquisition of a massive number of antennas or devices communicating in a coordinated and uncoordinated fashion, will remain a critical challenge. Such a challenge is magnified with the envisioned high mobility applications, such as high-speed rails, hyperloop communication, etc \cite{hedhly2021}. Accordingly, a trivial conclusion: a paradigm shift toward non-coherent communication, and the need to rethink the basics of existing non-coherent mechanisms, have become a necessity. Motivated by this, several research efforts were directed towards developing novel and efficient non-coherent waveform modulation schemes, that fit with the massive-scale networks vision of 6G. Among others, Grassmannian modulation and non-coherent tensor modulation has shown to be effective when applied in massive MIMO and massive multiple access scenarios, respectively. 

\subsection{Non-coherent Grassmannian Modulation}

At the moderate-high \ac{SNR} regime, Grassmannian signaling has demonstrated a close to optimal pilot-free non-coherent detection, compared to its coherent counterpart, particularly in highly dynamic scenarios, e.g., vehicular networks, in which channels experience short coherence time \cite{gohary2019noncoherent}. In Grassmannian modulation, transmitted information is conveyed over tall unitary matrices, each of which represents a particular binary sequence with a fixed length. The uniqueness of these matrices, and hence, the signals sequences, relies on the following two characteristics: i) each signal spans over a particular subspace, and  ii) when transmitted over block-fading channels, these matrices/messages experience rotation and scaling, yet their subspace remain unchanged.  Such behavior enables Grassmannian structure to realize the superior capacity performance at moderate and high \ac{SNR} values. It is worthy to highlight that the transmission data rate of Grassmannian signaling is determined based on the length of the information sequence, as well as the corresponding number of matrices. In this regard, Grassmannian signaling structure constitutes a natural choice in a large-scale implementation, such as massive \ac{MIMO} networks and massive machine-type communications, due to the significant reduction in the pilot overhead when Grassmannian modulation is employed in such scenarios \cite{cabrejas2016non}.

As a multi-dimensional scheme, constellation design in Grassmannian modulation represents a challenging task, due to the involvement of complex numerical optimization procedures. The increased complexity is attributed to the nature of the curved spaces in Grassmannian structure, as well as the significantly large number of unitary matrices (up to twenty-zeros numbers) required for such designs \cite{gohary2019noncoherent}. The interplay of machine learning algorithms and Grassmannian constellation designs has been recently tackled in the open literature, in which \ac{AE}-based constellation design is developed with the aim to achieve an improved diversity gain in the Grassmannian setup \cite{ fu2021grassmannian}. On the other hand, emphasizing on the large cardinality feature of Grassmannian modulation, symbols labeling/mapping is computationally challenging, and hence, the development of efficient, yet simple, labeling algorithms is of paramount importance. This can be achieved by the design of unconventional labeling methods, which do not rely on the distances between symbols only, but also on the geometry of the Grassmannian surface.

\subsection{Non-coherent Tensor Modulation}

Emphasizing on the fact that the merit of \ac{IoT} networks can be leveraged through the realization of efficient massive-scale connectivity, under strict on-board power resources and limited spectrum availability, \textit{unsourced random access} is considered as an intuitive choice in such networks \cite{shao2020}. In unsourced random access, a massive number of IoT devices are occasionally activated and allowed to access the spectrum simultaneously, in order to transmit short packets. It is further noted that such an activation process is random, i.e., the identity of the activated devices is unknown. While the total number of IoT devices is not of concern, the cardinality of the subset of the active devices is a key factor. Subsequently, the challenge of associating an active device with a particular transmitted waveform is intractable, rendering pilot-based CSI acquisition, and hence, grant-based random access a complicated task. This necessitates a radical departure from coordinated multiple access paradigms into grant-free solutions. 

A promising candidate, non-coherent \ac{TBM} has been recently deemed as an enabler for uncoordinated multiple access over the uplink communication, enabling reliable connectivity at a massive scale. In particular, with the aim to realize reliable spectrum access to a large number of IoT devices and signals detection, in \ac{TBM}, each user encodes its message in a rank-1 tensor, which is a generalized form of matrices with higher degrees of freedom. At the receiver end, Canonical Polyadic (CP) decomposition and successive interference cancellation are adopted to separate these tensors, and therefore, perform efficient signals detection \cite{phan2021canonical}. It was demonstrated through recent results that \ac{TBM} can successfully allow the concurrent activation of hundreds of devices, compared to coordinated time-division multiple access (TDMA), where the latter grants permit-to-access to a few numbers of devices at a given time \cite{decurninge2020tensor}. In addition to the increased number of activated users, \ac{TBM} plays an important role in relieving the users-base station coordination complexity, encountered in massive access scenarios. On the other hand, low-complex signals demapping and efficient constellation design in \ac{TBM} represent open research directions.

\subsection{Summary}
The high mobility and the dynamic nature of current and future wireless networks have reignited the interest in non-coherent schemes, as a mean to relieve the pressure imposed by \ac{CSI} acquisition. Within this context, Grassmannian signaling and non-coherent tensor modulation have been identified as promising approaches to modulate waveforms in a non-coherent fashion. While they showed superior performance when implemented in large-scale networks, in terms of throughput and coordination difficulty, there are several concerns that should be taken into account before the official adoption of these waveforms in 6G networks, including constellation design and symbols mapping/demapping.

\section{Conclusion} \label{conclusions}
%\hl{Lina}
In this article, we have revisited the basic foundations of the communication theory, and overviewed the key scientific challenges, that have surfaced with the recent revolutionary advancements in wireless technologies. In more detail, we have laid down a forward-looking vision for rethinking the basis of the electromagnetic theory, questioning the fundamentals of the physical limitations, and investigating the feasibility to approach the limits of infinite antennas. In our paper, through deep reconsideration of the key principles of the communication paradigms, it has been demonstrated that 6G networks will be designed on a large-scale, and this calls for a compelling need to revisit the existing non-coherent mechanisms, the integration of sensing and communication capabilities, as well as the efficient development of multi-agent learning systems. Furthermore, we investigated how incorporating the semantic aspect of transmitted messages will be an enabler for delivering experience-based applications. Also, we have sketched a roadmap towards understanding the interplay of queuing and information theory, and its implication on delivering \ac{URLLC} through the realization of efficient cross-layer solutions. In the thermodynamics of computation and communication theory, we narrowed the scope to focus on the energy efficiency of future 6G networks, where we highlighted the basic principles for striking a balance between communication performance and network energy consumption. In addition, we have studied the limitations of non-linear systems and the potential development of reliable signal processing algorithms that are capable of meeting the needs of such challenging systems. Furthermore, we shed light on the non-equilibrium information theory, where have explored the potentials of leveraging joint source-channel codes for enhanced performance in short-blocklength scenarios, and identified their performance limiting factors. Moreover, through a closer look into the limitation of wireless systems under double selective channels, we have deeply discussed the challenges and promising waveform designs within the time-varying systems theory. Finally, we have shed light on the problem of dealing with coarse scale information by digging into the fundamentals of the super-resolution theory.

\bibliographystyle{IEEEtran}
\bibliography{ref}

% Generated by IEEEtran.bst, version: 1.14 (2015/08/26)
\begin{thebibliography}{100}
\providecommand{\url}[1]{#1}
\csname url@samestyle\endcsname
\providecommand{\newblock}{\relax}
\providecommand{\bibinfo}[2]{#2}
\providecommand{\BIBentrySTDinterwordspacing}{\spaceskip=0pt\relax}
\providecommand{\BIBentryALTinterwordstretchfactor}{4}
\providecommand{\BIBentryALTinterwordspacing}{\spaceskip=\fontdimen2\font plus
\BIBentryALTinterwordstretchfactor\fontdimen3\font minus
  \fontdimen4\font\relax}
\providecommand{\BIBforeignlanguage}[2]{{%
\expandafter\ifx\csname l@#1\endcsname\relax
\typeout{** WARNING: IEEEtran.bst: No hyphenation pattern has been}%
\typeout{** loaded for the language `#1'. Using the pattern for}%
\typeout{** the default language instead.}%
\else
\language=\csname l@#1\endcsname
\fi
#2}}
\providecommand{\BIBdecl}{\relax}
\BIBdecl

\bibitem{tong20226g}
W.~Tong and P.~Zhu, ``6g: The next horizon,'' \emph{TITLES}, p.~54, 2022.

\bibitem{ji2021survey}
B.~Ji, Y.~Wang, K.~Song, C.~Li, H.~Wen, V.~G. Menon, and S.~Mumtaz, ``A survey
  of computational intelligence for {6G}: Key technologies, applications and
  trends,'' \emph{IEEE Trans. Industrial Informatics}, vol.~17, no.~10, pp.
  7145--7154, 2021.

\bibitem{jiang2021road}
W.~Jiang, B.~Han, M.~A. Habibi, and H.~D. Schotten, ``{The road towards 6G: A
  comprehensive survey},'' \emph{IEEE Open Journal of the Communications
  Society}, vol.~2, pp. 334--366, 2021.

\bibitem{alsabah20216g}
M.~Alsabah, M.~A. Naser, B.~M. Mahmmod, S.~H. Abdulhussain, M.~R. Eissa,
  A.~Al-Baidhani, N.~K. Noordin, S.~M. Sait, K.~A. Al-Utaibi, and F.~Hashim,
  ``6g wireless communications networks: A comprehensive survey,'' \emph{IEEE
  Access}, vol.~9, pp. 148\,191--148\,243, 2021.

\bibitem{uusitalo20216g}
M.~A. Uusitalo, P.~Rugeland, M.~R. Boldi, E.~C. Strinati, P.~Demestichas,
  M.~Ericson, G.~P. Fettweis, M.~C. Filippou, A.~Gati, M.-H. Hamon
  \emph{et~al.}, ``6g vision, value, use cases and technologies from european
  6g flagship project hexa-x,'' \emph{IEEE Access}, vol.~9, pp.
  160\,004--160\,020, 2021.

\bibitem{tomkos2020toward}
I.~Tomkos, D.~Klonidis, E.~Pikasis, and S.~Theodoridis, ``Toward the 6g network
  era: Opportunities and challenges,'' \emph{IT Professional}, vol.~22, no.~1,
  pp. 34--38, 2020.

\bibitem{tataria20216g}
H.~Tataria, M.~Shafi, A.~F. Molisch, M.~Dohler, H.~Sj{\"o}land, and
  F.~Tufvesson, ``6g wireless systems: Vision, requirements, challenges,
  insights, and opportunities,'' \emph{Proceedings of the IEEE}, vol. 109,
  no.~7, pp. 1166--1199, 2021.

\bibitem{tataria2022six}
H.~Tataria, M.~Shafi, M.~Dohler, and S.~Sun, ``Six critical challenges for 6g
  wireless systems: A summary and some solutions,'' \emph{IEEE Vehicular
  Technology Magazine}, vol.~17, no.~1, pp. 16--26, 2022.

\bibitem{bhat20216g}
J.~R. Bhat and S.~A. Alqahtani, ``6g ecosystem: Current status and future
  perspective,'' \emph{IEEE Access}, vol.~9, pp. 43\,134--43\,167, 2021.

\bibitem{ji2021several}
B.~Ji, Y.~Han, S.~Liu, F.~Tao, G.~Zhang, Z.~Fu, and C.~Li, ``Several key
  technologies for {6G}: challenges and opportunities,'' \emph{IEEE Commun.
  Standards Mag.}, vol.~5, no.~2, pp. 44--51, 2021.

\bibitem{Gabor}
D.~Gabor, ``Communication theory and physics,'' \emph{Transactions of the IRE
  Professional Group on Information Theory}, vol.~1, no.~1, pp. 48--59, 1953.

\bibitem{loyka2004information}
S.~Loyka, ``Information theory and electromagnetism: Are they related?'' in
  \emph{2004 10th International Symposium on Antenna Technology and Applied
  Electromagnetics and URSI Conference}.\hskip 1em plus 0.5em minus 0.4em\relax
  IEEE, 2004, pp. 1--5.

\bibitem{pierri1998information}
R.~Pierri and F.~Soldovieri, ``On the information content of the radiated
  fields in the near zone over bounded domains,'' \emph{Inverse Problems},
  vol.~14, no.~2, p. 321, 1998.

\bibitem{slepian1961prolate}
D.~Slepian and H.~O. Pollak, ``Prolate spheroidal wave functions, fourier
  analysis and uncertainty—i,'' \emph{Bell System Technical Journal},
  vol.~40, no.~1, pp. 43--63, 1961.

\bibitem{landau1961prolate}
H.~J. Landau and H.~O. Pollak, ``Prolate spheroidal wave functions, fourier
  analysis and uncertainty—ii,'' \emph{Bell System Technical Journal},
  vol.~40, no.~1, pp. 65--84, 1961.

\bibitem{linfoot1955information}
E.~Linfoot, ``Information theory and optical images,'' \emph{Josa}, vol.~45,
  no.~10, pp. 808--819, 1955.

\bibitem{bucci1989degrees}
O.~M. Bucci and G.~Franceschetti, ``On the degrees of freedom of scattered
  fields,'' \emph{IEEE transactions on Antennas and Propagation}, vol.~37,
  no.~7, pp. 918--926, 1989.

\bibitem{piestun2000electromagnetic}
R.~Piestun and D.~A. Miller, ``Electromagnetic degrees of freedom of an optical
  system,'' \emph{JOSA A}, vol.~17, no.~5, pp. 892--902, 2000.

\bibitem{bjornson2019massive}
E.~Bj{\"o}rnson, L.~Sanguinetti, H.~Wymeersch, J.~Hoydis, and T.~L. Marzetta,
  ``Massive mimo is a reality—what is next?: Five promising research
  directions for antenna arrays,'' \emph{Digital Signal Processing}, vol.~94,
  pp. 3--20, 2019.

\bibitem{marzetta2018spatially}
T.~L. Marzetta, ``Spatially-stationary propagating random field model for
  massive mimo small-scale fading,'' in \emph{2018 IEEE International Symposium
  on Information Theory (ISIT)}.\hskip 1em plus 0.5em minus 0.4em\relax IEEE,
  2018, pp. 391--395.

\bibitem{pizzo2020degrees}
A.~Pizzo, T.~L. Marzetta, and L.~Sanguinetti, ``Degrees of freedom of
  holographic mimo channels,'' in \emph{2020 IEEE 21st International Workshop
  on Signal Processing Advances in Wireless Communications (SPAWC)}.\hskip 1em
  plus 0.5em minus 0.4em\relax IEEE, 2020, pp. 1--5.

\bibitem{alghamdi2020intelligent}
R.~Alghamdi, R.~Alhadrami, D.~Alhothali, H.~Almorad, A.~Faisal, S.~Helal,
  R.~Shalabi, R.~Asfour, N.~Hammad, A.~Shams \emph{et~al.}, ``Intelligent
  surfaces for 6g wireless networks: A survey of optimization and performance
  analysis techniques,'' \emph{IEEE access}, 2020.

\bibitem{chen2016time}
Y.~Chen, B.~Wang, Y.~Han, H.-Q. Lai, Z.~Safar, and K.~R. Liu, ``Why time
  reversal for future 5g wireless?[perspectives],'' \emph{IEEE Signal
  Processing Magazine}, vol.~33, no.~2, pp. 17--26, 2016.

\bibitem{fink1992time}
M.~Fink, ``Time reversal of ultrasonic fields. i. basic principles,''
  \emph{IEEE transactions on ultrasonics, ferroelectrics, and frequency
  control}, vol.~39, no.~5, pp. 555--566, 1992.

\bibitem{guo2007reduced}
N.~Guo, B.~M. Sadler, and R.~C. Qiu, ``Reduced-complexity uwb time-reversal
  techniques and experimental results,'' \emph{IEEE Transactions on Wireless
  Communications}, vol.~6, no.~12, pp. 4221--4226, 2007.

\bibitem{han2012time}
F.~Han, Y.-H. Yang, B.~Wang, Y.~Wu, and K.~R. Liu, ``Time-reversal division
  multiple access over multi-path channels,'' \emph{IEEE Transactions on
  Communications}, vol.~60, no.~7, pp. 1953--1965, 2012.

\bibitem{lerosey2007focusing}
G.~Lerosey, J.~De~Rosny, A.~Tourin, and M.~Fink, ``Focusing beyond the
  diffraction limit with far-field time reversal,'' \emph{Science}, vol. 315,
  no. 5815, pp. 1120--1122, 2007.

\bibitem{linearization}
W.~Chen, S.~A. Bassam, X.~Li, Y.~Liu, K.~Rawat, M.~Helaoui, F.~M. Ghannouchi,
  and Z.~Feng, ``Design and linearization of concurrent dual-band doherty power
  amplifier with frequency-dependent power ranges,'' \emph{IEEE Transactions on
  Microwave Theory and Techniques}, vol.~59, no.~10, pp. 2537--2546, 2011.

\bibitem{arce2005nonlinear}
G.~R. Arce, \emph{Nonlinear signal processing: a statistical approach}.\hskip
  1em plus 0.5em minus 0.4em\relax John Wiley \& Sons, 2005.

\bibitem{Volterra}
M.~O. Franz and B.~Schölkopf, ``{A Unifying View of Wiener and Volterra Theory
  and Polynomial Kernel Regression},'' \emph{Neural Computation}, vol.~18,
  no.~12, pp. 3097--3118, 12 2006.

\bibitem{infeld2000nonlinear}
E.~Infeld and G.~Rowlands, \emph{Nonlinear waves, solitons and chaos}.\hskip
  1em plus 0.5em minus 0.4em\relax Cambridge university press, 2000.

\bibitem{RFtransceiver}
K.-L. Du and M.~N. Swamy, \emph{Wireless communication systems: from RF
  subsystems to 4G enabling technologies}.\hskip 1em plus 0.5em minus
  0.4em\relax Cambridge University Press, 2010.

\bibitem{rfImpairment}
L.~Smaini, \emph{{Rf analog impairments modeling for communication systems
  simulation: application to OFDM-based transceivers}}.\hskip 1em plus 0.5em
  minus 0.4em\relax John Wiley \& Sons, 2012.

\bibitem{RFmmWave}
I.~A. Hemadeh, K.~Satyanarayana, M.~El-Hajjar, and L.~Hanzo, ``{Millimeter-Wave
  Communications: Physical Channel Models, Design Considerations, Antenna
  Constructions, and Link-Budget},'' \emph{IEEE Communications Surveys and
  Tutorials}, vol.~20, no.~2, pp. 870--913, 2018.

\bibitem{non-linear-mimo}
O.~T. Demir and E.~Bjornson, ``The bussgang decomposition of nonlinear systems:
  Basic theory and mimo extensions [lecture notes],'' \emph{IEEE Signal
  Processing Magazine}, vol.~38, no.~1, pp. 131--136, 2021.

\bibitem{non-linear-FT1}
J.-W. Goossens, H.~Hafermann, and Y.~Jaouën, ``Data transmission based on
  exact inverse periodic nonlinear fourier transform, part i: Theory,''
  \emph{Journal of Lightwave Technology}, vol.~38, no.~23, pp. 6499--6519,
  2020.

\bibitem{non-linear-FT2}
J.-W. Goossens, H.~Hafermann, and Y.~Jaou\"en, ``Data transmission based on
  exact inverse periodic nonlinear fourier transform, part ii: Waveform design
  and experiment,'' \emph{Journal of Lightwave Technology}, vol.~38, no.~23,
  pp. 6520--6528, 2020.

\bibitem{DL_overair}
S.~Dörner, S.~Cammerer, J.~Hoydis, and S.~t. Brink, ``Deep learning based
  communication over the air,'' \emph{IEEE Journal of Selected Topics in Signal
  Processing}, vol.~12, no.~1, pp. 132--143, 2018.

\bibitem{wahab2021federated}
O.~A. Wahab, A.~Mourad, H.~Otrok, and T.~Taleb, ``Federated machine learning:
  Survey, multi-level classification, desirable criteria and future directions
  in communication and networking systems,'' \emph{IEEE Communications Surveys
  \& Tutorials}, vol.~23, no.~2, pp. 1342--1397, 2021.

\bibitem{dorri2018multi}
A.~Dorri, S.~S. Kanhere, and R.~Jurdak, ``Multi-agent systems: A survey,''
  \emph{IEEE Access}, vol.~6, pp. 28\,573--28\,593, 2018.

\bibitem{lazaridou2020emergent}
A.~Lazaridou and M.~Baroni, ``Emergent multi-agent communication in the deep
  learning era,'' \emph{arXiv preprint arXiv:2006.02419}, 2020.

\bibitem{simoes2019multi}
D.~Sim{\~o}es, N.~Lau, and L.~P. Reis, ``Multi-agent deep reinforcement
  learning with emergent communication,'' in \emph{2019 International Joint
  Conference on Neural Networks (IJCNN)}.\hskip 1em plus 0.5em minus
  0.4em\relax IEEE, 2019, pp. 1--8.

\bibitem{arnold2022maxray}
M.~Arnold, M.~Bauhofer, S.~Mandelli, M.~Henninger, F.~Schaich, T.~Wild, and
  S.~ten Brink, ``Maxray: A raytracing-based integrated sensing and
  communication framework,'' in \emph{2022 2nd IEEE International Symposium on
  Joint Communications \& Sensing (JC\&S)}.\hskip 1em plus 0.5em minus
  0.4em\relax IEEE, 2022, pp. 1--7.

\bibitem{candes2014towards}
E.~J. Cand{\`e}s and C.~Fernandez-Granda, ``Towards a mathematical theory of
  super-resolution,'' \emph{Communications on pure and applied Mathematics},
  vol.~67, no.~6, pp. 906--956, 2014.

\bibitem{sauer2018prony}
T.~Sauer, ``Pronys method: An old trick for new problems,'' 2018.

\bibitem{potts2010parameter}
D.~Potts and M.~Tasche, ``Parameter estimation for exponential sums by
  approximate prony method,'' \emph{Signal Processing}, vol.~90, no.~5, pp.
  1631--1642, 2010.

\bibitem{matrixPencil}
T.~Sarkar and O.~Pereira, ``Using the matrix pencil method to estimate the
  parameters of a sum of complex exponentials,'' \emph{IEEE Antennas and
  Propagation Magazine}, vol.~37, no.~1, pp. 48--55, 1995.

\bibitem{Li202204}
Z.~Li, A.~Nimr, P.~Schulz, and G.~Fettweis, ``Superresolution wireless
  multipath channel path delay estimation for cir-based localization,'' in
  \emph{IEEE Wireless Communications and Networking Conference (WCNC)}, Austin,
  USA, Apr 2022.

\bibitem{MUSIC}
R.~Roy, A.~Paulraj, and T.~Kailath, ``Estimation of signal parameters via
  rotational invariance techniques - esprit,'' in \emph{MILCOM 1986 - IEEE
  Military Communications Conference: Communications-Computers: Teamed for the
  90's}, vol.~3, 1986, pp. 41.6.1--41.6.5.

\bibitem{bazzi2016jaded}
A.~Bazzi, D.~T. Slock, and L.~Meilhac, ``{JADED-RIP: Joint Angle and Delay
  Estimator and Detector via Rotational Invariance Properties},'' in \emph{2016
  IEEE International Symposium on Signal Processing and Information Technology
  (ISSPIT)}.\hskip 1em plus 0.5em minus 0.4em\relax IEEE, 2016, pp. 160--165.

\bibitem{wax1994direction}
M.~Wax and J.~Sheinvald, ``{Direction finding of coherent signals via spatial
  smoothing for uniform circular arrays},'' \emph{IEEE transactions on antennas
  and propagation}, vol.~42, no.~5, pp. 613--620, 1994.

\bibitem{sparse}
B.~Diederichs and A.~Iske, ``Parameter estimation for bivariate exponential
  sums,'' in \emph{2015 International Conference on Sampling Theory and
  Applications (SampTA)}, 2015, pp. 493--497.

\bibitem{heckel2016super}
R.~Heckel, ``Super-resolution mimo radar,'' in \emph{2016 IEEE international
  symposium on information theory (ISIT)}.\hskip 1em plus 0.5em minus
  0.4em\relax IEEE, 2016, pp. 1416--1420.

\bibitem{6g_power}
P.~Skrimponis, S.~Dutta, M.~Mezzavilla, S.~Rangan, S.~H. Mirfarshbafan,
  C.~Studer, J.~Buckwalter, and M.~Rodwell, ``Power consumption analysis for
  mobile mmwave and sub-thz receivers,'' in \emph{2020 2nd 6G Wireless Summit
  (6G SUMMIT)}, 2020, pp. 1--5.

\bibitem{hexa_x_vision}
M.~A. Uusitalo \emph{et~al.}, ``6g vision, value, use cases and technologies
  from european 6g flagship project hexa-x,'' \emph{IEEE Access}, vol.~9, pp.
  160\,004--160\,020, 2021.

\bibitem{6g_platform}
X.~Qiao, Y.~Huang, S.~Dustdar, and J.~Chen, ``6g vision: An ai-driven
  decentralized network and service architecture,'' \emph{IEEE Internet
  Computing}, vol.~24, no.~4, pp. 33--40, 2020.

\bibitem{joint_com_comp}
M.~Merluzzi, C.~Battiloro, P.~Di~Lorenzo, and E.~C. Strinati,
  ``Energy-efficient classification at the wireless edge with reliability
  guarantees,'' \emph{arXiv preprint arXiv:2204.10399}, 2022.

\bibitem{joint_com_comp_latency}
J.~Ren, Y.~He, G.~Yu, and G.~Y. Li, ``Joint communication and computation
  resource allocation for cloud-edge collaborative system,'' in \emph{2019 IEEE
  Wireless Communications and Networking Conference (WCNC)}, 2019, pp. 1--6.

\bibitem{joint_com_comp_latency_energy}
D.~Wu, F.~Wang, X.~Cao, and J.~Xu, ``Joint communication and computation
  optimization for wireless powered mobile edge computing with d2d
  offloading,'' \emph{Journal of Communications and Information Networks},
  vol.~4, no.~4, pp. 72--86, 2019.

\bibitem{itu2014tactile}
I.~T.~U. ITU-T, ``The tactile internet,'' \emph{ITU-T Technology Watch Report},
  2014.

\bibitem{saifullah2014end}
A.~Saifullah, Y.~Xu, C.~Lu, and Y.~Chen, ``End-to-end communication delay
  analysis in industrial wireless networks,'' \emph{IEEE Transactions on
  Computers}, vol.~64, no.~5, pp. 1361--1374, 2014.

\bibitem{energyconsumption}
M.~H{\"o}yhty{\"a}, O.~Apilo, and M.~Lasanen, ``Review of latest advances in
  3gpp standardization: D2d communication in 5g systems and its energy
  consumption models,'' \emph{Future Internet}, vol.~10, no.~1, p.~3, 2018.

\bibitem{limitofcomput}
S.~Lloyd, ``Ultimate physical limits to computation,'' \emph{Nature}, vol. 406,
  no. 6799, pp. 1047--1054, 2000.

\bibitem{limitofcomp}
S.~Lloyd, V.~Giovannetti, and L.~Maccone, ``Physical limits to communication,''
  \emph{Physical review letters}, vol.~93, no.~10, p. 100501, 2004.

\bibitem{DL_limit}
N.~C. Thompson, K.~Greenewald, K.~Lee, and G.~F. Manso, ``The computational
  limits of deep learning,'' \emph{arXiv preprint arXiv:2007.05558}, 2020.

\bibitem{molisch2012wireless}
A.~F. Molisch, \emph{Wireless communications}.\hskip 1em plus 0.5em minus
  0.4em\relax John Wiley \& Sons, 2012.

\bibitem{bomfin2018theoretical}
R.~Bomfin, D.~Zhang, M.~Matth{\'e}, and G.~Fettweis, ``{Communication theory
  and physics},'' \emph{IEEE Communications Letters}, vol.~22, no.~11, pp.
  2394--2397, 2018.

\bibitem{bomfin2021robust}
R.~Bomfin, M.~Chafii, A.~Nimr, and G.~Fettweis, ``{A Robust Baseband
  Transceiver Design for Doubly-dispersive Channels},'' \emph{IEEE Transactions
  on Wireless Communications}, vol.~20, no.~8, pp. 4781--4796, 2021.

\bibitem{studer2011asic}
C.~Studer, S.~Fateh, and D.~Seethaler, ``{ASIC Implementation of soft-input
  soft-output MIMO Detection Using MMSE Parallel Interference Cancellation},''
  \emph{IEEE Journal of Solid-State Circuits}, vol.~46, no.~7, pp. 1754--1765,
  2011.

\bibitem{benvenuto2009single}
N.~Benvenuto, R.~Dinis, D.~Falconer, and S.~Tomasin, ``{Single Carrier
  Modulation with Nonlinear Frequency Domain Equalization: An Idea Whose Time
  Has Come-Again},'' \emph{Proceedings of the IEEE}, vol.~98, no.~1, pp.
  69--96, 2009.

\bibitem{nimr2018extended}
A.~Nimr, M.~Chafii, M.~Matth{\'e}, and G.~Fettweis, ``{Extended GFDM framework:
  OTFS and GFDM comparison},'' in \emph{2018 IEEE global communications
  conference (GLOBECOM)}.\hskip 1em plus 0.5em minus 0.4em\relax IEEE, 2018,
  pp. 1--6.

\bibitem{bomfin2020robust}
R.~Bomfin, A.~Nimr, M.~Chafii, and G.~Fettweis, ``A robust and low-complexity
  walsh-hadamard modulation for doubly-dispersive channels,'' \emph{IEEE
  Communications Letters}, vol.~25, no.~3, pp. 897--901, 2020.

\bibitem{bomfin2019performance}
R.~Bomfin, M.~Chafii, and G.~Fettweis, ``Performance assessment of orthogonal
  chirp division multiplexing in mimo space time coding,'' in \emph{2019 IEEE
  2nd 5G World Forum (5GWF)}.\hskip 1em plus 0.5em minus 0.4em\relax IEEE,
  2019, pp. 220--225.

\bibitem{bomfin2019low}
------, ``{Low-complexity iterative receiver for orthogonal chirp division
  multiplexing},'' in \emph{2019 IEEE Wireless Communications and Networking
  Conference Workshop (WCNCW)}.\hskip 1em plus 0.5em minus 0.4em\relax IEEE,
  2019, pp. 1--6.

\bibitem{thaj2020low}
T.~Thaj and E.~Viterbo, ``{Low Complexity Iterative Rake Decision Feedback
  Equalizer for Zero-padded OTFS Systems},'' \emph{IEEE transactions on
  vehicular technology}, vol.~69, no.~12, pp. 15\,606--15\,622, 2020.

\bibitem{enku2021two}
Y.~K. Enku, B.~Bai, F.~Wan, C.~U. Guyo, I.~N. Tiba, C.~Zhang, and S.~Li,
  ``{Two-Dimensional Convolutional Neural Network-Based Signal Detection for
  OTFS Systems},'' \emph{IEEE wireless communications letters}, vol.~10,
  no.~11, pp. 2514--2518, 2021.

\bibitem{naikoti2021low}
A.~Naikoti and A.~Chockalingam, ``{Low-complexity Delay-Doppler Symbol DNN for
  OTFS Signal Detection},'' in \emph{2021 IEEE 93rd Vehicular Technology
  Conference (VTC2021-Spring)}.\hskip 1em plus 0.5em minus 0.4em\relax IEEE,
  2021, pp. 1--6.

\bibitem{9813719}
A.~K. Gizzini and M.~Chafii, ``{A Survey on Deep Learning Based Channel
  Estimation in Doubly Dispersive Environments},'' \emph{IEEE Access}, vol.~10,
  pp. 70\,595--70\,619, 2022.

\bibitem{gizzini2021cnn}
A.~K. Gizzini, M.~Chafii, A.~Nimr, R.~M. Shubair, and G.~Fettweis, ``{CNN aided
  weighted interpolation for channel estimation in vehicular communications},''
  \emph{IEEE Transactions on Vehicular Technology}, vol.~70, no.~12, pp.
  12\,796--12\,811, 2021.

\bibitem{gizzini2020deep}
A.~K. Gizzini, M.~Chafii, A.~Nimr, and G.~Fettweis, ``{Deep learning based
  channel estimation schemes for IEEE 802.11 p standard},'' \emph{IEEE Access},
  vol.~8, pp. 113\,751--113\,765, 2020.

\bibitem{gizzini2021temporal}
A.~K. Gizzini, M.~Chafii, S.~Ehsanfar, and R.~M. Shubair, ``{Temporal Averaging
  LSTM-based Channel Estimation Scheme for IEEE 802.11 p Standard},'' in
  \emph{2021 IEEE Global Communications Conference (GLOBECOM)}.\hskip 1em plus
  0.5em minus 0.4em\relax IEEE, 2021, pp. 01--07.

\bibitem{ehsanfar2020uw}
S.~Ehsanfar, M.~Chafii, and G.~P. Fettweis, ``On uw-based transmission for mimo
  multi-carriers with spatial multiplexing,'' \emph{IEEE Transactions on
  Wireless Communications}, vol.~19, no.~9, pp. 5875--5890, 2020.

\bibitem{fettweis20216g}
G.~P. Fettweis and H.~Boche, ``{6G: The Personal Tactile Internet-and Open
  Questions for Information Theory},'' \emph{IEEE BITS the Information Theory
  Magazine}, 2021.

\bibitem{neuhaus2021enabling}
P.~Neuhaus, M.~Schl{\"u}ter, C.~Jans, M.~D{\"o}rpinghaus, and G.~Fettweis,
  ``Enabling energy-efficient tbit/s communications by 1-bit quantization and
  oversampling,'' in \emph{2021 Joint European Conference on Networks and
  Communications \& 6G Summit (EuCNC/6G Summit)}.\hskip 1em plus 0.5em minus
  0.4em\relax IEEE, 2021, pp. 84--89.

\bibitem{shi2021semantic}
G.~Shi, Y.~Xiao, Y.~Li, and X.~Xie, ``From semantic communication to
  semantic-aware networking: Model, architecture, and open problems,''
  \emph{IEEE Commun. Mag.}, vol.~59, no.~8, pp. 44--50, 2021.

\bibitem{lan2021semantic}
Q.~Lan, D.~Wen, Z.~Zhang, Q.~Zeng, X.~Chen, P.~Popovski, and K.~Huang, ``What
  is semantic communication? a view on conveying meaning in the era of machine
  intelligence,'' \emph{Journal of Communications and Information Networks},
  vol.~6, no.~4, pp. 336--371, 2021.

\bibitem{baudot2015homological}
P.~Baudot and D.~Bennequin, ``The homological nature of entropy,''
  \emph{Entropy}, vol.~17, no.~5, pp. 3253--3318, 2015.

\bibitem{vigneaux2019topology}
J.~P. Vigneaux, ``Topology of statistical systems: a cohomological approach to
  information theory,'' Ph.D. dissertation, Universit{\'e} Sorbonne Paris
  Cit{\'e}, 2019.

\bibitem{5776640}
C.~Sturm and W.~Wiesbeck, ``Waveform design and signal processing aspects for
  fusion of wireless communications and radar sensing,'' \emph{Proceedings of
  the IEEE}, vol.~99, no.~7, pp. 1236--1259, 2011.

\bibitem{9497736}
W.~Li, M.~J. Bocus, C.~Tang, R.~J. Piechocki, K.~Woodbridge, and K.~Chetty,
  ``On csi and passive wi-fi radar for opportunistic physical activity
  recognition,'' \emph{IEEE Transactions on Wireless Communications}, vol.~21,
  no.~1, pp. 607--620, 2022.

\bibitem{9154212}
H.~Sun, P.~Wang, M.~Pajovic, T.~Koike-Akino, P.~V. Orlik, A.~Taira, and
  K.~Nakagawa, ``Fingerprinting-based outdoor localization with 28-ghz channel
  measurement: A field study,'' in \emph{2020 IEEE 21st International Workshop
  on Signal Processing Advances in Wireless Communications (SPAWC)}, 2020, pp.
  1--5.

\bibitem{bazzi2022outage}
A.~Bazzi and M.~Chafii, ``{On Outage-based Beamforming Design for
  Dual-Functional Radar-Communication 6G Systems},'' \emph{IEEE Transactions on
  Wireless Communications}, pp. 1--1, 2023.

\bibitem{liu2021cramer}
F.~Liu, Y.-F. Liu, A.~Li, C.~Masouros, and Y.~C. Eldar, ``Cram{\'e}r-rao bound
  optimization for joint radar-communication beamforming,'' \emph{IEEE
  Transactions on Signal Processing}, vol.~70, pp. 240--253, 2021.

\bibitem{chiriyath2015inner}
A.~R. Chiriyath, B.~Paul, G.~M. Jacyna, and D.~W. Bliss, ``Inner bounds on
  performance of radar and communications co-existence,'' \emph{IEEE
  Transactions on Signal Processing}, vol.~64, no.~2, pp. 464--474, 2015.

\bibitem{he2018performance}
Q.~He, Z.~Wang, J.~Hu, and R.~S. Blum, ``Performance gains from cooperative
  mimo radar and mimo communication systems,'' \emph{IEEE Signal Processing
  Letters}, vol.~26, no.~1, pp. 194--198, 2018.

\bibitem{cui2018perspective}
Y.~Cui, V.~Koivunen, and X.~Jing, ``A perspective on degrees of freedom for
  radar in radar-communication interference channel,'' in \emph{2018 52nd
  Asilomar Conference on Signals, Systems, and Computers}.\hskip 1em plus 0.5em
  minus 0.4em\relax IEEE, 2018, pp. 403--408.

\bibitem{fortunati2020massive}
S.~Fortunati, L.~Sanguinetti, F.~Gini, M.~S. Greco, and B.~Himed, ``Massive
  mimo radar for target detection,'' \emph{IEEE Transactions on Signal
  Processing}, vol.~68, pp. 859--871, 2020.

\bibitem{kobayashi2018joint}
M.~Kobayashi, G.~Caire, and G.~Kramer, ``Joint state sensing and communication:
  Optimal tradeoff for a memoryless case,'' in \emph{2018 IEEE International
  Symposium on Information Theory (ISIT)}.\hskip 1em plus 0.5em minus
  0.4em\relax IEEE, 2018, pp. 111--115.

\bibitem{ahmadipour2022information}
M.~Ahmadipour, M.~Kobayashi, M.~Wigger, and G.~Caire, ``An
  information-theoretic approach to joint sensing and communication,''
  \emph{IEEE Transactions on Information Theory}, 2022.

\bibitem{joudeh2022joint}
H.~Joudeh and F.~M. Willems, ``Joint communication and binary state
  detection,'' \emph{IEEE Journal on Selected Areas in Information Theory},
  vol.~3, no.~1, pp. 113--124, 2022.

\bibitem{han2018propagation}
C.~Han and Y.~Chen, ``Propagation modeling for wireless communications in the
  terahertz band,'' \emph{IEEE Communications Magazine}, vol.~56, no.~6, pp.
  96--101, 2018.

\bibitem{saleh1987statistical}
A.~A. Saleh and R.~Valenzuela, ``A statistical model for indoor multipath
  propagation,'' \emph{IEEE Journal on selected areas in communications},
  vol.~5, no.~2, pp. 128--137, 1987.

\bibitem{zwick2002stochastic}
T.~Zwick, C.~Fischer, and W.~Wiesbeck, ``A stochastic multipath channel model
  including path directions for indoor environments,'' \emph{IEEE journal on
  Selected Areas in Communications}, vol.~20, no.~6, pp. 1178--1192, 2002.

\bibitem{li2021integrated}
X.~Li, J.~He, Z.~Yu, G.~Wang, and P.~Zhu, ``Integrated sensing and
  communication in 6g: the deterministic channel models for thz imaging,'' in
  \emph{2021 IEEE 32nd Annual International Symposium on Personal, Indoor and
  Mobile Radio Communications (PIMRC)}.\hskip 1em plus 0.5em minus 0.4em\relax
  IEEE, 2021, pp. 1--6.

\bibitem{blender2018blender}
O.~Blender, ``Blender—a 3d modelling and rendering package,''
  \emph{Retrieved. represents the sequence of Constructs1 to}, vol.~4, 2018.

\bibitem{almers2007survey}
P.~Almers, E.~Bonek, A.~Burr, N.~Czink, M.~Debbah, V.~Degli-Esposti,
  H.~Hofstetter, P.~Ky{\"o}sti, D.~Laurenson, G.~Matz \emph{et~al.}, ``Survey
  of channel and radio propagation models for wireless mimo systems,''
  \emph{EURASIP Journal on Wireless Communications and Networking}, vol. 2007,
  pp. 1--19, 2007.

\bibitem{su2020secure}
N.~Su, F.~Liu, and C.~Masouros, ``Secure radar-communication systems with
  malicious targets: Integrating radar, communications and jamming
  functionalities,'' \emph{IEEE Transactions on Wireless Communications},
  vol.~20, no.~1, pp. 83--95, 2020.

\bibitem{jiao2021openwifi}
X.~Jiao, M.~Mehari, W.~Liu, M.~Aslam, and I.~Moerman, ``openwifi csi fuzzer for
  authorized sensing and covert channels,'' in \emph{Proceedings of the 14th
  ACM Conference on Security and Privacy in Wireless and Mobile Networks},
  2021, pp. 377--379.

\bibitem{haneda2010measurement}
K.~Haneda, E.~Kahra, S.~Wyne, C.~Icheln, and P.~Vainikainen, ``Measurement of
  loop-back interference channels for outdoor-to-indoor full-duplex radio
  relays,'' in \emph{Proceedings of the Fourth European Conference on Antennas
  and Propagation}.\hskip 1em plus 0.5em minus 0.4em\relax IEEE, 2010, pp.
  1--5.

\bibitem{ahmed2015all}
E.~Ahmed and A.~M. Eltawil, ``All-digital self-interference cancellation
  technique for full-duplex systems,'' \emph{IEEE Transactions on Wireless
  Communications}, vol.~14, no.~7, pp. 3519--3532, 2015.

\bibitem{snow2011transmit}
T.~Snow, C.~Fulton, and W.~J. Chappell, ``Transmit--receive duplexing using
  digital beamforming system to cancel self-interference,'' \emph{IEEE
  Transactions on Microwave Theory and Techniques}, vol.~59, no.~12, pp.
  3494--3503, 2011.

\bibitem{griffiths2009passive}
H.~Griffiths, ``Passive bistatic radar and waveform diversity,'' DEFENCE
  ACADEMY OF THE UNITED KINGDOM SHRIVENHAM (UNITED KINGDOM), Tech. Rep., 2009.

\bibitem{bazzi2022integrated}
A.~Bazzi and M.~Chafii, ``{On Integrated Sensing and Communication Waveforms
  with Tunable PAPR},'' \emph{arXiv preprint arXiv:2210.02892}, 2022.

\bibitem{bazzi2022ris}
------, ``{RIS-Enabled Passive Radar towards Target Localization},''
  \emph{arXiv preprint arXiv:2210.11887}, 2022.

\bibitem{futatsumori2016design}
S.~Futatsumori, K.~Morioka, A.~Kohmura, K.~Okada, and N.~Yonemoto, ``Design and
  field feasibility evaluation of distributed-type 96 ghz fmcw millimeter-wave
  radar based on radio-over-fiber and optical frequency multiplier,''
  \emph{Journal of Lightwave Technology}, vol.~34, no.~20, pp. 4835--4843,
  2016.

\bibitem{shin2016distributed}
D.-H. Shin, D.-H. Jung, D.-C. Kim, J.-W. Ham, and S.-O. Park, ``A distributed
  fmcw radar system based on fiber-optic links for small drone detection,''
  \emph{IEEE Transactions on Instrumentation and Measurement}, vol.~66, no.~2,
  pp. 340--347, 2016.

\bibitem{couillet2011a}
R.~Couillet and M.~Debbah, \emph{Random matrix methods for wireless
  communications}.\hskip 1em plus 0.5em minus 0.4em\relax Cambridge University
  Press, 2011.

\bibitem{dupuy2011}
F.~Dupuy and P.~Loubaton, ``On the capacity achieving covariance matrix for
  frequency selective {MIMO} channels using the asymptotic approach,''
  \emph{IEEE Trans. Info. Theory}, vol.~57, no.~9, pp. 5737--5753, 2011.

\bibitem{8320821}
L.~Liu and W.~Yu, ``Massive connectivity with massive {MIMO—Part II}:
  Achievable rate characterization,'' \emph{IEEE Tran. Signal Process.},
  vol.~66, no.~11, pp. 2947--2959, 2018.

\bibitem{bianchi2009}
P.~Bianchi, J.~Najim, M.~Maida, and M.~Debbah, ``Performance analysis of some
  eigen-based hypothesis tests for collaborative sensing,'' in \emph{2009
  IEEE/SP 15th Workshop on Statistical Signal Processing}.\hskip 1em plus 0.5em
  minus 0.4em\relax IEEE, 2009, pp. 5--8.

\bibitem{geraci2013}
G.~Geraci, R.~Couillet, J.~Yuan, M.~Debbah, and I.~B. Collings, ``Large system
  analysis of linear precoding in {MISO} broadcast channels with confidential
  messages,'' \emph{IEEE J. Sel. Areas Commun.}, vol.~31, no.~9, pp.
  1660--1671, 2013.

\bibitem{couillet2011}
R.~Couillet, M.~Debbah, and J.~W. Silverstein, ``A deterministic equivalent for
  the analysis of correlated {MIMO} multiple access channels,'' \emph{IEEE
  Trans. Info. Theory}, vol.~57, no.~6, pp. 3493--3514, 2011.

\bibitem{sifaou2016}
H.~Sifaou, A.~Kammoun, L.~Sanguinetti, M.~Debbah, and M.-S. Alouini,
  ``Max--{Min} {SINR} in large-scale single-cell {MU-MIMO}: Asymptotic analysis
  and low-complexity transceivers,'' \emph{IEEE Trans. Signal Process.},
  vol.~65, no.~7, pp. 1841--1854, 2016.

\bibitem{mestre2008}
X.~Mestre and M.~{\'A}. Lagunas, ``Modified subspace algorithms for {DoA}
  estimation with large arrays,'' \emph{IEEE Trans. Signal Process.}, vol.~56,
  no.~2, pp. 598--614, 2008.

\bibitem{adlam2019}
B.~Adlam, J.~Levinson, and J.~Pennington, ``A random matrix perspective on
  mixtures of nonlinearities for deep learning,'' \emph{arXiv preprint
  arXiv:1912.00827}, 2019.

\bibitem{ge2021}
J.~Ge, Y.-C. Liang, Z.~Bai, and G.~Pan, ``Large-dimensional random matrix
  theory and its applications in deep learning and wireless communications,''
  \emph{Random Matrices: Theory and Applications}, p. 2230001, 2021.

\bibitem{xin2020}
R.~Xin, S.~Kar, and U.~A. Khan, ``Decentralized stochastic optimization and
  machine learning: A unified variance-reduction framework for robust
  performance and fast convergence,'' \emph{IEEE Signal Proc. Mag.}, vol.~37,
  no.~3, pp. 102--113, 2020.

\bibitem{bianchi2011}
P.~Bianchi and J.~Jakubowicz, ``Distributed stochastic approximation for
  constrained and unconstrained optimization,'' \emph{arXiv preprint
  arXiv:1104.2773}, 2011.

\bibitem{xin2019}
R.~Xin, S.~Kar, and U.~A. Khan, ``An introduction to decentralized stochastic
  optimization with gradient tracking,'' \emph{arXiv preprint
  arXiv:1907.09648}, 2019.

\bibitem{cichocki2016}
A.~Cichocki, N.~Lee, I.~V. Oseledets, A.-H. Phan, Q.~Zhao, and D.~Mandic,
  ``Low-rank tensor networks for dimensionality reduction and large-scale
  optimization problems: Perspectives and challenges part 1,'' \emph{arXiv
  preprint arXiv:1609.00893}, 2016.

\bibitem{cichocki2014tensor}
A.~Cichocki, ``Tensor networks for big data analytics and large-scale
  optimization problems,'' \emph{arXiv preprint arXiv:1407.3124}, 2014.

\bibitem{decurninge2020}
A.~Decurninge, I.~Land, and M.~Guillaud, ``Tensor-based modulation for
  unsourced massive random access,'' \emph{IEEE Wireless Commun. Lett.},
  vol.~10, no.~3, pp. 552--556, 2020.

\bibitem{HC}
F.~P. Miller, A.~F. Vandome, and J.~McBrewster, \emph{Hamming Code: Parity Bit,
  Two- out- of- Five Code, Hamming(7,4), Reed?Muller Code, Reed?Solomon Error
  Correction, Turbo Code, Low- Density Parity- Check Code, Telecommunication,
  Linear Code}.\hskip 1em plus 0.5em minus 0.4em\relax Alpha Press, 2009.

\bibitem{paterson2000generalized}
K.~G. Paterson, ``Generalized reed-muller codes and power control in {OFDM}
  modulation,'' \emph{IEEE Trans. Info. Theory}, vol.~46, no.~1, pp. 104--120,
  2000.

\bibitem{chien1964cyclic}
R.~Chien, ``Cyclic decoding procedures for bose-chaudhuri-hocquenghem codes,''
  \emph{IEEE Transactions on information theory}, vol.~10, no.~4, pp. 357--363,
  1964.

\bibitem{wicker1999reed}
S.~B. Wicker and V.~K. Bhargava, \emph{Reed-Solomon codes and their
  applications}.\hskip 1em plus 0.5em minus 0.4em\relax John Wiley \& Sons,
  1999.

\bibitem{viterbi1971convolutional}
A.~Viterbi, ``Convolutional codes and their performance in communication
  systems,'' \emph{IEEE Transactions on Communication Technology}, vol.~19,
  no.~5, pp. 751--772, 1971.

\bibitem{vucetic2012turbo}
B.~Vucetic and J.~Yuan, \emph{Turbo codes: principles and applications}.\hskip
  1em plus 0.5em minus 0.4em\relax Springer Science \& Business Media, 2012,
  vol. 559.

\bibitem{tal2013construct}
I.~Tal and A.~Vardy, ``How to construct polar codes,'' \emph{IEEE Transactions
  on Information Theory}, vol.~59, no.~10, pp. 6562--6582, 2013.

\bibitem{gallager1962low}
R.~Gallager, ``Low-density parity-check codes,'' \emph{IRE Transactions on
  information theory}, vol.~8, no.~1, pp. 21--28, 1962.

\bibitem{arora2020survey}
K.~Arora, J.~Singh, and Y.~S. Randhawa, ``A survey on channel coding techniques
  for {5G} wireless networks,'' \emph{Telecommunication Systems}, vol.~73,
  no.~4, pp. 637--663, 2020.

\bibitem{lu20206g}
Y.~Lu and X.~Zheng, ``{6G}: A survey on technologies, scenarios, challenges,
  and the related issues,'' \emph{Journal of Industrial Information
  Integration}, vol.~19, p. 100158, 2020.

\bibitem{cocskun2019efficient}
M.~C. Co{\c{s}}kun, G.~Durisi, T.~Jerkovits, G.~Liva, W.~Ryan, B.~Stein, and
  F.~Steiner, ``Efficient error-correcting codes in the short blocklength
  regime,'' \emph{Physical Communication}, vol.~34, pp. 66--79, 2019.

\bibitem{huang2019ai}
L.~Huang, H.~Zhang, R.~Li, Y.~Ge, and J.~Wang, ``{AI} coding: Learning to
  construct error correction codes,'' \emph{IEEE Transactions on
  Communications}, vol.~68, no.~1, pp. 26--39, 2019.

\bibitem{dong2022joint}
Y.~Dong, J.~Dai, K.~Niu, S.~Wang, and Y.~Yuan, ``{Joint source-channel coding
  for 6G communications},'' \emph{China Communications}, vol.~19, no.~3, pp.
  101--115, 2022.

\bibitem{guyader2001joint}
A.~Guyader, E.~Fabre, C.~Guillemot, and M.~Robert, ``Joint source-channel turbo
  decoding of entropy-coded sources,'' \emph{IEEE Journal on Selected Areas in
  Communications}, vol.~19, no.~9, pp. 1680--1696, 2001.

\bibitem{pu2007ldpc}
L.~Pu, Z.~Wu, A.~Bilgin, M.~W. Marcellin, and B.~Vasic, ``{LDPC}-based
  iterative joint source-channel decoding for {JPEG2000},'' \emph{IEEE
  Transactions on Image Processing}, vol.~16, no.~2, pp. 577--581, 2007.

\bibitem{grangetto2005joint}
M.~Grangetto, P.~Cosman, and G.~Olmo, ``Joint source/channel coding and {MAP}
  decoding of arithmetic codes,'' \emph{IEEE Transactions on Communications},
  vol.~53, no.~6, pp. 1007--1016, 2005.

\bibitem{garcia2003ldpc}
J.~Garcia-Frias and W.~Zhong, ``{LDPC} codes for compression of multi-terminal
  sources with hidden {Markov} correlation,'' \emph{IEEE Communications
  Letters}, vol.~7, no.~3, pp. 115--117, 2003.

\bibitem{bhattad2006decision}
K.~Bhattad and K.~R. Narayanan, ``A decision feedback based scheme for
  {Slepian-Wolf} coding of sources with hidden {Markov} correlation,''
  \emph{IEEE communications letters}, vol.~10, no.~5, pp. 378--380, 2006.

\bibitem{telatar1995combining}
I.~E. Telatar and R.~G. Gallager, ``Combining queueing theory with information
  theory for multiaccess,'' \emph{IEEE Journal on Selected Areas in
  Communications}, vol.~13, no.~6, pp. 963--969, 1995.

\bibitem{lau2006channel}
V.~K. Lau and Y.-K.~R. Kwok, \emph{Channel-adaptive technologies and
  cross-layer designs for wireless systems with multiple antennas: theory and
  applications}.\hskip 1em plus 0.5em minus 0.4em\relax John Wiley \& Sons,
  2006, vol.~85.

\bibitem{Berry}
R.~Berry and E.~Yeh, ``Cross-layer wireless resource allocation,'' \emph{IEEE
  Signal Processing Magazine}, vol.~21, no.~5, pp. 59--68, 2004.

\bibitem{she}
C.~She, C.~Sun, Z.~Gu, Y.~Li, C.~Yang, H.~V. Poor, and B.~Vucetic, ``A tutorial
  on ultrareliable and low-latency communications in 6g: Integrating domain
  knowledge into deep learning,'' \emph{Proceedings of the IEEE}, vol. 109,
  no.~3, pp. 204--246, 2021.

\bibitem{zhang2021}
J.~Zhang, X.~Xu, K.~Zhang, S.~Han, X.~Tao, and P.~Zhang, ``Learning based
  flexible cross-layer optimization for ultra-reliable and low latency
  applications in iot scenarios,'' \emph{IEEE Internet of Things Journal},
  2021.

\bibitem{ephremides1998}
A.~Ephremides and B.~Hajek, ``Information theory and communication networks: An
  unconsummated union,'' \emph{IEEE Transactions on Information Theory},
  vol.~44, no.~6, pp. 2416--2434, 1998.

\bibitem{gallager1985}
R.~Gallager, ``A perspective on multiaccess channels,'' \emph{IEEE Transactions
  on Information Theory}, vol.~31, no.~2, pp. 124--142, 1985.

\bibitem{fattah2002overview}
H.~Fattah and C.~Leung, ``An overview of scheduling algorithms in wireless
  multimedia networks,'' \emph{IEEE Wireless Communications}, vol.~9, no.~5,
  pp. 76--83, 2002.

\bibitem{amraoui2003coding}
A.~Amraoui, G.~Kramer, and S.~Shamai, ``Coding for the mimo broadcast
  channel,'' in \emph{IEEE International Symposium on Information Theory, 2003.
  Proceedings.}\hskip 1em plus 0.5em minus 0.4em\relax IEEE, 2003, pp.
  296--296.

\bibitem{wang2021deepnetqoe}
R.~Wang, M.~Chen, N.~Guizani, Y.~Li, H.~Gharavi, and K.~Hwang, ``Deepnetqoe:
  Self-adaptive qoe optimization framework of deep networks,'' \emph{IEEE
  Network}, vol.~35, no.~3, pp. 161--167, 2021.

\bibitem{chiang2007layering}
M.~Chiang, S.~H. Low, A.~R. Calderbank, and J.~C. Doyle, ``Layering as
  optimization decomposition: A mathematical theory of network architectures,''
  \emph{Proceedings of the IEEE}, vol.~95, no.~1, pp. 255--312, 2007.

\bibitem{chowdhury2015}
M.~Chowdhury, A.~Manolakos, and A.~J. Goldsmith, ``Coherent versus noncoherent
  massive {SIMO} systems: Which has better performance?'' in \emph{IEEE
  International Conference on Communications (ICC)}.\hskip 1em plus 0.5em minus
  0.4em\relax IEEE, 2015, pp. 1691--1696.

\bibitem{hedhly2021}
W.~Hedhly, O.~Amin, B.~Shihada, and M.-S. Alouini, ``Hyperloop communications:
  Challenges, advances, and approaches,'' \emph{IEEE Open J. Commun. Society},
  vol.~2, pp. 2413--2435, 2021.

\bibitem{gohary2019noncoherent}
R.~H. Gohary and H.~Yanikomeroglu, ``Noncoherent {MIMO} signaling for
  block-fading channels: Approaches and challenges,'' \emph{IEEE veh. technol.
  mag.}, vol.~14, no.~1, pp. 80--88, 2019.

\bibitem{cabrejas2016non}
J.~Cabrejas, S.~Roger, D.~Calabuig, Y.~M. Fouad, R.~H. Gohary, J.~F. Monserrat,
  and H.~Yanikomeroglu, ``Non-coherent open-loop {MIMO} communications over
  temporally-correlated channels,'' \emph{IEEE Access}, vol.~4, pp. 6161--6170,
  2016.

\bibitem{fu2021grassmannian}
X.~Fu and D.~L. Ruyet, ``Grassmannian constellation design for noncoherent
  {MIMO} systems using autoencoders,'' \emph{arXiv preprint arXiv:2109.00621},
  2021.

\bibitem{shao2020}
X.~Shao, X.~Chen, D.~W.~K. Ng, C.~Zhong, and Z.~Zhang, ``Cooperative activity
  detection: Sourced and unsourced massive random access paradigms,''
  \emph{IEEE Trans. Signal Process.}, vol.~68, pp. 6578--6593, 2020.

\bibitem{phan2021canonical}
A.-H. Phan, P.~Tichavsk{\`y}, K.~Sobolev, K.~Sozykin, D.~Ermilov, and
  A.~Cichocki, ``Canonical polyadic tensor decomposition with low-rank factor
  matrices,'' in \emph{IEEE International Conference on Acoustics, Speech and
  Signal Processing (ICASSP)}.\hskip 1em plus 0.5em minus 0.4em\relax IEEE,
  2021, pp. 4690--4694.

\bibitem{decurninge2020tensor}
A.~Decurninge, I.~Land, and M.~Guillaud, ``Tensor-based modulation for
  unsourced massive random access,'' \emph{IEEE Wireless Commun. Lett.},
  vol.~10, no.~3, pp. 552--556, 2020.

\end{thebibliography}

\begin{IEEEbiography}[{\includegraphics[width=1in,height=1.25in,clip,keepaspectratio]{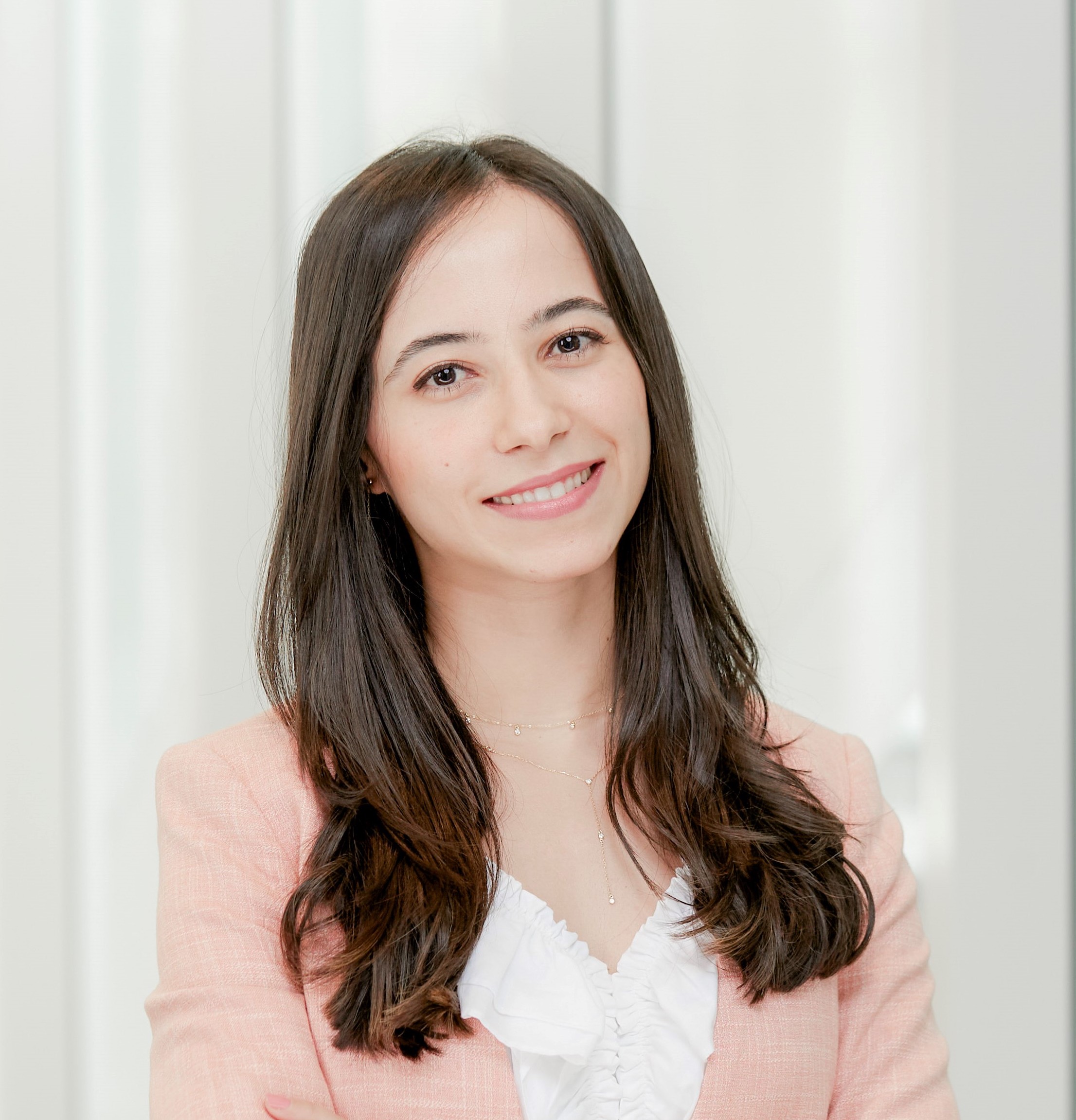}}]{Marwa Chafii}
received her Ph.D. degree in electrical engineering in 2016, and her Master's degree in the field of advanced wireless communication systems (SAR) in 2013, both from CentraleSupélec, France. Between 2014 and 2016, she has been a visiting researcher at Poznan University of Technology (Poland), University of York (UK), Yokohama National University (Japan), and University of Oxford (UK). She joined the Technical University of Dresden, Germany, in 2018 as a research group leader, and ENSEA, France, in 2019 as an associate professor where she held a Chair of Excellence on Artificial Intelligence from CY Initiative. Since September 2021, she has been an associate professor at New York University (NYU) Abu Dhabi, and NYU WIRELESS, NYU Tandon School of Engineering. Her research interests include advanced waveform design, integrated sensing and communication, and machine learning for wireless communications.

She received the IEEE ComSoc Best Young Researcher Award for the Europe Middle East and Africa (EMEA) region, the prize of the best Ph.D. in France in the fields of Signal, Image \& Vision, and she has been nominated in the top 10 Rising Stars in Computer Networking and Communications by N2Women in 2020.  She served as Associate Editor at IEEE Communications Letters 2019-2021, where she received the Best Editor Award in 2020. Between 2018 and 2021, she was the research lead of the Women in AI organization. She is currently Associate Editor at IEEE Transactions on Communications, serving as vice-chair of the IEEE ComSoc ETI on Machine Learning for Communications and leading the Education working group of the IEEE ComSoc ETI on Integrated Sensing and Communications.
\end{IEEEbiography}

\begin{IEEEbiography}[{\includegraphics[width=1in,height=1.25in,clip,keepaspectratio]{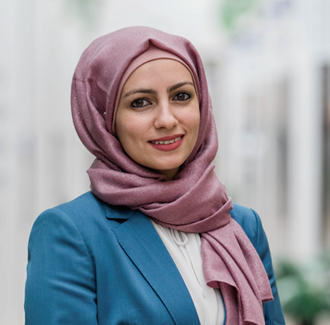}}]{Lina Barieh}
(Senior Member, IEEE) received the M.Sc. and Ph.D. degrees in communications engineering from Khalifa University, Abu Dhabi, UAE, in 2015 and 2018, respectively. She was a Visiting Researcher with the Department of Systems and Computer Engineering, Carleton University, Ottawa, ON, Canada, in 2019, and an affiliate research fellow, James Watt School of Engineering, University of Glasgow, UK. She is currently a Senior Researcher at the technology Innovation institute, a visiting research scientist at Khalifa University, and an affiliate researcher in the University at Albany, USA. Dr. Bariah is a senior member of the IEEE, IEEE Communications Society, IEEE Vehicular Technology Society, and IEEE Women in Engineering. She is currently an Associate Editor for the IEEE Communication Letters, an Associate Editor for the IEEE Open Journal of the Communications Society, and an Area Editor for Physical Communication (Elsevier). She is a Guest Editor in IEEE Network Magazine, and the RS Open Journal on Innovative Communication Technologies (RS-OJICT). Dr. Bariah was a member of the technical program committee of a number of IEEE conferences, such as ICC and Globecom. She is currently organizing/chairing a number of workshops. She serves as a session chair and an active reviewer for numerous IEEE conferences and journals.
\end{IEEEbiography}

\begin{IEEEbiography}[{\includegraphics[width=1in,height=1.25in,clip,keepaspectratio]{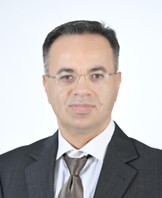}}]{Sami Muhaidat}
(S’01–M’07–SM’11) received the Ph.D. degree in Electrical and Computer Engineering from the University of Waterloo, Waterloo, Ontario, in 2006. From 2007 to 2008, he was an NSERC Postdoctoral Fellow in the Department of Electrical and Computer Engineering, University of Toronto, Canada. From 2008-2012, he was an Assistant Professor in the School of Engineering Science, Simon Fraser University, BC, Canada. He is currently a Professor at Khalifa University and an Adjunct Professor with Carleton University, Ontario, Canada.  Sami's research interests focus on advanced digital signal processing techniques for wireless communications, intelligent surfaces, MIMO, optical communications, massive multiple access techniques, backscatter communications, and machine learning for communications.  He is currently an Area Editor of the IEEE Transactions on Communications, a Guest Editor of the IEEE Network “Native Artificial Intelligence in Integrated Terrestrial and Non-Terrestrial Networks in 6G” special issue, and a Guest Editor of the IEEE OJVT “Recent Advances in Security and Privacy for 6G Networks” special issue.  He served as a Senior Editor and Editor of the IEEE Communications Letters, an Editor of the IEEE Transactions on Communications, and an Associate Editor of the IEEE Transactions on Vehicular Technology. 
\end{IEEEbiography}

\begin{IEEEbiography}[{\includegraphics[width=1in,height=1.25in,clip,keepaspectratio]{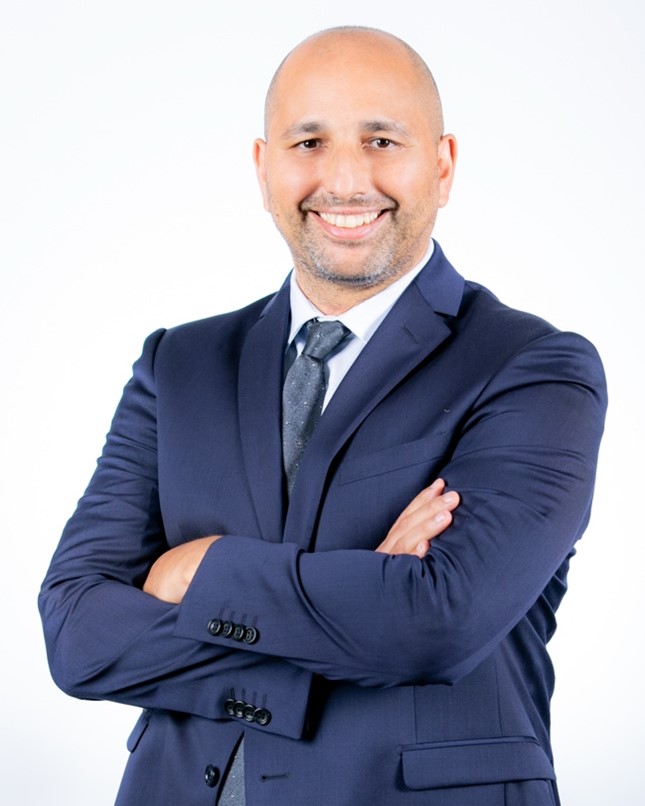}}]{Merouane Debbah}
(Fellow, IEEE) received the M.Sc. and Ph.D. degrees from the Ecole Normale Supérieure Paris-Saclay, France. He was with Motorola Labs, Saclay, France, from 1999 to 2002, and also with the Vienna Research Center for Telecommunications, Vienna, Austria, until 2003. From 2003 to 2007, he was an Assistant Professor with the Mobile Communications Department, Institut Eurecom, Sophia Antipolis, France. In 2007, he was appointed Full Professor at CentraleSupelec, Gif-sur-Yvette, France. From 2007 to 2014, he was the Director of the Alcatel-Lucent Chair on Flexible Radio. From 2014 to 2021, he was Vice-President of the Huawei France Research Center. He was jointly the director of the Mathematical and Algorithmic Sciences Lab as well as the director of the Lagrange Mathematical and Computing Research Center. Since 2021, he is Chief Research Officer at the Technology Innovation Institute in Abu Dhabi. He leads jointly the AI and Telecommunication centers. He has managed 8 EU projects and more than 24 national and international projects. He is an IEEE Fellow, a WWRF Fellow, a Eurasip Fellow, an Institut Louis Bachelier Fellow and a Membre émérite SEE. He was a recipient of the ERC Grant (Advanced Mathematical Tools for Complex Network Engineering) from 2012 to 2017.
\end{IEEEbiography}

\end{document}